\documentclass[a4paper,11pt]{article}
\pdfoutput=1 

\usepackage{jheppub} 

\usepackage{amsmath}
\usepackage{graphicx,tabularx}
\usepackage{url}
\usepackage{footmisc}
\usepackage{amsfonts}
\usepackage{cancel}
\usepackage{color}
\usepackage{multirow}  
\usepackage{amssymb}
\usepackage{pifont}

\usepackage{array}
\newcolumntype{C}[1]{>{\centering\arraybackslash}p{#1}}

\newcommand{\mathsym}[1]{{}}
\newcommand{\unicode}[1]{{}}

\newcommand{\h}{h}
\newcommand{\HH}{H}
\newcommand{\A}{A}
\newcommand{\hpm}{H^\pm}

\title{Anatomy of Exotic Higgs Decays in 2HDM}
\author[a,b]{Felix Kling,}
\author[c]{Jose Miguel No,}
\author[a]{Shufang Su}
\affiliation[a]{Department of Physics, University of Arizona,  Tucson, Arizona  85721, USA}
\affiliation[b]{Fermilab, P.O. Box 500,  Batavia, IL 60510, USA}
\affiliation[c]{Department of Physics and Astronomy,   University of Sussex,  Brington,  BN1 9QH, UK }

\emailAdd{kling@email.arizona.edu}
\emailAdd{J.M.No@sussex.ac.uk}
\emailAdd{shufang@email.arizona.edu}

\preprint{
\begin{flushright}
FERMILAB-PUB-16-093-T
\end{flushright}
}
 
\abstract{Large mass splittings between new scalars in two-Higgs-doublet models (2HDM) open a key avenue to search 
for these  new states via exotic heavy Higgs decays. We discuss in detail the different search channels for these new scalars at the LHC in the presence of 
a sizable mass splitting, {\it i.e.}~a hierarchical 2HDM scenario, taking into account the theoretical and experimental constraints. 
We provide benchmark planes to exploit the complementarity among these searches,
analyzing their potential to probe the hierarchical 2HDM parameter space during LHC Run 2.}
\begin{document}

\titlepage

\maketitle

\newpage


\section{Introduction}
\label{sec:intro}

Analyses of the results from the LHC 7-8 TeV run by both ATLAS and CMS show that the properties of the Higgs particle at $m_h\sim$~125~GeV
are close to those expected for the Standard Model (SM) Higgs boson $h_{\mathrm{SM}}$~\cite{Aad:2015gba,Khachatryan:2014jba}. 
The complete nature of the scalar sector responsible for electroweak (EW) symmetry-breaking,  however,  remains to be determined, and 
it is particularly interesting to ascertain whether the Higgs sector consists of only one $\mathrm{SU(2)}_L$ scalar doublet or
has a richer structure containing additional states. Addressing this question is a key task for present and future studies at the Large Hadron Collider (LHC).

Two Higgs doublet models (2HDM) constitute the prime example of a well-motivated extended Higgs sector, appearing in many extensions 
of the SM such as the MSSM~\cite{Djouadi:2005gj}, composite Higgs models~\cite{Composite} and viable EW baryogenesis 
scenarios~\cite{EWBG}. In addition to the SM-like CP-even Higgs boson, the 2HDM spectrum contains one more CP-even Higgs, a CP-odd Higgs and a pair of charged 
ones\footnote[1]{Here we take the assumption of a CP-conserving 2HDM. In the case of CP-violation, the three neutral Higgses are mixed together to form three mass eigenstates
without definite CP properties.}. 
In recent years, its allowed parameter space has been scrutinized in light of ATLAS/CMS   
Higgs coupling measurements and searches for extra Higgses at the 
LHC~\cite{Celis:2013rcs,Grinstein:2013npa,Coleppa:2013dya,Chen:2013rba,Eberhardt:2013uba,Dumont:2014wha,Bernon:2014vta,Craig:2015jba,Bernon:2015qea,Dorsch:2016tab}. 

A key avenue to probe the 2HDM heavy Higgs bosons at the LHC which has started to attract attention 
recently is the search for exotic decays of the heavy Higgses in the presence of a sizable mass splitting among 
them~\cite{Coleppa:2013xfa,Coleppa:2014hxa,Dorsch:2014qja,Coleppa:2014cca,Li:2015lra,Kling:2015uba} 
(see also~\cite{Dorsch:2016tab}). These sizable splittings are difficult to realize in the MSSM, 
while in more general 2HDM scenarios they may lead to important physical consequences\footnote[2]{{\it e.g.} it 
has been shown in~\cite{Dorsch:2014qja} that sizable mass splittings between the 
2HDM new scalars favour a strong EW phase transition that could lead to baryogenesis.}. While the conventional decay channels of a heavy Higgs 
into two SM quarks, leptons or gauge bosons have been the focus of most of the existing searches, the exotic (non-SM) modes of a heavy Higgs decaying into 
two light Higgses, or one light Higgs with one SM gauge boson quickly dominate once they are kinematically open. The current exclusion bounds on extra Higgses based on 
their conventional decays only will be therefore significantly relaxed. On the other hand, the exotic decay modes offer new discovery channels,    
which have already shown exclusion power during the 8 TeV LHC run~\cite{CMS:2015mba,Khachatryan:2016are}, and yield very promising prospects for the 13 TeV LHC run. 
In this work, we aim to provide a comprehensive categorization and analysis of the exotic search channels for the new 2HDM scalars, 
highlighting the complementarity among them, and provide guiding benchmark planes for Run 2 of the LHC at 13 TeV.

After a review of the 2HDM in Section~\ref{Sec1}, we present the constraints on the 2HDM parameter space coming from theoretical considerations 
(stability of the EW minimum, perturbativity and tree-level unitarity) and experimental measurements in Section~\ref{sec:constraints}, where we also 
introduce the salient features of our benchmark scenarios for exotic 2HDM Higgs decays (Section~\ref{Sec3_summary}) motivated 
by the theoretical and experimental constraints. In Section~\ref{sec:pro_decay} we discuss the production and decay of non-SM Higgses at the LHC, 
and then analyze in depth our different benchmark scenarios in Section~\ref{sec:bp}, before concluding in Section~\ref{sec:conclusion}.

\section{Two Higgs Doublet Models: A Review}
\label{Sec1}

\subsection{2HDM Lagrangian and Higgs Potential}
 
In the 2HDM, we introduce two ${\rm SU}(2)_L$ doublets $\Phi_{i}$ ($i=1,2$):
\begin{equation}
\Phi_{i}=\begin{pmatrix} 
  \phi_i^{+}    \\ 
  (v_i+\phi^{0}_i+i\varphi_i)/\sqrt{2}  
\end{pmatrix},
\label{eq:doublet}
\end{equation}
where $v_i$ are the vacuum expectation values (vev) of the neutral components, satisfying $v_1^2+v_2^2= v^2$, with $v=$ 246 {\rm GeV}. The 
ratio of vevs is defined as $\tan\beta\equiv v_2/v_1$. The 2HDM Lagrangian for $\Phi_i$ can be written as
\begin{equation}
 \mathcal{L} =   \sum_i |D_\mu \Phi_i|^2 - V(\Phi_1,\Phi_2) + \mathcal{L}_{\mathrm{Yuk}}, 
 \label{eq:lagrangian}
\end{equation}
where the first term denotes the kinetic term for the two Higgs doublets, $V(\Phi_1, \Phi_2)$ is the Higgs potential and the 
last term denotes the Yukawa interactions between $\Phi_i$ and the SM fermions. 
Assuming CP conservation and a soft $\mathbb{Z}_2$ symmetry breaking, the 2HDM Higgs potential can be written down as\footnote[3]{The most general scalar potential 
also contains the terms $\left[ \lambda_6(\Phi^\dagger_1 \Phi_1)+\lambda_7 (\Phi^\dagger_2 \Phi_2)  \right](\Phi^\dagger_1 \Phi_2 )+\textrm{h.c.}$ (leading to 
potentially dangerous flavour changing neutral currents), which can however be forbidden by imposing a $\mathbb{Z}_2$ symmetry, softly broken by the $m^2_{12}$ term.}:
\begin{equation}
\begin{aligned}
V(\Phi_1, \Phi_2)&=  m^2 _{11} \Phi^\dagger _1 \Phi_1 + m^2 _{22} \Phi^\dagger _2 \Phi_2 - m^2 _{12}( \Phi^\dagger _1 \Phi_2 + \mathrm{h.c.}) 
+ \frac{\lambda_1}{2} (\Phi^\dagger _1 \Phi_1)^2 + \frac{\lambda_2}{2} (\Phi^\dagger _2 \Phi_2 )^2\\
&   + \lambda_3 (\Phi^\dagger _1 \Phi_1) (\Phi^\dagger _2 \Phi_2) + 
\lambda_4 (\Phi^\dagger_1 \Phi_2 ) (\Phi^\dagger _2 \Phi_1)  + \frac{1}{2} \left[  \lambda_5(\Phi^\dagger _1 \Phi_2 )^2 + \mathrm{h.c.} \right] \, .\\
\end{aligned}
\label{eq:potential}
\end{equation}
After EW symmetry breaking, the physical 2HDM scalar spectrum consists of five states: two CP-even Higgses $\h$, $\HH$ with $m_h<m_{H}$, 
a CP-odd scalar $\A$ and a charged scalar pair $\hpm$~\cite{Gunion:1989we}, which may be written as
\begin{eqnarray}
\label{eq:fields}
\left(\begin{array}{c}
H\\
h
\end{array}\right) = 
\left(
\begin{array}{cc}
c_{\alpha}& s_{\alpha}\\
-s_{\alpha} & c_{\alpha}
\end{array}
\right)
\,\,
\left(\begin{array}{c}
\phi_1^0\\
\phi_2^0
\end{array}\right), \quad\quad \quad \quad \quad \quad \quad\nonumber \\
\\
\left(\begin{array}{c}
G\\
A
\end{array}\right) = 
\left(
\begin{array}{cc}
c_{\beta}& s_{\beta}\\
-s_{\beta} & c_{\beta}
\end{array}
\right)
\,\,
\left(\begin{array}{c}
\varphi_1\\
\varphi_2
\end{array}\right)\,\,,\quad  
\left(\begin{array}{c}
G^{\pm}\\
H^{\pm}
\end{array}\right) = 
\left(
\begin{array}{cc}
c_{\beta}& s_{\beta}\\
-s_{\beta} & c_{\beta}
\end{array}
\right)
\,\,
\left(\begin{array}{c}
\phi^{\pm}_1\\
\phi^{\pm}_2
\end{array}\right), \nonumber
\end{eqnarray}
with the angle $\alpha$ parametrizing the mixing between the neutral CP-even components (we use the shorthand notation 
$s_{x}\equiv\sin\, x$, $c_{x}\equiv\cos\,x$, $t_{x} \equiv \tan \,x $). 
The Goldstone bosons $G$ and $G^\pm$ are absorbed as longitudinal components of the $Z$ and $W^\pm$ bosons. In the limit $c_{\beta-\alpha} = 0$ 
(the {\it alignment limit} for $h$), the state $h$ can be identified with the SM Higgs, 
its couplings to fermions and gauge bosons being precisely those predicted by the SM\footnote[4]{We note that if the heavier neutral 
CP even Higgs $H$ is identified with the observed 125 GeV SM-like Higgs, the alignment limit is instead described by $s_{\beta-\alpha}=0$~\cite{Bernon:2015wef}.}. 
It is thus convenient to describe the model in terms of 
$t_\beta,\, c_{\beta-\alpha}$, the physical scalar masses $m_h, m_H, m_A, m_{H^\pm}$, the soft $\mathbb{Z}_2$ symmetry breaking parameter $m_{12}^2$ and the 
vev $v$.
The quartic couplings in Eq.~(\ref{eq:potential}) can be expressed in terms of the physical masses and mixing angles as (see e.g.~\cite{Gunion:2002zf})
\begin{eqnarray}
 &&v^2 \lambda_1  = \frac{m_H^2 c_\alpha^2 + m_h^2 s_\alpha^2 - m_{12}^2 t_\beta}{ c_\beta^2}, \ \ \ 
v^2 \lambda_2 = \frac{m_H^2 s_\alpha^2 + m_h^2 c_\alpha^2 - m_{12}^2 t_\beta^{-1}}{s_\beta^2},  \nonumber \\  
&&v^2 \lambda_3 =  \frac{(m_H^2-m_h^2) s_\alpha c_\alpha + 2 m_{H^{\pm}}^2 s_\beta c_\beta - m_{12}^2}{ s_\beta c_\beta }, \ \ \ 
v^2 \lambda_4 = \frac{(m_A^2-2m_{H^{\pm}}^2) s_\beta c_\beta + m_{12}^2}{ s_\beta c_\beta },  \nonumber \\
 &&v^2 \lambda_5=  \frac{ - m_A^2 s_\beta c_\beta  + m_{12}^2}{ s_\beta c_\beta }\,. 
 \label{eq:lambdas}
\end{eqnarray}

\subsection{Interactions in the 2HDM}

The couplings of the CP-even scalars to a pair of gauge bosons, arising from the Higgs kinetic term in Eq.~(\ref{eq:lagrangian}), are~\cite{Gunion:1989we}
\begin{equation}
\begin{aligned}
g_{hZZ} = \frac{2i m_Z^2}{v} s_{\beta-\alpha},\ \  g_{HZZ} = \frac{2i  m_Z^2}{v} c_{\beta-\alpha}, \ \ 
g_{hWW} = \frac{2i  m_W^2}{v} s_{\beta-\alpha},\ \  g_{HWW} = \frac{2i   m_W^2}{v} c_{\beta-\alpha}.
\end{aligned}
\label{eq:HVVcoupling}
\end{equation}
The CP-odd scalar $A$ does not couple to pairs of vector bosons, while the charged scalar $H^{\pm}$ only couples to pair of vector bosons at loop level. 
In addition, the couplings of two scalars and one vector boson read
\begin{equation}
\begin{aligned}
g_{hAZ} 			&=\phantom{-} \frac{\;m_Z\;}{v} c_{\beta-\alpha} (p_A^\mu-p_h^\mu),\ \ \ \ \ \  \ \ \ \,
g_{HAZ} 			= -\frac{\;m_Z\;}{v} s_{\beta-\alpha} (p_A^\mu-p_H^\mu), \\
g_{hH^\pm W^\mp} 	&= \pm \frac{i m_W}{v} c_{\beta-\alpha} (p_{H^+}^\mu-p_h^\mu),\ \ \ 
g_{HH^\pm W^\mp} 	= \mp \frac{im_W}{v} s_{\beta-\alpha} (p_{H^+}^\mu-p_H^\mu),\\
g_{AH^\pm W^\mp} 	&= \phantom{-} \frac{\;m_W}{v}  (p_{H^+}^\mu-p_A^\mu), \ \ \ 				 \\
\end{aligned}
\label{eq:HHVcoupling}
\end{equation}
in which $p^\mu$ are the outgoing momentum for the corresponding particle. The $hHZ$-coupling is absent due to CP conservation.
We note that, considering $h$ ($H$) to be the SM-like 125 GeV Higgs with $c_{\beta-\alpha}=0$ ($s_{\beta-\alpha}=0$), gauge boson couplings to two non-SM like Higgses 
are unsuppressed, while the gauge boson couplings to $h$ ($H$) and one non-SM like Higgs are suppressed by $c_{\beta-\alpha}$ ($s_{\beta-\alpha}$).

\vspace{2mm}

Regarding the cubic couplings among scalars arising from the 2HDM scalar potential Eq.~(\ref{eq:potential}), the relevant ones for our analysis are
\begin{eqnarray}
g_{Hhh} &=& -\frac{1}{4\,s_{2\beta}\,v} \Big(  \frac{4m_{12}^2}{s_\beta c_\beta} (c_{\beta-\alpha}^2 s_{\beta+\alpha} -
2 s_{\beta-\alpha} c_{\beta-\alpha} c_{\beta+\alpha})-(2 m_h^2+m_H^2) (s_{3\alpha-\beta}+s_{\alpha+\beta})\Big ), \nonumber \\
\label{eq:HHH coupling}
g_{HAA} &=&-\frac{1}{4\,s_{2\beta}\,v}  \Big ( \frac{4m_{12}^2}{s_\beta c_\beta} s_{\beta+\alpha} - 
8 m_A^2 c_{\beta-\alpha}s_\beta c_\beta- m_H^2 (s_{\alpha-3\beta}+3s_{\alpha+\beta}) \Big ),  \nonumber \\
g_{HH^+H^-}&=&- \frac{1}{4\,s_{2\beta}\,v} \Big ( \frac{4m_{12}^2}{s_\beta c_\beta} s_{\beta+\alpha} - 
8 m_{H^\pm}^2 c_{\beta-\alpha}s_\beta c_\beta- m_H^2 (s_{\alpha-3\beta}+3s_{\alpha+\beta}) \Big ),  
\end{eqnarray}
which could mediate decays with $\HH$ being the parent scalar: $\HH \to \h\h$, $H \to \A\A$ and $H \to H^{+} H^{-}$. As seen directly from Eq.~(\ref{eq:HHH coupling}), 
these couplings depend not only on the mass spectrum, but also on the soft $\mathbb{Z}_2$ symmetry breaking 
term $m_{12}^2$ (we note here that the couplings shown in \cite{Gunion:1989we} assume the MSSM relation $m_{12}^2 = m_A^2 s_\beta c_\beta$).
We also stress that for a light CP-odd scalar $A$ with $m_{A}<m_h/2$, the decay channel $h\rightarrow AA$ 
could be open, being however very constrained experimentally\footnote[5]{The possibility of a light charged scalar with $m_{H^{\pm}}<m_h/2$ 
has been ruled out experimentally by LEP, which puts a lower bound $m_{H^{\pm}} > 80$ GeV for Type II ($m_{H^{\pm}} > 72$ GeV for Type I) 2HDM~\cite{Abbiendi:2013hk}, 
thus forbidding the decay $h\rightarrow H^+H^-$.} (see~\cite{Bernon:2014nxa} for a discussion of this region of the 2HDM parameter space). 

\begin{table}[h!]
\vspace{4mm}
\begin{center}
\begin{tabular}{|c|c|c|}
\hline
State & Up-type fermions & Down-type fermions \cr
\hline
 $h$ & $c_\alpha/s_\beta=s_{\beta-\alpha} +c_{\beta-\alpha}/t_\beta$ & $-s_\alpha/c_\beta=s_{\beta-\alpha} -c_{\beta-\alpha}\ t_\beta$   \cr
\hline
 $H$ & $s_\alpha/s_\beta=c_{\beta-\alpha}-s_{\beta-\alpha}/t_\beta$ & \phantom{$-$}$c_\alpha/c_\beta=c_{\beta-\alpha}+ s_{\beta-\alpha}\ t_\beta$ \cr
\hline
 $A$ & $1/t_\beta$  & $t_\beta$ \cr
\hline 
\end{tabular}
\end{center}
\caption{Tree-level couplings to up-type fermions and down-type fermions normalized to their SM values for $h,\ H$ and $A$ in the Type~II 2HDM.   }
\label{table:couplings}
\end{table}

Finally, as is well-known the couplings of the 2HDM scalars to SM fermions, contained in $\mathcal{L}_{\mathrm{Yuk}}$ in Eq.~(\ref{eq:lagrangian}) 
are not univocally determined by the gauge structure of the model. In the presence of a $\mathbb{Z}_2$ symmetry guaranteeing the absence of tree-level 
flavour changing neutral currents~\cite{ Glashow:1976nt}, four possible 2HDM types exist (see~\cite{Branco:2011iw} for a discussion).
The couplings of the neutral scalar states to SM fermions, normalized to their SM values, can be expressed in terms of functions of $\alpha$ and $\beta$, 
shown in Table~\ref{table:couplings} for the particular case of a Type II 2HDM (one Higgs doublet $\Phi_2$ couples to the up-type quarks, while the other Higgs doublet 
$\Phi_1$ couples to the down-type quarks and leptons).

\subsection{The Alignment Limit and the Role of $m_{12}^2$}

It is useful to cast the relations between the quartic couplings and the physical masses Eq.~(\ref{eq:lambdas}) in terms of $c_{\beta-\alpha}$, which 
characterizes the departure from the alignment limit for $h$ 
\begin{eqnarray}
v^2 \lambda_1 &=&  m_h^2 - \frac{t_\beta\,(m_{12}^2 -m_H^2  s_\beta c_\beta ) }{ c_\beta^2}   + 
(m_h^2 - m_H^2)\left[c_{\beta-\alpha}^2(t_{\beta}^2-1)-2t_{\beta}s_{\beta-\alpha}c_{\beta-\alpha}\right],\nonumber \\
v^2 \lambda_2 &=& m_h^2 - \frac{ (m_{12}^2 -m_H^2  s_\beta c_\beta) }{ t_\beta s_\beta^2 }  +
(m_h^2 - m_H^2)\left[c_{\beta-\alpha}^2(t_{\beta}^{-2}-1)+2t^{-1}_{\beta}s_{\beta-\alpha}c_{\beta-\alpha}\right],\nonumber\\
\label{eq:alignment_lambda}
v^2 \lambda_3 &=&  m_h^2 + 2 m_{H^{\pm}}^2 - 2m_H^2 -  \frac{(m_{12}^2 -m_H^2  s_\beta c_\beta)}{  s_\beta c_\beta } -
(m_h^2 - m_H^2)\left[2c_{\beta-\alpha}^2+s_{\beta-\alpha}c_{\beta-\alpha}(t_{\beta}-t_{\beta}^{-1})\right],\nonumber\\
v^2 \lambda_4 &=&  m_A^2-  2 m_{H^{\pm}}^2 + m_H^2+  \frac{ (m_{12}^2 -m_H^2  s_\beta c_\beta)}{  s_\beta c_\beta }\,,\nonumber\\
v^2 \lambda_5 &=&  m_H^2 - m_A^2+  \frac{ (m_{12}^2 -m_H^2  s_\beta c_\beta)}{  s_\beta c_\beta } \,. 
\end{eqnarray}
Current data from LHC Run 1 favour the alignment limit $c_{\beta-\alpha}=0$~\cite{Aad:2015pla} 
(see also~\cite{Celis:2013rcs,Grinstein:2013npa,Coleppa:2013dya,Eberhardt:2013uba,Dumont:2014wha,Bernon:2014vta,Bernon:2015qea}).
For a Type II 2HDM the only other allowed possibility 
is the {\it wrong-sign} scenario~\cite{Ferreira:2014naa} $s_{\beta+\alpha} \simeq 1$ (compatible with measurements of Higgs signal strengths for $t_{\beta} > 3$).
For $c_{\beta-\alpha}=0$,  the relations Eq.~(\ref{eq:alignment_lambda}) simply become
\begin{eqnarray}
v^2 \lambda_1 &=&  m_h^2 - \frac{t_\beta\,(m_{12}^2 -m_H^2  s_\beta c_\beta ) }{ c_\beta^2}\,,\nonumber \\
v^2 \lambda_2 &=& m_h^2 - \frac{ (m_{12}^2 -m_H^2  s_\beta c_\beta) }{ t_\beta s_\beta^2 }\,,\nonumber\\
\label{eq:alignment2_lambda}
v^2 \lambda_3 &=&  m_h^2 + 2 m_{H^{\pm}}^2 - 2m_H^2 -  \frac{(m_{12}^2 -m_H^2  s_\beta c_\beta)}{  s_\beta c_\beta }\,,\nonumber\\
v^2 \lambda_4 &=&  m_A^2-  2 m_{H^{\pm}}^2 + m_H^2+  \frac{ (m_{12}^2 -m_H^2  s_\beta c_\beta)}{  s_\beta c_\beta }\,,\nonumber\\
v^2 \lambda_5 &=&  m_H^2 - m_A^2+  \frac{ (m_{12}^2 -m_H^2  s_\beta c_\beta)}{  s_\beta c_\beta } \,. 
\end{eqnarray}
The combination $m_{12}^2 -m_H^2  s_\beta c_\beta$ in Eq.~(\ref{eq:alignment2_lambda}) will play a key role in the following discussion: 
the value of $m^2_{12}$ is not fixed by the mass spectrum or the scalar couplings to gauge bosons and fermions, only entering the trilinear scalar couplings 
Eq.~(\ref{eq:HHH coupling}). Its possible allowed values are dictated by theoretical constraints on the 2HDM parameter space, namely the boundedness from 
below of the scalar potential  Eq.~(\ref{eq:potential}) and the stability of the EW minimum, and the requirements of perturbativity and tree-level unitarity 
on the quartic couplings $\lambda_i$, 
as shown in the next section. These have a large impact on the allowed values of masses $m_{H}$, $m_{A}$, $m_{H^{\pm}}$, $m_{12}^2$  and $t_{\beta}$ 
(and $c_{\beta-\alpha}$ away from alignment), 
as the absence of a value of $m^2_{12}$ satisfying the theoretical constraints for a given set of values for $m_{H}$, $m_{A}$, $m_{H^{\pm}}$ and $t_{\beta}$, indicates 
that such set of values is not physically viable (see e.g.~\cite{Dorsch:2016tab}).

\section{2HDM Theoretical and Experimental Constraints}
\label{sec:constraints}

\subsection{Vacuum Stability}
\label{sec:stability}

\begin{figure}[h!]
\centering
\includegraphics[width=0.495\textwidth]{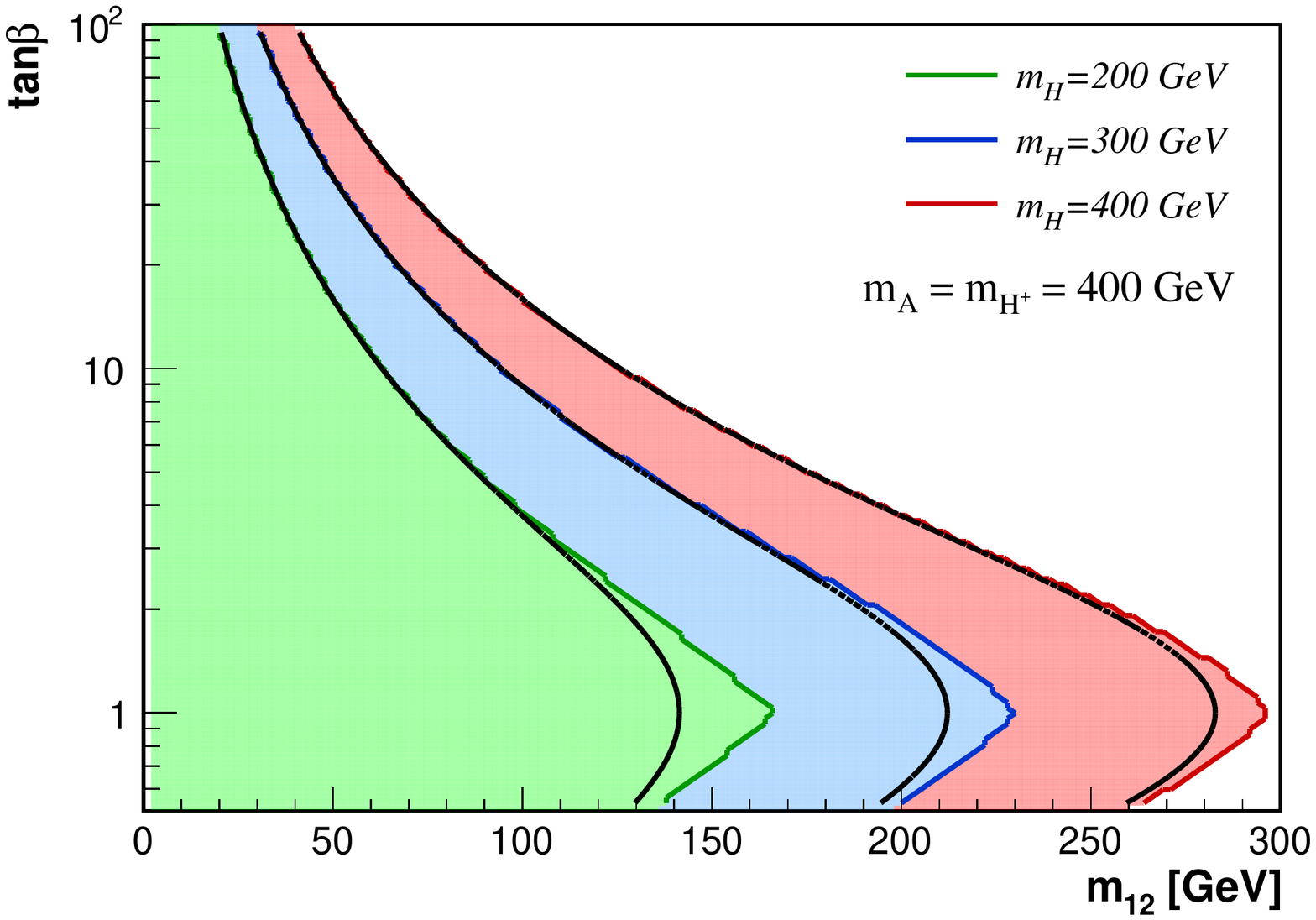}
\includegraphics[width=0.495\textwidth]{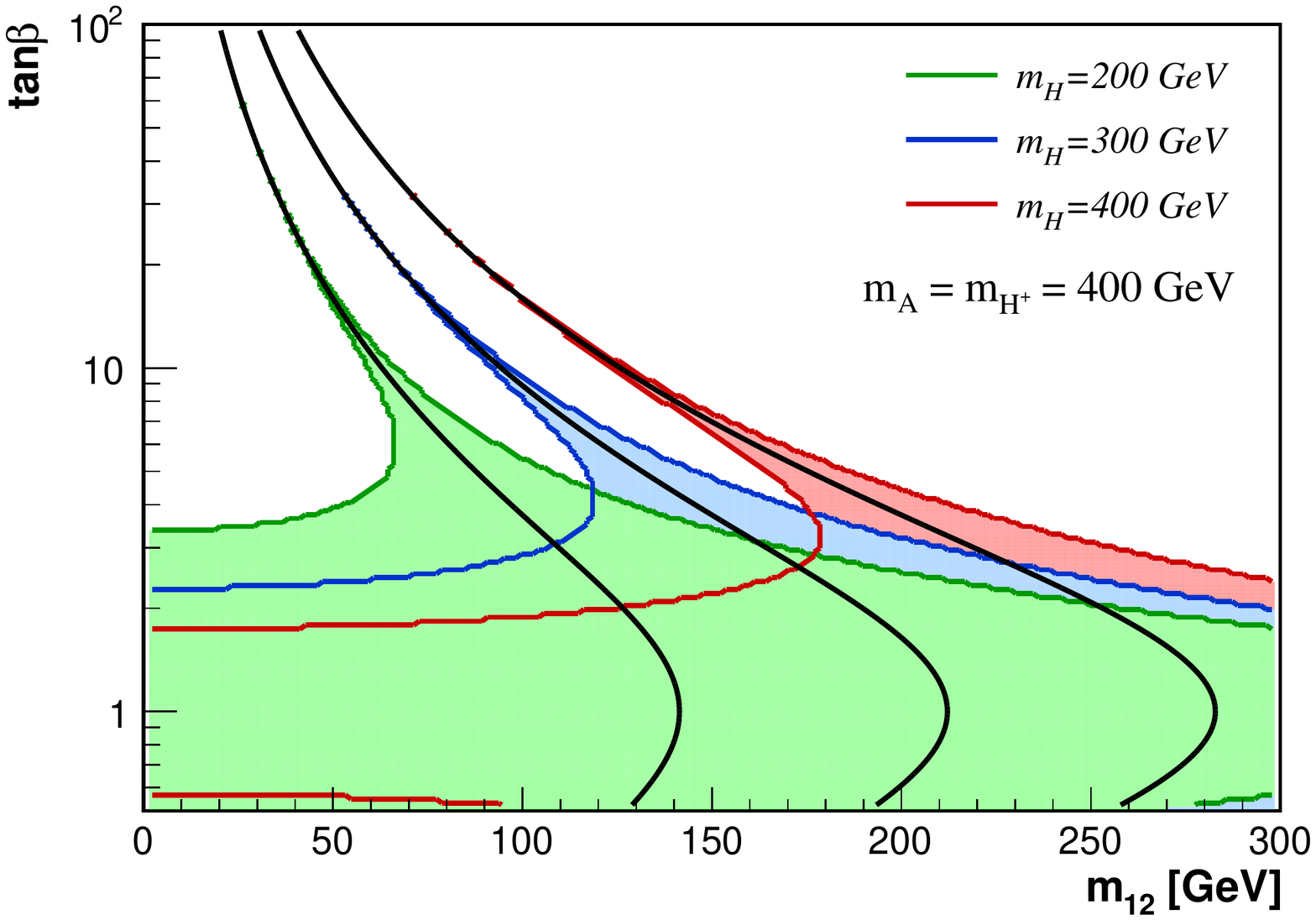}

\vspace{-2mm}
\caption{Allowed region in the ($m_{12}$, $t_\beta$) plane from vacuum stability (left panel) and unitarity $|\Lambda_i|<8\pi$ (right panel) for $m_H$ = 400 GeV (red), 300 GeV (blue) 
and 200 GeV (green), assuming $c_{\beta-\alpha} = 0$ and $m_{A} = m_{H^{\pm}} =$ 400 GeV. The black lines denote the relation $m_{12}^2 = m_H^2 s_\beta c_\beta $.}
\label{fig:perturbativity}
\end{figure}

In order to have a stable vacuum, the following conditions need to be fulfilled~\cite{Gunion:2002zf}
\begin{equation}
 \lambda_1>0\,, \quad \,\, \lambda_2>0\,,  \quad \,\,
\lambda_3>-\sqrt{\lambda_1 \lambda_2}\,, \quad\,\, \lambda_3+\lambda_4-|\lambda_5|>-\sqrt{\lambda_1 \lambda_2}\,.
 \label{eq:vacstability}
\end{equation}
For $c_{\beta-\alpha}=0$, satisfying the first two conditions requires $m_{12}^2- m_{H}^2  s_\beta c_\beta \lesssim 0$ for 
either $t_{\beta} > 1$ or $t_{\beta} < 1$, as seen from Eq.~(\ref{eq:alignment2_lambda}). Moreover, 
Eq.~(\ref{eq:alignment_lambda}) shows that a departure from alignment generically has a negative impact on the first two stability
conditions. Focusing on the alignment limit, the first two requirements are automatically satisfied for $m_{12}^2- m_{H}^2  s_\beta c_\beta = 0$, with the last two given by
\begin{equation}
m_h^2 + m_{H^{\pm}}^2 - m_H^2 > 0 \quad , \quad \quad m_h^2 + m_{A}^2 - m_H^2  > 0\,.
\label{eq:case1_stab}
\end{equation}
This implies that for $m_H>m_A, m_{H^\pm}$, the mass splittings between the heavy CP-even Higgs $H$ and the other heavy scalars $A$ and $H^{\pm}$ have to be small,  
such that the decays of $\HH$ into $A Z$, $A A$, $H^+ H^{-}$ or $H^\pm W^{\mp}$ are not kinematically allowed. 
For $m^2_{12} = 0$ all four stability conditions of Eq.~(\ref{eq:vacstability}) are automatically fulfilled. 
The allowed region in the $m_{12}$ vs. $t_\beta$ plane is shown in the left panel of Figure~\ref{fig:perturbativity} for 
$m_A=m_{H^\pm}=400$ GeV and $m_H=200,\ 300,\ 400$ GeV as an illustration. As seen from Figure~\ref{fig:perturbativity}, the regions $m_{12}^2 < m_{H}^2 s_\beta c_\beta$ are 
generically allowed by the vacuum stability requirement.

\subsection{Perturbativity and Unitarity}

Upon imposing the perturbativity condition $|\lambda_i|\leq 4\pi$, the strongest constraints in the alignment limit come respectively from 
$v^2\lambda_1\sim t_\beta^3 (m_{12}^2 - m_H^2 s_\beta c_\beta)$ for $t_\beta \gg 1$ and $v^2\lambda_2\sim t_\beta^{-3} (m_{12}^2 - m_H^2 s_\beta c_\beta)$ 
for $t_\beta \ll 1$. Thus, perturbativity requires $\left|m_{12}^2 - m_H^2 s_\beta c_\beta\right| \lesssim v^2$ unless 
$t_{\beta} \sim 1$. Moreover, even for $m_{12}^2 = m_H^2 s_\beta c_\beta$, perturbativity of $\lambda_{3-5}$ imposes constraints on the size of the mass splittings among 
the new scalars.

Even stronger constraints are found when requiring tree-level unitarity\footnote[6]{An analysis of unitarity constraints at one-loop level has been performed in \cite{Grinstein:2015rtl}.} 
of the scattering matrix in the 2HDM scalar sector~\cite{Ginzburg:2005dt}. The eigenvalues of the scattering matrix read
\begin{eqnarray}
\label{eq:unitarity1}
\Lambda_{1,2} &=& \lambda_3 \pm \lambda_4,\nonumber \\  
\Lambda_{3,4} &=& \lambda_3 \pm \lambda_5,\nonumber \\
\Lambda_{5,6} &=& \lambda_3 +2\lambda_4\pm 3 \lambda_5, \nonumber\\
\Lambda_{7,8} &=&\frac{1}{2} \left(\lambda_1+\lambda_2 \pm \sqrt{ (\lambda_1-\lambda_2)^2 + 4 \lambda_4^2} \right ),\nonumber\\
\Lambda_{9,10} &=&\frac{1}{2} \left(\lambda_1+\lambda_2 \pm \sqrt{ (\lambda_1-\lambda_2)^2 + 4 |\lambda_5|^2} \right),\nonumber \\
\Lambda_{11,12} &=&\frac{1}{2} \left(3(\lambda_1+\lambda_2) \pm \sqrt{ 9(\lambda_1-\lambda_2)^2 + 4 (2\lambda_3+\lambda_4)^2} \right ).\end{eqnarray}
Performing a partial-wave expansion of the scattering amplitudes yields limits 
on the partial wave amplitudes, which for the $J=0$ case translate into the constraint 
$\left|\mathrm{Re}(\Lambda_i)\right| < 8 \pi$ (see {\it e.g.}~\cite{Grinstein:2015rtl}), which we consider here.
A quick inspection of Eq.~(\ref{eq:unitarity1}) shows that for $t_{\beta} \gg 1$ the scattering matrix eigenvalues scale as
$\Lambda_{7,9,11} \sim \lambda_1$ (particularly $\Lambda_{11} \simeq 3\, \lambda_1$), which again imposes 
$\left|m_{12}^2 - m_H^2 s_\beta c_\beta\right| \lesssim v^2$ (and yields an even stronger constraint than the perturbativity one). 
A similar argument follows for $t_{\beta} \ll 1$, this time with $\Lambda_{7,9,11} \sim \lambda_2$.
As a result, $m_{12}^2 \approx m_H^2 s_\beta c_\beta$ is strongly preferred unless $t_{\beta} \sim 1$, as shown explicitly in the right panel 
of Figure~\ref{fig:perturbativity} (for $m_{A} = m_{H^{\pm}}$). 
In the limit $m_{12}^2 = m_H^2 s_\beta c_\beta$,  the scattering matrix eigenvalues from Eq.~(\ref{eq:unitarity1}) become independent of 
$t_{\beta}$ (in alignment $c_{\beta-\alpha} = 0$) and read
\begin{equation}
\label{eq:unitarity2}
 \begin{array}{ll}
 \Lambda_{1(9),10} v^2 = m_h^2 \mp m_H^2 \pm m_A^2, &\quad \Lambda_{2} v^2 =  m_h^2 - 3  m_H^2 - m_A^2 + 4  m_{H^{\pm}}^2, \\
 \Lambda_{3} v^2  =  m_h^2 - m_H^2 - m_A^2 + 2  m_{H^{\pm}}^2, &\quad \Lambda_{4,5}v^2 =  m_h^2 \mp 3  m_H^2 \pm m_A^2 \pm 2 m_{H^{\pm}}^2, \\
 \Lambda_{6}v^2 =  m_h^2 - 3  m_H^2 +  5  m_A^2 - 2   m_{H^{\pm}}^2, &\quad \Lambda_{7,8} v^2 =  m_h^2 \pm m_H^2 \pm m_A^2 \mp 2   m_{H^{\pm}}^2,  \\
 \Lambda_{11}v^2 = 5  m_h^2 - 3\  m_H^2 + m_A^2 +2  m_{H^{\pm}}^2, &\quad  \Lambda_{12}v^2 = m_h^2 + 3   m_H^2 -  m_A^2 - 2   m_{H^{\pm}}^2,
 \end{array}
\end{equation}
such that $\left|\Lambda_i\right| < 8 \pi$ (note that $\Lambda_i$ are real) impose upper limits on the mass splittings (although not on the masses themselves). We also note that for 
$m_{12}^2 = 0$, $\Lambda_{1-6}$ are independent of $t_{\beta}$ (depending only on the scalar masses) while $\Lambda_{7-12}$ do depend on $t_\beta$, which once again results in 
$t_\beta \approx 1$ being the only accessible region for large mass splittings in this case.

\vspace{-2mm}

\subsection{Electroweak Precision Measurements}
\label{sec:delta_rho}

\begin{figure}[h!]
\centering
\includegraphics[width=0.495\textwidth]{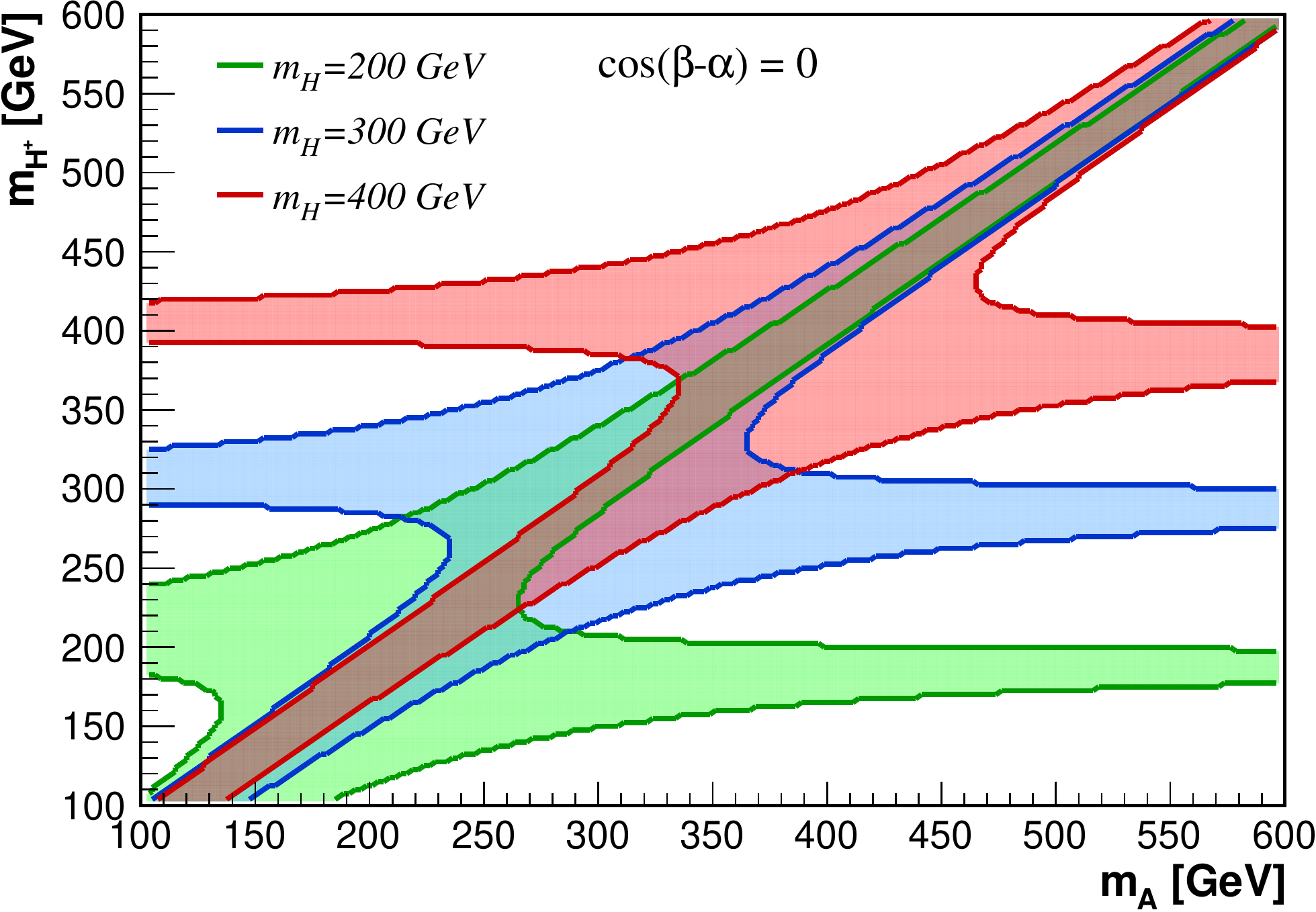}
\includegraphics[width=0.495\textwidth]{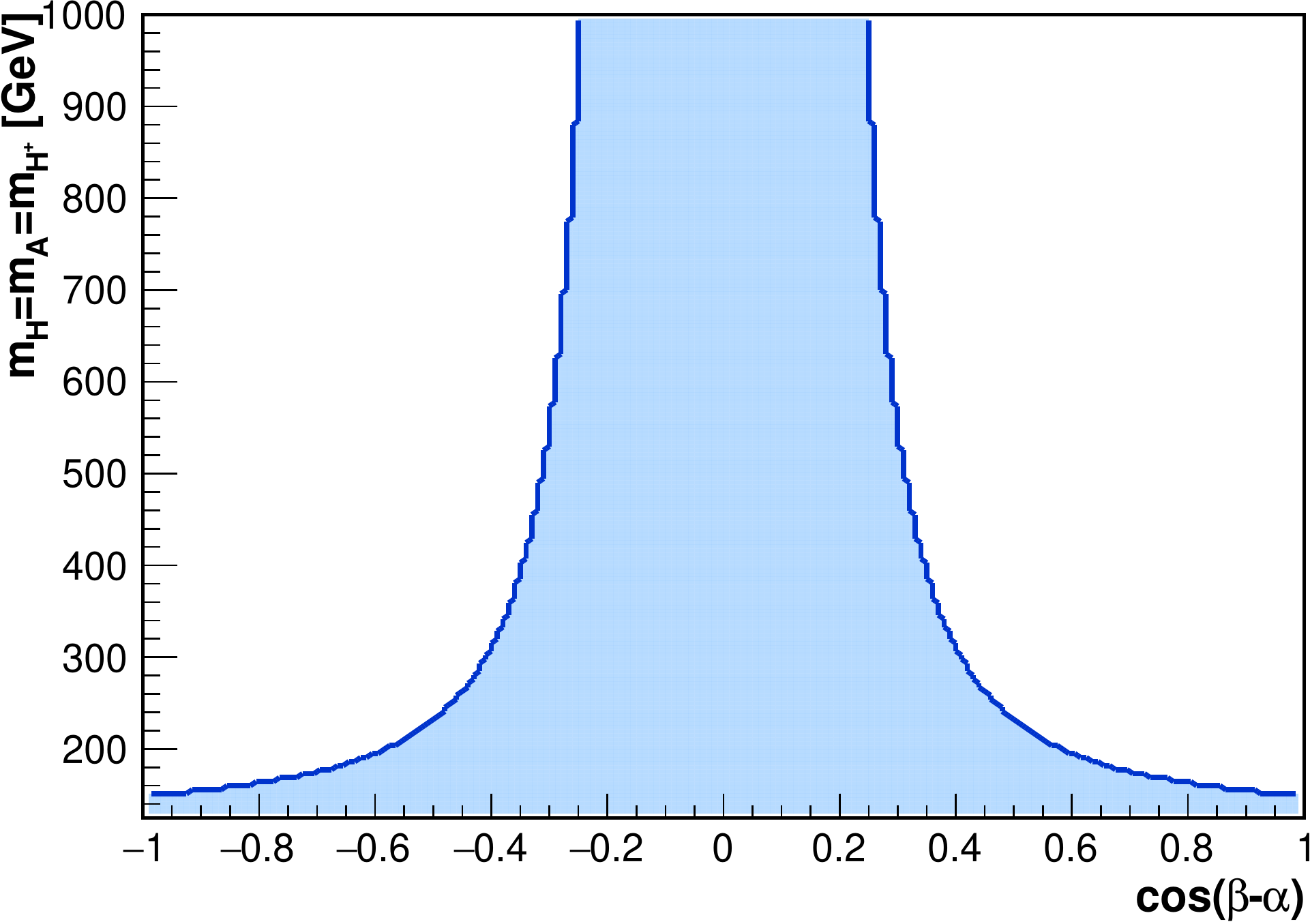}
\caption{Left: 2HDM parameter space in the ($m_A$, $m_{H^\pm}$) plane allowed at 95\% C.L. by $S$ and $T$ measurements~\cite{Baak:2014ora}, for $m_H=400$ GeV (red), 
$m_H = 300$ GeV (blue) and $m_H = 200$ GeV (green), assuming $c_{\beta-\alpha} = 0$. Right:  $S-T$ constraints in the ($c_{\beta-\alpha}$, $m_H$) 
plane for $m_H=m_A=m_{H^\pm}$.}
\label{fig:st}
\end{figure}

Measurements of EW precision observables (EWPO) impose strong constraints on the 2HDM mass spectrum. Adopting the current 95\% C.L. constraints on 
the $S$ and $T$ oblique parameters (with $U=0$)~\cite{Baak:2014ora}, the allowed region of parameter space in the ($m_A$, $m_{H^\pm}$) plane is shown, 
for $c_{\beta-\alpha} = 0$ (neither $t_\beta$ nor $m_{12}^2$ affect $S$ and $T$),
in the left panel of Figure~\ref{fig:st} respectively for $m_H=400$ GeV (red), $m_H=300$ GeV (blue) and $m_H=200$ GeV (green). Satisfying EWPO constraints requires 
the charged scalar mass to be close to one of the heavy neutral scalar masses: $m_{H^{\pm}} \approx m_H$ or $m_{H^{\pm}} \approx m_A$.

Away from the alignment limit, additional contributions to $S$ and $T$ proportional to $c_{\beta-\alpha}$ appear~\cite{Gunion:1989we} (see also~\cite{Gorbahn:2015gxa}), 
such that the scenario $m_H=m_A=m_{H^{\pm}}$ is only allowed for small $|c_{\beta -\alpha}|$ once $m_{H} \gg v$ is realized, as shown in the right panel of 
Figure~\ref{fig:st}. The departure from alignment also allows for mild mass splittings among all the new scalars (e.g. $m_A > m_H + m_Z$ and 
$m_H \gtrsim m_{H^{\pm}} + m_W$), which however does not significantly alter the phenomenology of exotic Higgs decays at the LHC, discussed in detail 
in Section~\ref{sec:bp}.

\subsection{Flavour Constraints}
\label{sec:flavour}

\begin{figure}[h!]
\centering
\includegraphics[width=0.495\textwidth]{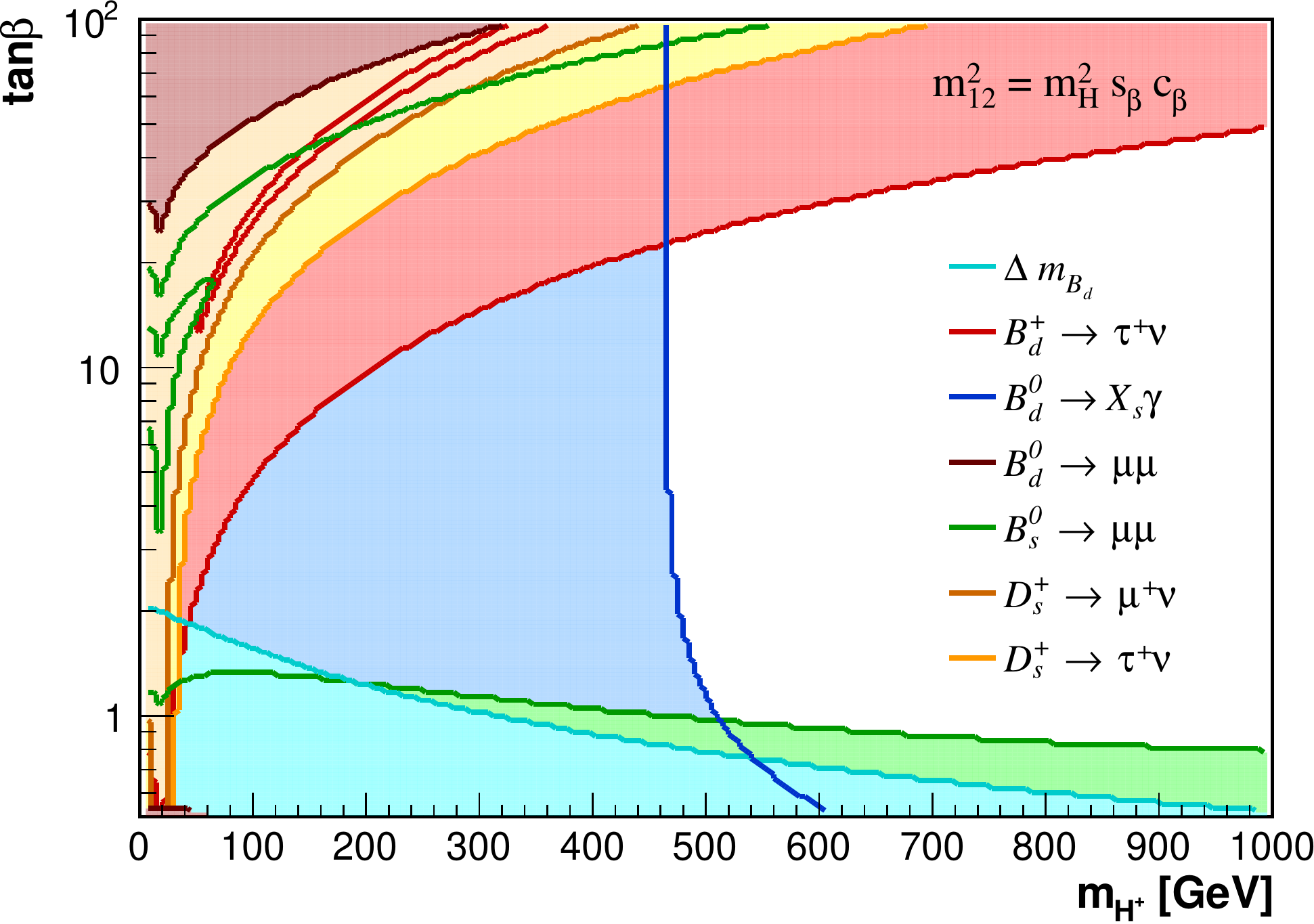}
\includegraphics[width=0.495\textwidth]{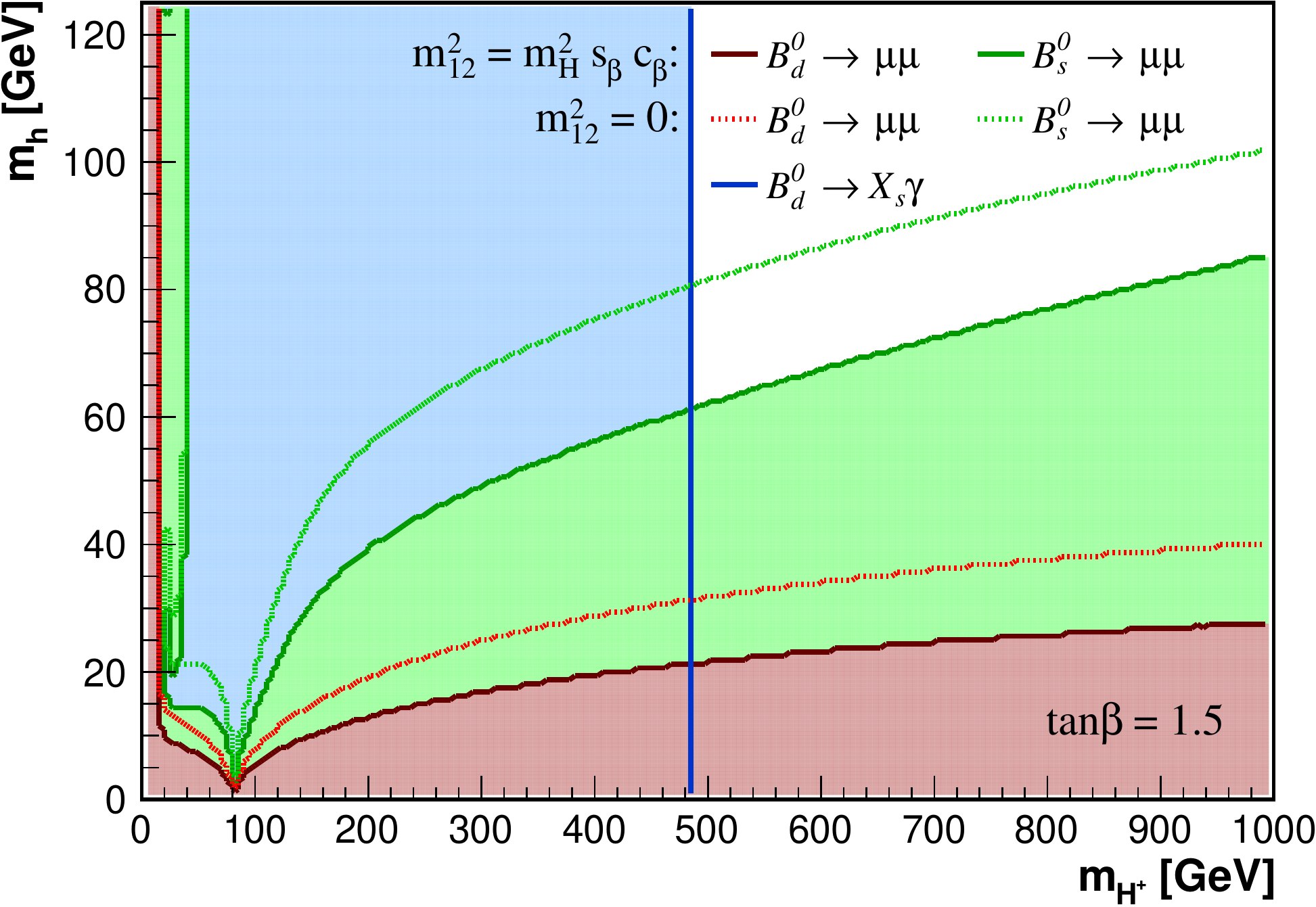}

\vspace{-3mm}
\caption{Type II 2HDM parameter space excluded by flavour constraints (see text for details).}
\label{fig:flavor}
\end{figure}

Various flavour measurements~\cite{Amhis:2014hma} provide indirect constraints on the charged scalar mass $m_{H^{\pm}}$ as a function of 
$t_{\beta}$. The different limits are computed for the case of a Type II 2HDM with {\sc SuperIso}~\cite{Mahmoudi:2008tp,Mahmoudi:2009zz}, 
and shown in the left panel of Figure~\ref{fig:flavor}.
The most stringent of these\footnote[7]{We note here that the recent measurement from the BaBar Collaboration of the ratios of $B \to D^* \tau \nu$ to $B \to D^* \ell \nu$ decays
and $B \to D \tau \nu$ to $B \to D \ell \nu$ decays cannot be accommodated within the Type II 2HDM~\cite{Lees:2012xj}. However, a new measurement of the former ratio 
by the Belle Collaboration~\cite{Huschle:2015rga,Abdesselam:2016cgx} is in tension with this conclusion. Since this matter is not settled yet, we choose not to include 
these flavour measurements in our discussion.} comes from the measurement of the 
branching fraction (BR) of $b\to s\gamma$ ($B^0_d \to X_s \gamma$), which sets a limit $m_{H^{\pm}} > 480$ GeV at 95 \% C.L.~\cite{Misiak:2015xwa} 
(we note that the limit is even stronger for $t_{\beta} < 2$). For large $t_{\beta} \gtrsim 20$, the lower limit on $m_{H^{\pm}}$ set by 
the measurement of the branching fraction $B^+_d \to \tau^+\nu$ is significantly stronger, with $m_{H^{\pm}} \gtrsim 700$ GeV for $t_{\beta} = 30$.
Similarly, the region $t_{\beta} \lesssim 1$ is very strongly constrained by $B^0_s \to \mu\mu$ and $\Delta m_{B_d}$. 

For $m_H =$ 125 GeV and $s_{\beta-\alpha}=0$, when the heavy CP-even scalar $H$ is the SM-like Higgs,  
the mass of the light CP-even Higgs $h$ is constrained by flavour measurements as well, as shown in the right panel of 
Figure~\ref{fig:flavor} for Type II 2HDM. The strongest constraint in this case comes from $B^0_s\rightarrow \mu^+\mu^-$, which can exclude up to $m_h < 100$ GeV 
(the precise bound depending   on $m^2_{12}$ and $t_{\beta}$) for masses $m_{H^{\pm}}$ satisfying the $b\to s\gamma$ constraint. 

\vspace{2mm}

Note that flavour constraints are typically very model dependent. Contributions from additional sectors in the model could relax the constraints, 
as has e.g. been studied in the MSSM framework for $b \rightarrow s \gamma$~\cite{Han:2013mga}. Being mostly focused on the collider aspects of 2HDM Higgses, 
we will not consider flavour as a hard constraint in the following, however indicating its effect on the parameter space under consideration. 


\vspace{-2mm}

\subsection{LHC and LEP Constraints}
\label{sec:LEP_LHC}

We now review the constraints from direct searches of the new scalars. Besides the LEP bound $m_{H^{\pm}} > 80$ GeV (72 GeV) for Type II (I) 
2HDM~\cite{Abbiendi:2013hk}, LEP searches for $e^+e^-\rightarrow A\,H\, (H\to bb/\tau\tau, \,A\to bb/\tau\tau)$ constrain the sum of the masses
$m_A+m_H \gtrsim 209$ GeV~\cite{Schael:2006cr}. At the LHC, the searches for $A/H$ in $bb$-associated production and decaying to 
$\tau\tau$ by ATLAS/CMS~\cite{Aad:2014vgg,Khachatryan:2014wca} constrain the high $t_\beta$ region in  the Type II 2HDM.
Away from alignment, searches by ATLAS/CMS for $H\to W^{+}W^{-}, ZZ$~\cite{ATLAS:2013nma,ATLAS:2014aga,Khachatryan:2015cwa}, 
$A\to h Z \,(h \to b b)$~\cite{Aad:2015wra, Khachatryan:2015lba} and 
$H\to h h \to b b \gamma\gamma, b b b b$~\cite{Aad:2014yja,CMS:2014ipa,Khachatryan:2015yea}
yield strong constraints on the ($c_{\beta-\alpha},\,t_{\beta}$) plane as a function of the respective mass $m_{H}/m_A$
(see e.g.~\cite{Craig:2015jba,Dorsch:2016tab}). 
We however stress that the limits summarized above can be significantly weakened once exotic Higgs decay channels are
open~\cite{Dorsch:2016tab,Coleppa:2013xfa,Coleppa:2014hxa,Coleppa:2014cca,Kling:2015uba}). 
Searches for these new channels, e.g. via $A/H \to H Z /A Z$ are then crucial for probing 2HDM scenarios with large mass splittings among the new 
states (i.e. hierarchical 2HDM scenarios), and there is already ongoing effort by CMS in this direction~\cite{CMS:2015mba,Khachatryan:2016are}.

Finally, ATLAS/CMS searches impose constraints on the charged scalar~\cite{Aad:2014kga,Khachatryan:2015qxa} beyond those of LEP. A light charged scalar
$m_{\hpm} \lesssim m_t$ is mostly excluded by the non-observation of the decay $t \to H^{+} b  \to \tau \nu b$ where the top is produced in top pair production.   
For $m_{H^{\pm}} > m_t$, the current limit is very weak and only constrains the high $t_\beta$ region for $m_{H^{\pm}}$ not much above the top mass 
(see~\cite{Coleppa:2014cca} for a detailed discussion).

\subsection{From Constraints to 2HDM Benchmarks}
\label{Sec3_summary}

The combination of previous constraints provides a key guideline to the design of simplified 2HDM benchmark scenarios for LHC Run 2 searches at 13 TeV.   
EWPO measurements require the mass of the charged scalar to be close to the 
mass of one of the neutral scalars, and so we fix $m_{\hpm} = m_H$ or $m_{\hpm} = m_A$ in the following. In addition, measurements of Higgs signal strengths at the LHC
favour the alignment limit ($c_{\beta-\alpha} = 0$ if $h$ is the 125 GeV SM-like Higgs), particularly 
for Type II 2HDM. We then focus our analysis mostly on the alignment limit, and only consider deviations from alignment when discussing possible decays of the new scalars into 
the SM-like Higgs $h$. 

Regarding the impact of theoretical constraints on the 2HDM parameter space, the previous discussion shows that satisfying unitarity/perturbativity and vacuum stability 
bounds (close to the alignment limit) for arbitrary values of $t_{\beta}$ requires $m_{12}^2 = m_H^2  s_\beta c_\beta$ and 
$m_{H} \lesssim m_A,m_{H^{\pm}}$. Alternatively, stability is satisfied for any mass ordering if $m_{12}^2 = 0$, while unitarity requires in this case a low value of 
$t_{\beta}$. We thus consider these two scenarios as benchmark cases for our analysis:
\begin{itemize}
\item \textbf{Case 1:} $m_{12}^2 =m_H^2  s_\beta c_\beta$\footnote[8]{The authors of \cite{Das:2015mwa} have shown that this case is preferred when requiring the 2HDM to be stable up 
to the Planck scale.}  
	\begin{itemize}
	\item From Eq.~(\ref{eq:case1_stab}), vacuum stability requires $m_H \lesssim m_A$ and $m_H \lesssim m_{H^+}$, and thus
	the exotic decays $H \to A\,Z $ and $H \to H^\pm \,W^\mp$ are not kinematically allowed. 
	
	\item Unitarity requires $|\Lambda_i| < 8\pi$, constraining the mass differences among the new scalar states (but not the absolute mass values). 
	In particular, using Eq.~(\ref{eq:unitarity2}) we obtain the 
	bound $\left|5(m_A^2 - m_H^2) + m_h^2\right| < 8 \pi v^2$ if $m_{\hpm} = m_H$, and the bound $\left|3(m_A^2 - m_H^2) + 5\,m_h^2\right| < 8 \pi  v^2$ if $m_{\hpm} = m_A$. 
	
	\item The cubic scalar couplings Eq.~(\ref{eq:HHH coupling}) now read 
	\begin{equation}
\begin{aligned}
g_{Hhh} 
&=-\frac{c_{\beta-\alpha}}{s_{2\beta}\,v} \left[2\, (m_H^2+m_h^2)\, s_{2\alpha}-  m_H^2\, s_{2\beta}  \right] \\
g_{HAA} 
&= -\frac{c_{\beta-\alpha}}{2\,v} (m_H^2 - 2\,m_A^2)  \\
g_{HH^+H^-}
&= -\frac{c_{\beta-\alpha}}{2\,v}  (m_H^2 - 2\,m_{H^\pm}^2)  
\end{aligned}
\label{eq:case1_coupling}
\end{equation}
In the alignment limit $c_{\beta-\alpha}=0$ all these couplings vanish, and therefore the decays $H \to A A$, $H \to H^+ H^-$ and $H \to h\,h$ are absent 
($H \to A\,A$ and $H \to H^+ H^-$ are also not kinematically allowed for $m_{12}^2 =m_H^2  s_\beta c_\beta$). 

\end{itemize}


\item \textbf{Case 2:} $m_{12}^2=0$ and $t_\beta \sim 1$\footnote[9]{This case has also been analyzed in \cite{Das:2015qva}.} 

\begin{itemize}
	\item Vacuum stability does not constrain the parameter space. In particular $m_H > m_A, \,m_{H^+}$ is now possible, allowing the decays 
	$H \to A\,Z $ and $H \to H^\pm\, W^\mp$ (and potentially also $H \to A A$ and $H \to H^+ H^-$).

	\item  Unitarity imposes an upper bound on the scalar masses (not only on the mass splittings). This bound 
	scales as $t_\beta^{-2}$ for $t_\beta>1$ and as $t_\beta^2$ for $t_\beta<1$, such that only the region $t_{\beta} \sim 1$ is allowed (we recall that in Type II 2HDM, 
	at least one of the neutral scalars needs to be heavy due to the combination of EWPO and flavour constraints).

	\item The cubic scalar couplings Eq.~(\ref{eq:HHH coupling}) now read
		\begin{equation}
\begin{aligned}
g_{Hhh} 
&=\frac{c_{\beta-\alpha}}{2\,s_{2\beta}\,v}  (2 m_h^2+m_H^2)  \left[ (c_{\beta-\alpha}^2 -s_{\beta-\alpha}^2) \,s_{2\beta} - 2\,s_{\beta-\alpha}\,c_{\beta-\alpha}\, c_{2\beta} \right]\\
g_{HAA} 
&= -\frac{1}{2\,s_{2\beta}\,v} \left[ (m_H^2 - 2m_A^2)\, c_{\beta-\alpha}\, s_{2\beta} + 2\, m_H^2 \,s_{\beta-\alpha}\, c_{2\beta} \right]\\
g_{HH^+H^-}
&=  -\frac{1}{2\,s_{2\beta}\,v}  \left[ (m_H^2 - 2m_{H^{\pm}}^2)\, c_{\beta-\alpha}\, s_{2\beta} + 2 \,m_H^2 \,s_{\beta-\alpha}\, c_{2\beta} \right]
\end{aligned}
\label{eq:case2_coupling}
\end{equation}
In the alignment limit the coupling $g_{Hhh}$ vanishes and thus the decay $H \to hh$ is absent. However,
the couplings $g_{HAA}$ and $g_{HH^+H^-}$ are non-vanishing as long as $t_\beta \neq 1$. 
	\end{itemize}
\end{itemize}

For our analysis of benchmark scenarios away from alignment, which focus on the decays of $A$, $H$, $H^{\pm}$ into the SM-like Higgs $h$, we 
consider the same two cases above for consistency (even though these cases are motivated by theoretical constraints for $c_{\beta-\alpha} = 0$).

\section{LHC Production and Decay of 2HDM Higgses}
\label{sec:pro_decay}

We now discuss the salient features of the production and decay of the new 2HDM scalars at the LHC.     
The production of the CP-even and CP-odd neutral scalars $H$, $A$ at the 13 TeV LHC occurs via gluon fusion ($gg\to H/A$) and 
$bb$-associated production. Gluon fusion is the dominant production mechanism for small and moderate values of $t_{\beta}$, while for 
Type II 2HDM, $bb$-associated production dominates at large $t_{\beta}$. In both cases, we compute the production cross section for 
$H$ and $A$ at NNLO in QCD via {\sc SusHi}~\cite{Harlander:2012pb} (for $H$, the cross section does depend on $c_{\beta-\alpha}$, and in that case we consider the 
alignment limit $c_{\beta-\alpha} = 0$).
For the charged scalar $H^{\pm}$, the dominant production mode for $m_{H^{\pm}}>m_t$ is in association with a $tb$ pair, 
and we use the NLO cross section values provided by the Higgs Cross Section Working Group (HXSWG) for $m_{H^{\pm}} > 200$ GeV~\cite{ChargedHiggsXS}. 
A light charged scalar ($m_{H^\pm}<m_t$) is mainly produced through top quark decays $t \to H^+ b$, and we  
use {\sc Top++2.0}~\cite{Czakon:2011xx} to compute the top pair production to NNLO in QCD, assuming a top-quark mass $m_t = 172.5$ GeV.
The LHC production cross sections for $H$, $A$ and $H^+$ at 13 TeV are shown in Appendix~\ref{sec:pro_decay2}.  


Regarding the decays of the new scalars, in the alignment limit $c_{\beta-\alpha} = 0$ the conventional (SM-like) decays of $A$ and $H$ are into $tt$ (if kinematically accessible), 
$bb$, $cc$, $\tau\tau$, and with a highly suppressed branching fraction into $gg$, $\gamma\gamma$ and $\mu\mu$. When open, the decay into $tt$ is dominant for 
low and moderate $t_{\beta}$, followed by the decay into $bb$.  At high $t_{\beta}$, for Type II 2HDM, the decay into $\tau\tau$ becomes important, where the decay into 
$bb$ can dominate even above the $tt$ threshold. For the CP-even Higgs $H$, the decay into massive gauge bosons $W^+W^-$ and $ZZ$ is present away from 
the alignment limit, and dominates as soon as the departure from alignment is sizeable. 
For the charged scalar, the decay $H^\pm \rightarrow tb$ dominates once it is kinematically open, followed by $H^\pm \rightarrow \tau\nu$, $cs$ and $cb$.  
In the following, we compute all 2HDM branching fractions using {\sc 2HDMC}~\cite{Eriksson:2009ws}.

 \begin{table}[h!]
 \centering
 {\small 
 \begin{tabular}{c|c|l|l} \hline
Parent Scalar&Decay&$\quad\,\,\, \quad\,\,\, \quad\,\,\,\,$Possible Final States&$\,\,\,\,\,$Channels in 2HDM  \\ \hline  
 &$H_i H_i$ & $(bb/\tau\tau/WW/ZZ/\gamma\gamma)\,(bb/\tau\tau/WW/ZZ/\gamma\gamma)$ & $H \rightarrow A A,\, hh$ \\   \cline{2-4}
Neutral&$H_i Z$  & $(bb/\tau\tau/WW/ZZ/\gamma\gamma)\,(\ell\ell/qq/\nu\nu)$ & $H \rightarrow A Z, A \rightarrow  H Z,\, hZ$  \\   \cline{2-4}
$H$, $A$& $H^+H^-$  & $(tb/\tau\nu/cs)\,(tb/\tau\nu/cs)$ & $H  \rightarrow H^+H^-$ \\  \cline{2-4}
&$H^\pm W^\mp$  & $(tb/\tau\nu/cs)\,(\ell\nu/qq^\prime)$ & $H/A \rightarrow H^\pm W^\mp$\\  \hline
Charged $H^{\pm}$&$ H_i W^\pm$ & $(bb/\tau\tau/WW/ZZ/\gamma\gamma)\,(\ell\nu/qq^\prime)$ & $H^\pm \rightarrow  hW^{\pm}, H W^{\pm}, A W^{\pm}$  \\  \hline
 \end{tabular}
}
 \caption{Summary of exotic decay modes for non-SM Higgs bosons. For each type of exotic decays (second column), we present possible final states (third column) and 
 relevant channels in 2HDM (fourth column). In the second column, $H_i = h, \,H,\, A$.}
 \label{tab:decay_exo}
 \end{table}
 
\begin{figure}[h!].
\centering
\hspace{-6mm}
\includegraphics[width=0.49\textwidth]{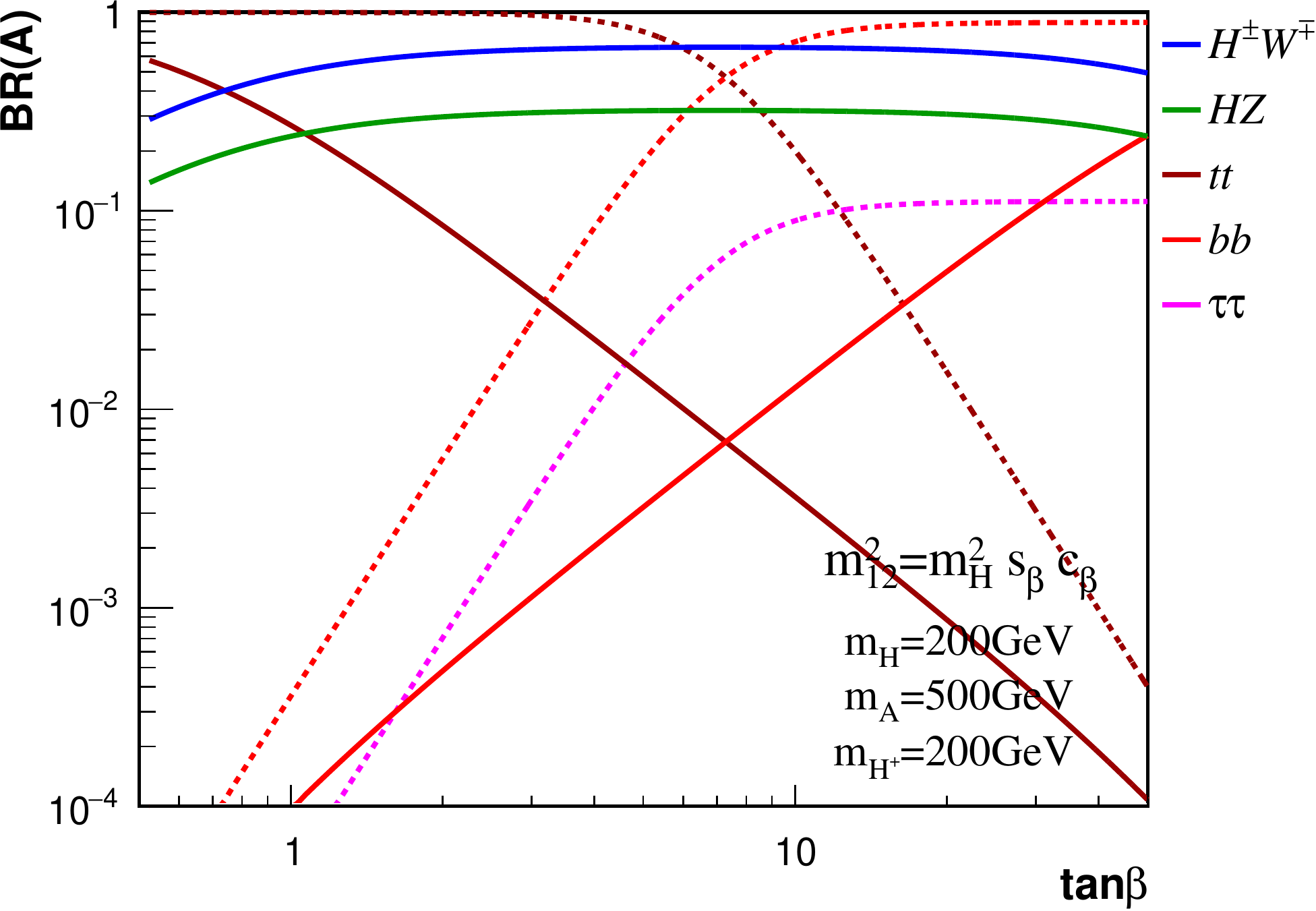}
\hspace{0.2mm}
\includegraphics[width=0.472\textwidth]{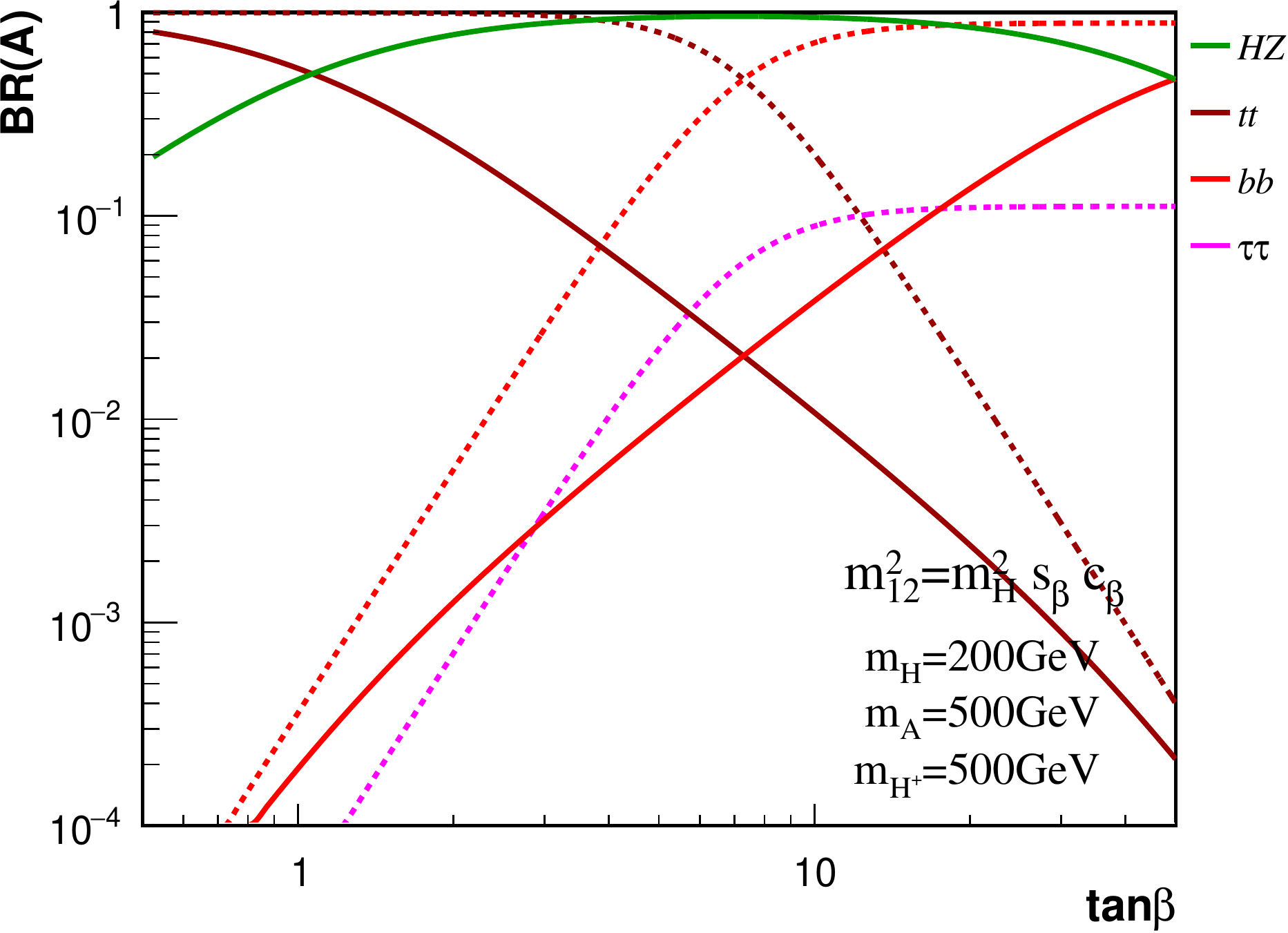}
\includegraphics[width=0.484\textwidth]{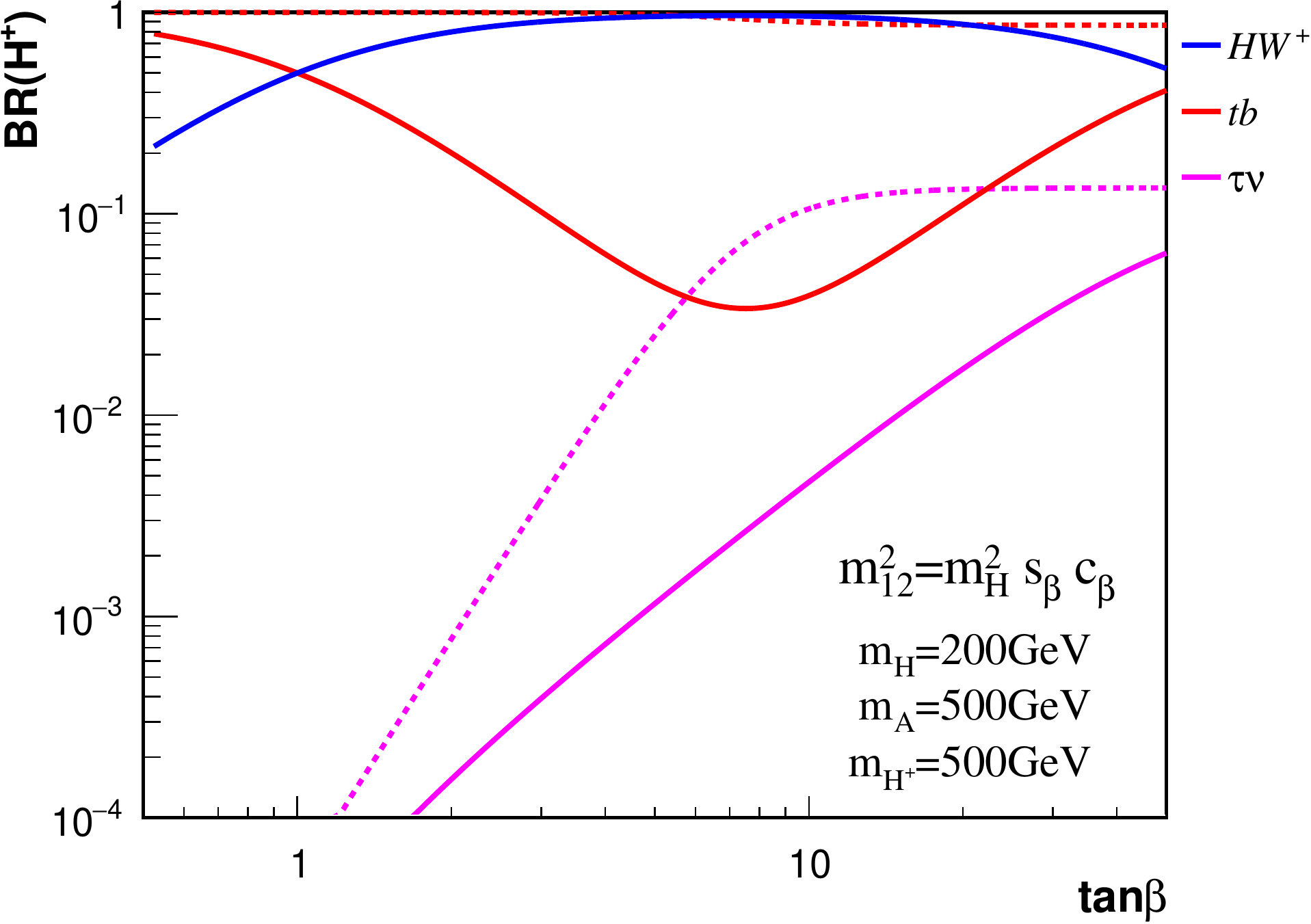}
\hspace{1.2mm}
\includegraphics[width=0.493\textwidth]{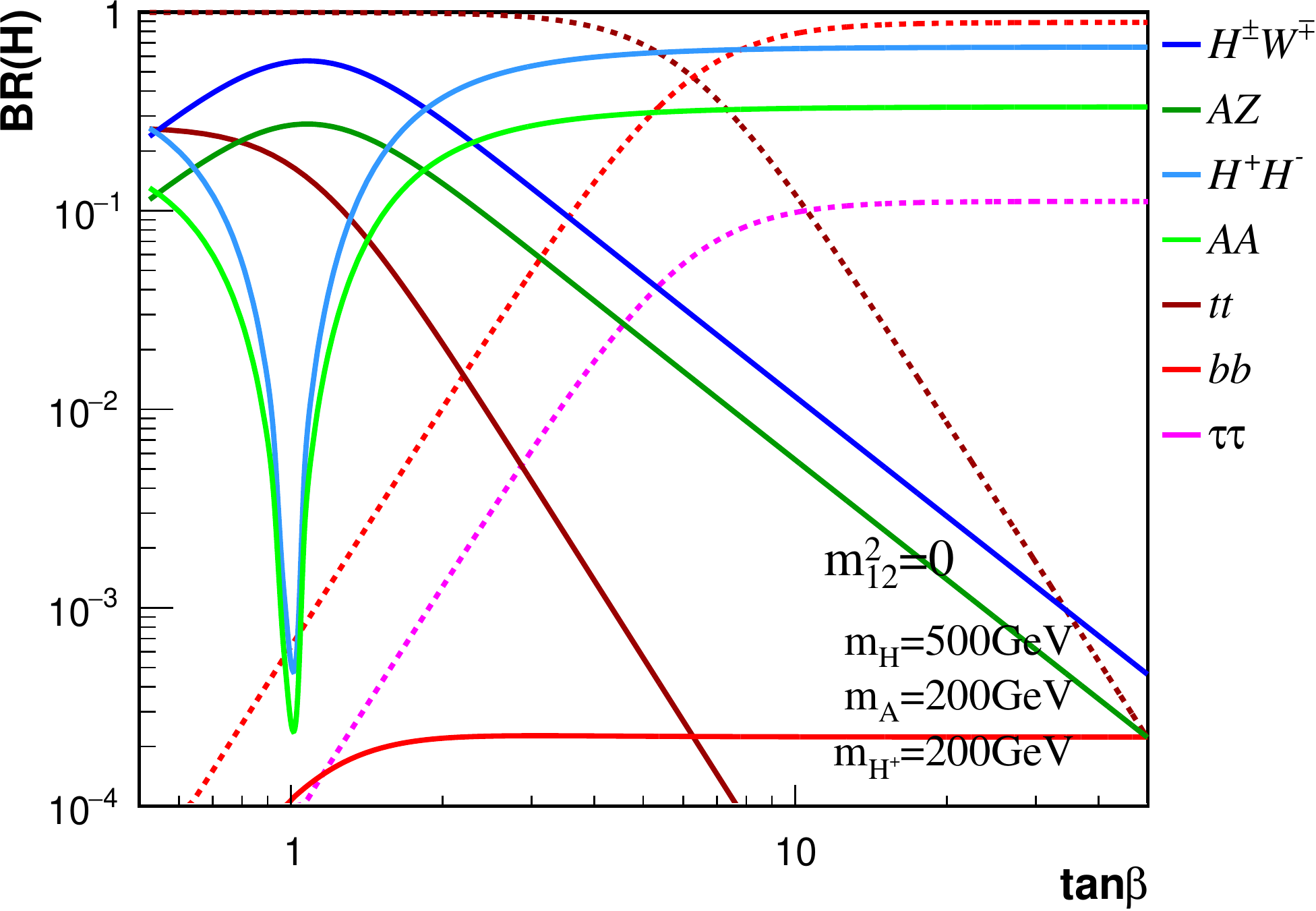}
 \caption{Branching fractions in Type II 2HDM as a function of $t_{\beta}$ for $c_{\beta-\alpha} = 0$, with parent and daughter scalar masses fixed to 500 GeV and 200 GeV, respectively. 
 Top:  Branching fractions for $A$ with $m_{H}=m_{H^{\pm}}<m_A$ (left) and with $m_{H}<m_A=m_{H^{\pm}}$ (right), in both cases for $m_{12}^2=m_{H}^2 s_\beta c_\beta$.
 Bottom: Branching fractions for $H^\pm$ with $m_{H} < m_{H^{\pm}}= m_A$ and $m_{12}^2=m_{H}^2 s_\beta c_\beta$ (left) 
 and for $H$ with $m_{H} > m_{H^{\pm}}= m_A$ and $m_{12}^2=0$ (right). In all cases, dashed lines indicate the branching 
 fractions to SM fermion pairs when exotic decay modes are absent.   }
 \label{fig:decay}
 \end{figure}

We here stress that the above conventional decays of the new 2HDM scalars
become suppressed once exotic (non SM-like) decay modes open up.
These can be decays involving several states among $H$, $A$, $H^{\pm}$, in the presence of a large mass splitting among the new scalars
(see e.g.~\cite{Coleppa:2013xfa,Coleppa:2014hxa,Dorsch:2014qja,Coleppa:2014cca,Li:2015lra,Kling:2015uba} for existing studies on individual channels), 
and/or decays into the SM-like Higgs boson $h$, namely $H \to hh$, $A \to h Z$, $H^{\pm} \to h W^{\pm}$, which are possible for $c_{\beta-\alpha} \neq 0$ 
and are also considered in the following as exotic (despite involving SM decay products) as they don't occur in the SM. 
In the former case, we can further distinguish between the decay of a new scalar into another one and a gauge boson, and the potential decays of 
$H$ into either $AA$ or $H^{+}H^{-}$. The different types of exotic decay modes for the 2HDM are summarized in Table~\ref{tab:decay_exo}.

The impact of the presence of exotic Higgs decay modes on the branching ratios is shown in Figure~\ref{fig:decay} for $c_{\beta-\alpha} = 0$. 
The top two panels show the relevant branching fractions of $A$ with $m_{H}=m_{H^\pm}<m_A$ (left), and $m_{H}<m_A=m_{H^\pm}$ (right) for 
$m_A = 500$ GeV and $m_H = 200$ GeV, with $m_{12}^2=m_{H}^2 s_\beta c_\beta$.
In the former case, the decays $A \rightarrow H^\pm W^\mp$ (solid blue) and $A \rightarrow H Z$ (solid green) completely dominate over the 
SM decays $A \to tt, \,bb,\, \tau\tau$ for most values of $t_{\beta}$, with ${\rm BR}(A \rightarrow H^\pm W^\mp) \sim$ 50$-$60\% and 
${\rm BR}(A \rightarrow HZ) \sim$ 20$-$30\%, while in the latter with $A \rightarrow H^\pm W^\mp$ being absent, the branching fraction of $A \rightarrow H Z$ is more than 50\%. 
Decays of $A \to tt,\,bb$ are only important for very small or very large $t_{\beta}$.
The dashed lines show for comparison the branching fractions into the conventional SM states when the exotic decays are absent, 
which highlights the suppression the SM channels suffer in the presence of the exotic decays.
The bottom left panel in Figure~\ref{fig:decay} shows the branching fractions of $H^\pm$ for $m_{H}<m_A=m_{H^\pm}$ (with $m_A = 500$ GeV and $m_H = 200$ GeV) 
and $m_{12}^2=m_{H}^2 s_\beta c_\beta$. 
The decay $H^\pm \rightarrow H W^\pm$ (solid blue) dominates with ${\rm BR}(H^\pm \rightarrow H W^\pm)\gtrsim 50\%$, particularly for a not too heavy state $H^{\pm}$. In that case, 
$H^\pm \rightarrow tb$ is suppressed to be about few percent for intermediate $t_{\beta}$, and only reaches about 50\% in the very 
small and very large $t_{\beta}$ region. 
Finally, the bottom right panel in Figure~\ref{fig:decay} shows the branching fractions of $H$ for $m_H > m_{H^\pm}=m_A$ 
(with $m_H = 500$ GeV and $m_A = 200$ GeV) and $m_{12}^2 = 0$. In this case the decays $H\rightarrow AA$ and $H \rightarrow H^+H^-$ are allowed and dominate 
over most of the $t_\beta$ region, except for $t_\beta\sim 1$, where $H \rightarrow AZ$ and $H\rightarrow H^\pm W^\mp$ become dominant due to the accidental 
suppression of the $H H^+H^-$ and $HAA$ couplings at $t_\beta\sim 1$. Note however that for $m_{12}^2 = 0$ the theoretical constraints 
do not allow a significant departure from $t_{\beta} \sim 1$, such that a large branching fraction for $H \rightarrow AZ$ and $H \rightarrow H^\pm W^\mp$ 
is expected. Decays to SM fermions are highly suppressed in this scenario.

\section{2HDM Planes for Exotic Higgs Decays}
\label{sec:bp}

Our analysis of exotic Higgs decays in the 2HDM focuses on a few key benchmark planes which show the complementarity among different LHC search channels for the new scalars.
We first focus on the alignment limit: $c_{\beta-\alpha} = 0$ for $m_{H} > m_h = 125$ GeV and 
$s_{\beta-\alpha} = 0$ for $m_h < m_H = 125$ GeV. In this context, we consider two possible mass planes: $m_A$ vs. $m_H = m_{H^\pm}$ (Plane I) and 
$m_H$ vs. $m_A = m_{H^\pm}$ (Plane II). These two choices are motivated by EWPO constraints (recall the discussion in Section~\ref{sec:delta_rho}). 
This is in contrast to a potential $m_{H^\pm}$ vs. $m_H = m_A$ plane, highly constrained by EWPO to a small mass splitting $m_{H^{\pm}} - m_{H/A}$ which     
closes the phase space needed for on-shell exotic Higgs decays\footnote[10]{As discussed in Section~\ref{sec:delta_rho}, a sizable departure from alignment 
could allow for a mass hierarchy $m_A > m_H + m_Z$ (such that $A \to HZ$ is kinematically allowed) and $m_H \gtrsim m_{H^{\pm}} + m_W$ 
(such that $H \to H^{\pm}W^{\mp}$ is kinematically allowed, but nevertheless phase space suppressed). The phenomenology of this kind of scenario is however largely 
contained in Planes I-II, and so we do not consider it separately.}, so that we don't consider such benchmark plane in our current study.
Finally let us remark that, while we do not impose the flavour bounds as hard constraints on our 2HDM benchmark planes (recall the discussion in Section~\ref{sec:flavour}),
we do show them as indicative in the following. 
Our 2HDM benchmark plane (BP) scenarios in alignment are then:

\vspace{2mm}

\noindent \underline{\textbf{Plane I:  $m_A$ vs. $m_H=m_{H^\pm}$}}

\begin{itemize}
\item{BP IA}: $m_A > m_H = m_{H^\pm}$.

As discussed in Section~\ref{Sec3_summary}, this mass ordering is allowed for Case 1 ($m_{12}^2 = m_H^2 s_\beta c_\beta$ and all $t_{\beta}$ values) and 
Case 2 ($m_{12}^2 = 0$ and $t_{\beta} \sim 1$). 
We thus consider four scenarios: Case 1 with $t_\beta = 1.5,\, 7,\,30$ and Case 2 with $t_\beta=1.5$. 

\item{BP IB}: $m_A< m_H=m_{H^\pm} $.

This mass ordering is not compatible with Case 1 due to vacuum stability (see Sections~\ref{sec:stability} and~\ref{Sec3_summary}). Thus, we only consider
Case 2 with $t_\beta=1.5$\footnote[11]{$t_\beta\neq 1$ is chosen for the exotic decays into two 
lighter new scalars ($H\to A A$ in this case) not to vanish.}.

\end{itemize}


\noindent \underline{\textbf{Plane II:  $m_H$ vs. $m_A=m_{H^\pm}$}}

\begin{itemize}
\item{BP IIA}: $m_H> m_A=m_{H^\pm}$.

As for BP IB, this mass ordering is not compatible with Case 1, and so we only consider Case 2 with $t_\beta=1.5$.  

\item{BP IIB}: $m_H< m_A=m_{H^\pm}  $.

As for BP IA, we consider Case 1 with $t_\beta = 1.5,\, 7,\,30$ and Case 2 with $t_\beta=1.5$.  

\end{itemize}

In order to study the decays of the new scalars into the SM-like Higgs, we also consider a plane in which the departure from alignment is explored, assuming 
$h$ is the 125 GeV SM-like Higgs (Plane III). We set $m_H = m_A = m_{H^{\pm}}$ for simplicity, and define the plane as $m_H = m_A = m_{H^{\pm}}$ vs. $c_{\beta-\alpha}$:
 
\vspace{2mm}

\noindent \underline{\textbf{Plane III:  $m_A=m_H=m_{H^\pm}$ vs. $c_{\beta-\alpha}$}}

\begin{itemize}
\item{BP III}:

We consider Case 1 with $t_\beta = 1.5,\, 7,\,30$ and Case 2 with $t_\beta=1.5$.
\end{itemize}

A summary of the different benchmark planes considered and the relevant exotic decay modes is shown in Table~\ref{tab:Summary}.
In all cases, we present the $\sigma \times$ BR of each characteristic decay channel at the 13 TeV LHC, together with 
a detailed analysis of the regions disfavoured by theoretical and experimental constraints
(including flavour constraints, shown for reference only). 
The results for Planes I and II ($c_{\beta-\alpha} = 0$) are presented in Section~\ref{sec:align}, while the 
results for decays to SM-like Higgs away from alignment, corresponding to Plane III, are presented in Section~\ref{sec:nonalign}.
Further details on the cross sections and decay branching fractions for the non-SM like Higgses can be found in Appendix~\ref{sec:pro_decay2}.

\begin{table}[h]
\begin{center}
\begin{tabular}{|c | c | c | c | c | c |}
\hline
&Mass Planes & decays & $m_{12}^2$ &$\tan\beta$ & Figures \\
\hline
\hline
 	BP IA& $m_A > m_H=m_{H^{\pm}}  $&
   $A \to H^\pm W^\mp$ &
   $m_H^2 s_\beta c_\beta$	&	1.5, 7, 30	& {\ref{fig:bp-CH-vs-A}, \ref{fig:bp-HC-vs-A}} \\   \cline{4-5}   
  	&& $A \to HZ$
   &  
   0					&	1.5	& \\ 
\hline 
BP IB & $m_A < m_H=m_{H^{\pm}} $	&$H\to AZ $, $H\to AA$ &
    0					&	1.5	& \ref{fig:BPIB}    \\ 
 &  	& $H^\pm \to A W^\pm$&
     					&	 	&   \\ 
\hline 
BP IIA & $m_H > m_A=m_{H^{\pm}} $	&$H\to AZ$, $H\to AA$ &
    0					&	1.5	& \ref{fig:BPIIA}     \\ 
 &  & $H\to H^+H^-$, $H\to H^\pm W^\mp$&
  					&	 	&    \\ 
\hline 
BP IIB & $ m_H < m_A=m_{H^{\pm}}$	&$A\to HZ$&
    $m_H^2 s_\beta c_\beta$					&	1.5, 7, 30	&  \ref{fig:bp-H-vs-AC}, \ref{fig:bp-H-vs-CA}  \\   \cline{4-5}
  &  	& $H^\pm \to HW^\pm$&
    0					&	1.5	&   \\ 
\hline 
BP III & $ m_A=m_H=m_{H^{\pm}}$ 	&$A\to hZ$, $H^\pm \rightarrow h W^\pm$&
    $m_H^2 s_\beta c_\beta$					&	1.5, 7, 30	& \ref{fig:bp-ACH-vs-hsm}, \ref{fig:bp-CHA-vs-hsm}, \ref{fig:bp-HAC-vs-hsm}   \\   \cline{4-5}
  &  vs. $c_{\beta-\alpha}$	& $H\to hh$&
    0					&	1.5	&    \\ 
\hline\hline
\end{tabular}
\caption{Summary Table of the different 2HDM benchmark planes.}
\label{tab:Summary}
\end{center}
\end{table}

Before we move on to discuss in detail our different 2HDM planes for LHC searches at 13 TeV, let us comment on the comparison of these benchmark scenarios with others 
proposed in the literature. In particular, our Planes I and II have a substantial overlap with the 2HDM ``{\it short cascade}" scenario D from~\cite{Haber:2015pua}, 
while our specific BP IA and BP IIB have similarities with the $A \to H Z$ benchmarks for $c_{\beta-\alpha} = 0$ in~\cite{Dorsch:2016tab} (see also~\cite{Dorsch:2014qja}). 
As compared to~\cite{Haber:2015pua}, the present analysis explores the full mass plane, not restricted to specific benchmark lines with fixed relations\footnote[12]{In 
particular, we note that the fixed relations in~\cite{Haber:2015pua} result in the exotic Higgs decays being largely subdominant above the $t\bar{t}$ threshold, 
which may not be the case in general (see e.g. Figure~\ref{fig:decay}).} among 
$m_H$, $m_A$ and $m_{H^{\pm}}$. We also explore the dependence on $t_{\beta}$, which has a significant impact on the allowed 
2HDM parameter space for Planes I and II. Moreover, our analysis includes the 8 TeV experimental constraint from the CMS $H \to AZ$/$A\to HZ$ 
search~\cite{CMS:2015mba,Khachatryan:2016are}, precisely tailored to probe these 2HDM scenarios and thus a key ingredient in a study of 2HDM exotic Higgs decays.  
 
\begin{figure}[h!]
 \centering
 \includegraphics[width=0.495\textwidth]{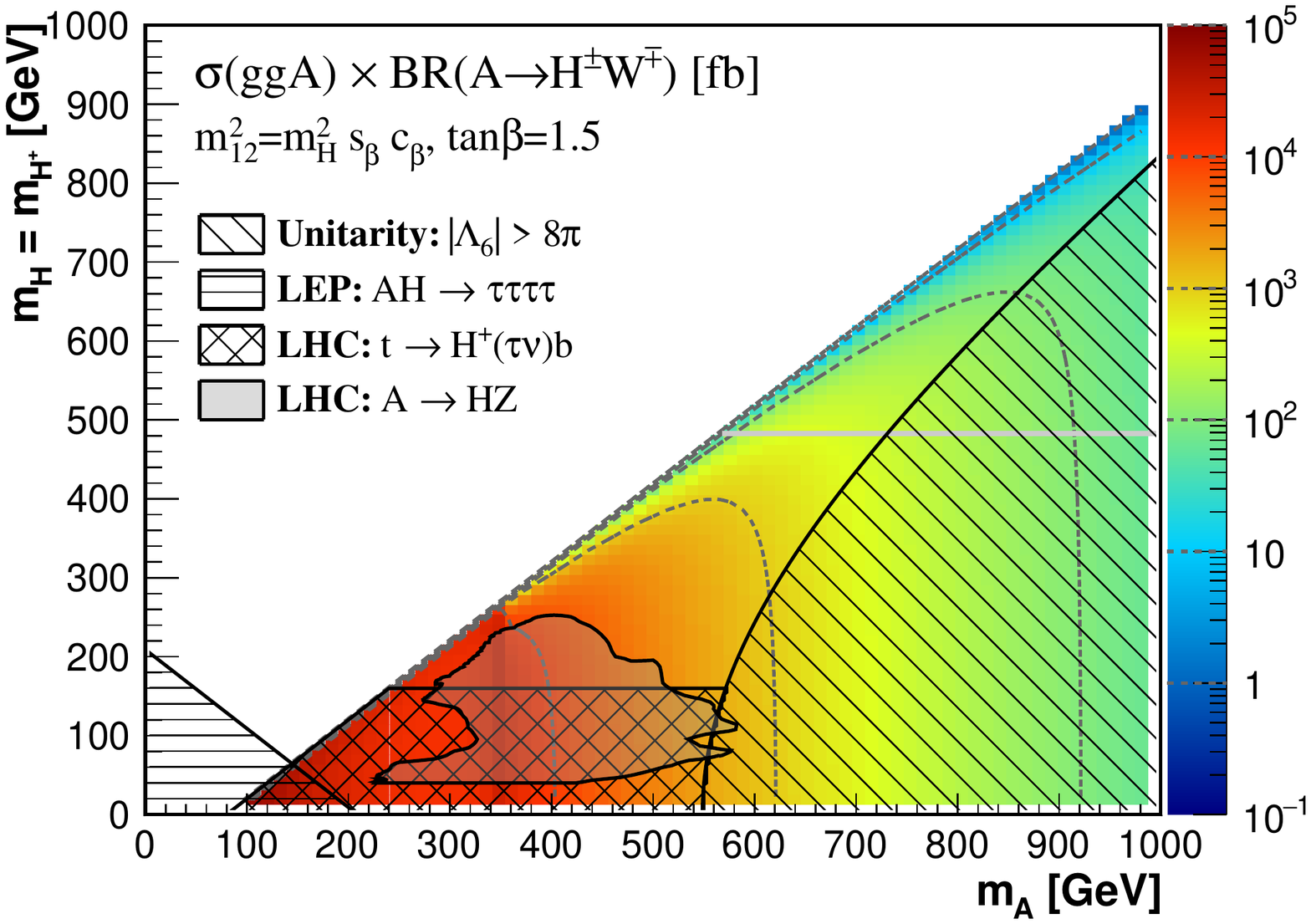}
	\includegraphics[width=0.495\textwidth]{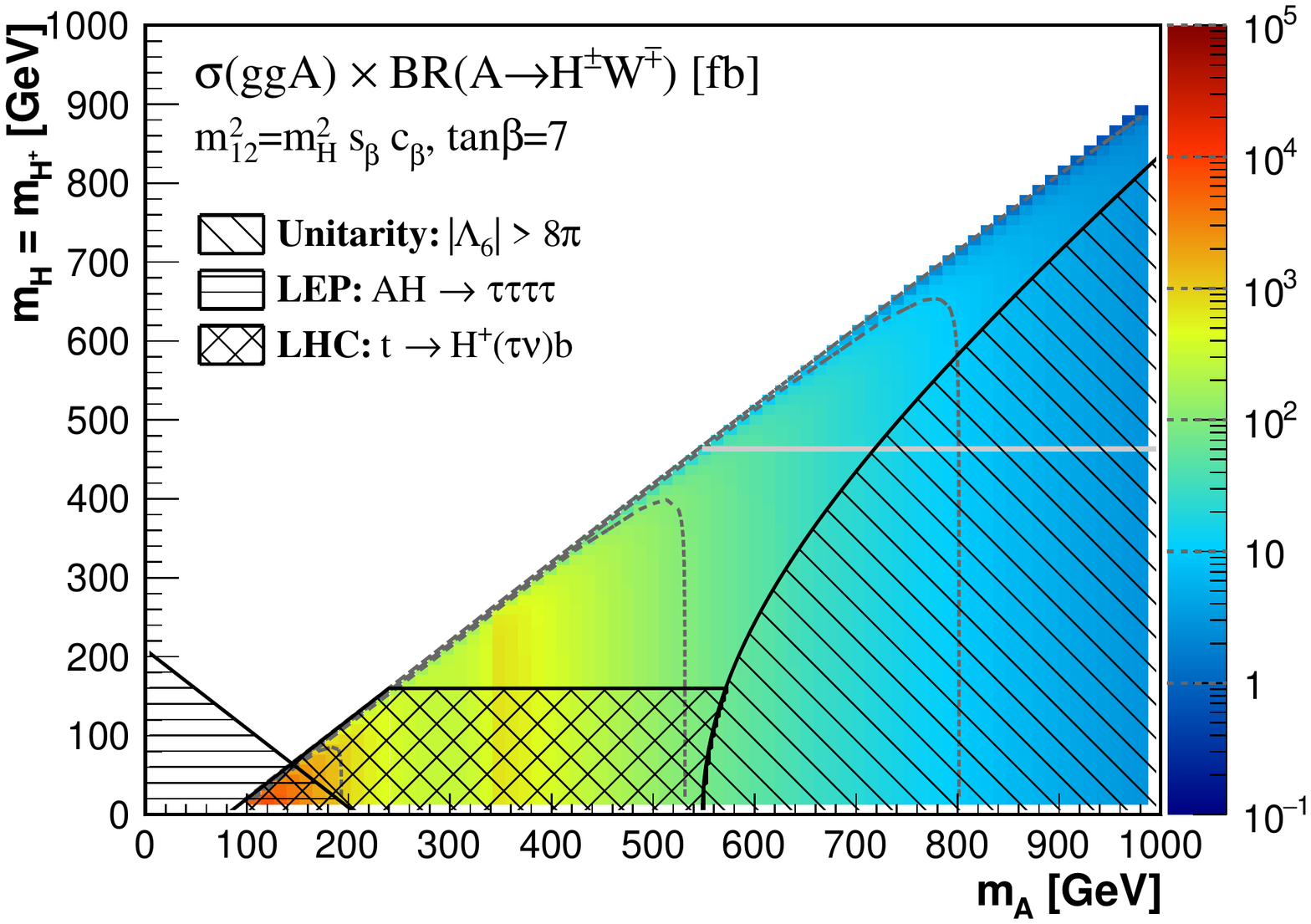}
	\includegraphics[width=0.495\textwidth]{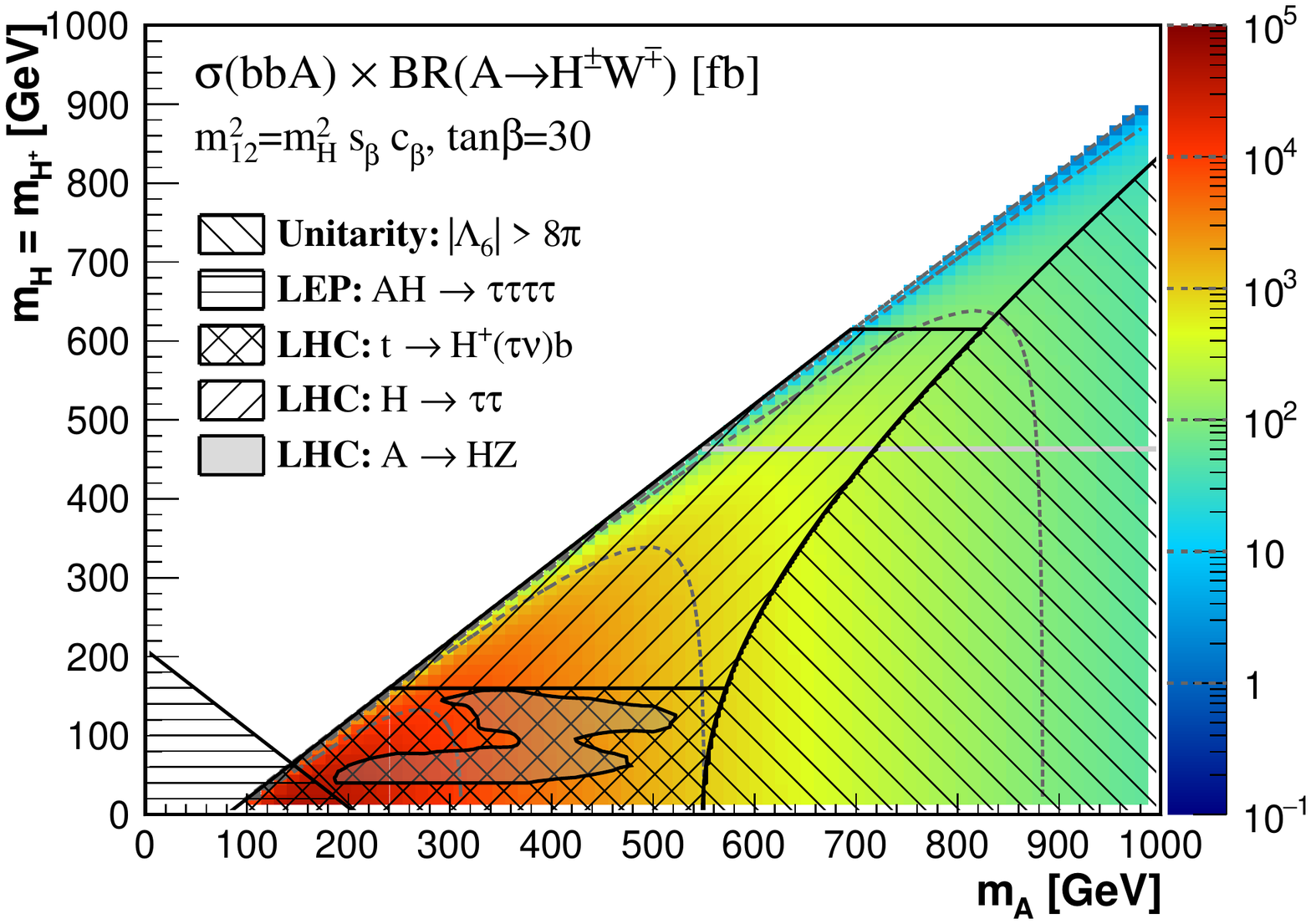}
	\includegraphics[width=0.495\textwidth]{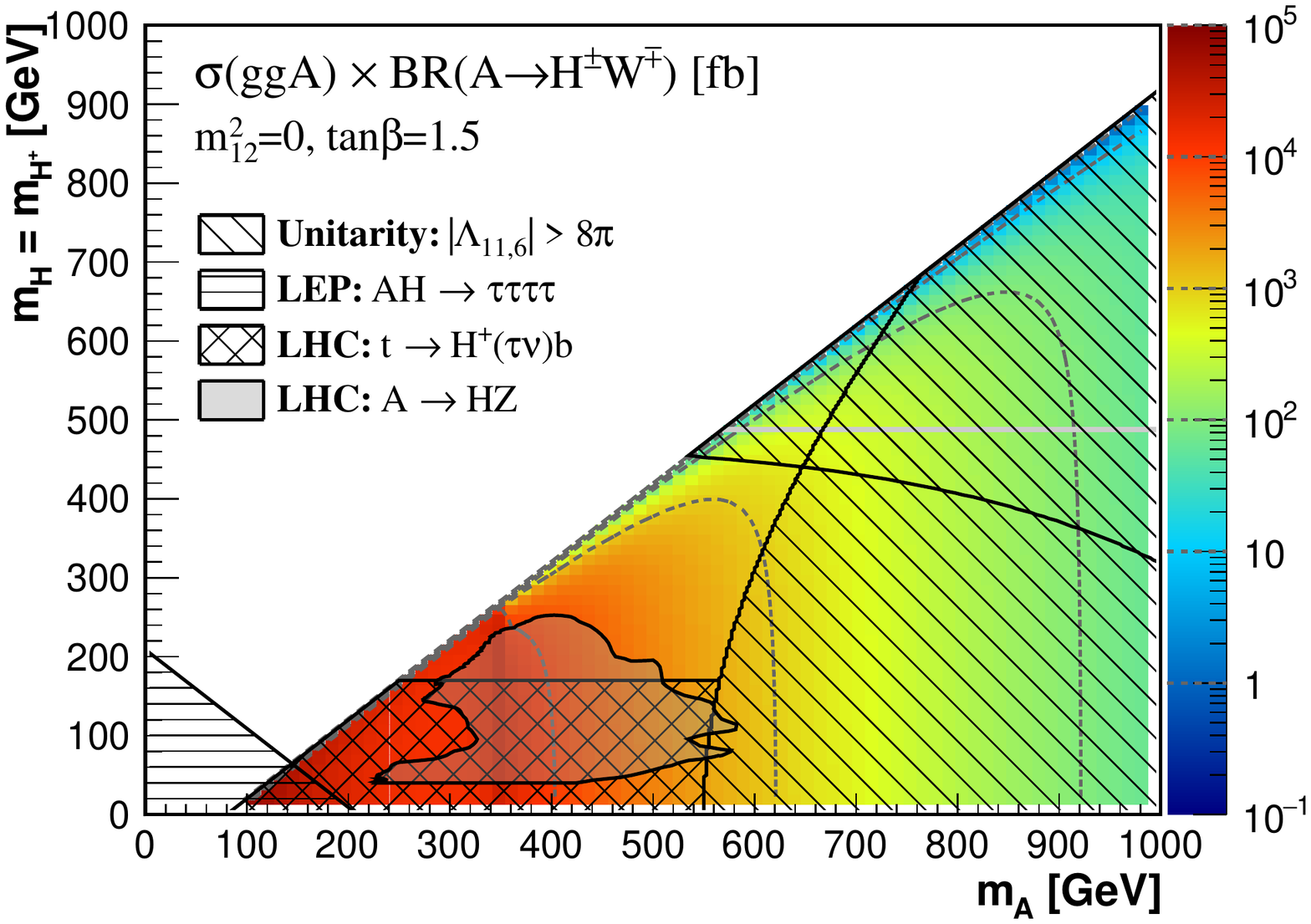}
\caption{$\sigma\times{\rm BR}$ for the exotic decay $A \rightarrow H^\pm W^\mp$ in BP IA: $m_A$ vs $m_H=m_{H^{\pm}}$ plane, 
for Case 1 with $t_{\beta}=1.5$ (upper left), 7 (upper right), 30 (lower left) and Case 2 with $t_{\beta}=1.5$ (lower right).  
Contour lines of  10, $10^2$, $10^3$ and $10^4$ fb are drawn as light grey dashed curves to guide the eye. 
The shaded areas enclosed by an irregular curve and hatched regions are ruled out by theoretical and experimental constraints (see text for details). 
The solid horizontal grey line indicates the flavour constraint $m_{H^{\pm}} > 480$ GeV.}
\label{fig:bp-CH-vs-A}
\end{figure} 

It is worth discussing here the extent to which a departure from our Benchmark Plane assumptions 
(Case 1 with $m_{12}^2 =m_H^2  s_\beta c_\beta$ or Case 2 with $m_{12}^2=0$, with certain mass degeneracy 
relations between $m_A$, $m_H$ and $m_{H^\pm}$) may lead to modifications of the 2HDM phenomenology 
w.r.t. the scenarios we consider in this work: 
 
 \begin{itemize}
\item \textbf{Negative $m_{12}^2$:} Extending $m_{12}^2$ to be negative does not introduce further constraints regarding vacuum stability. 
In addition, the variation of unitarity constraints for $m_{12}^2<0$ w.r.t. the $m_{12}^2=0$ case is mild, such that $m_{12}^2=0$ effectively captures this 
scenario. 
 
\item \textbf{Deviation from $m_{12}^2 = m_H^2 s_\beta c_\beta $:} Assuming the alignment limit $c_{\beta-\alpha} = 0$, at high $t_\beta$ 
the soft $\mathbb{Z}_2$ breaking term $m_{12}^2$ is fixed to be around $ m_H^2 s_\beta c_\beta $ by unitarity. 
Only when departing significantly from the alignment limit are deviations from $m_{12}^2 = m_H^2 s_\beta c_\beta$ at high $t_\beta$ possible, 
yielding a larger allowed region in parameter space. 

At low values of $t_\beta$ deviations from $m_{12}^2 = m_H^2 s_\beta c_\beta $ can be easily accommodated in the alignment limit. We however stress 
that changing the value of $m_{12}^2$ only affects cubic and quartic Higgs 
couplings. In particular, $HH^+H^-$ and $HAA$ coupling vanish for $m_{12}^2 = m_H^2 s_\beta c_\beta$ in the alignment limit (see Section~\ref{Sec3_summary}). Deviations 
from  $m_{12}^2 = m_H^2 s_\beta c_\beta$ will then increase the branching fractions of $H\rightarrow AA$ and 
$H\rightarrow H^+H^-$ (if allowed by kinematics) with the branching fractions to other final states decreasing accordingly. 
For $m_{12}^2$ far away from $m_H^2 s_\beta c_\beta $, the branching fraction dependence on $m_{12}^2$ is mild, with Case 2 
($m_{12}^2=0$) being a representative scenario.

\item \textbf{Deviations from the alignment limit $c_{\beta-\alpha} = 0$:} In our discussion we mainly focus on the alignment 
limit, preferred by Higgs coupling measurements at the LHC. 
In particular, we find that in the alignment limit, unitarity and vacuum stability cannot be satisfied simultaneously 
if  $m_H>m_A$ or $m_{H^{\pm}}$ at high $t_\beta$ and therefore decays of the type $H\to AZ$, $H\to H^\pm W^\mp$ are not permitted for large $t_\beta$. This 
statement can be relaxed  when deviating from the alignment limit. For moderate values of $|c_{\beta-\alpha}|$ around 0.2$-$0.6, there are regions of 
parameter space with $m_H>m_A, \ m_{H^\pm}$, which are allowed by both unitarity and vacuum stability. 
Note however that once away from the alignment limit, 
channels like $H\to AZ$, $H^\pm W^\mp$ will generically have reduced branching fractions. There have been 
studies in the literature for exotic Higgs decays in those non-alignment region~\cite{Dorsch:2014qja,Gao:2016ued,Bauer:2015fxa}. 

\item \textbf{Deviation from mass degeneracy:}  The mass degeneracy relation is mainly imposed by the electroweak precision measurements. As shown in the 
left panel of Fig.~\ref{fig:st}, small deviations from $m_{H^{\pm}}\approx m_{A/H}$ are allowed. In principle,  there are small regions of parameter space 
corresponding to a mass hierarchy $m_H=m_A<m_{H^{\pm}}$. However,  these region of parameter space are already in almost 2$\sigma$ tension with observation. 
Other deviations from mass degeneracy will not lead to a phenomenology that would differ significantly from that of BP I and BP II. 
 
\end{itemize}

\subsection{Exotic Decays in the Alignment Limit}
\label{sec:align}
 
\subsubsection{BP IA: $m_A > m_H=m_{H^{\pm}}$ }

\begin{figure}[t!]
 \centering
 	\includegraphics[width=0.495\textwidth]{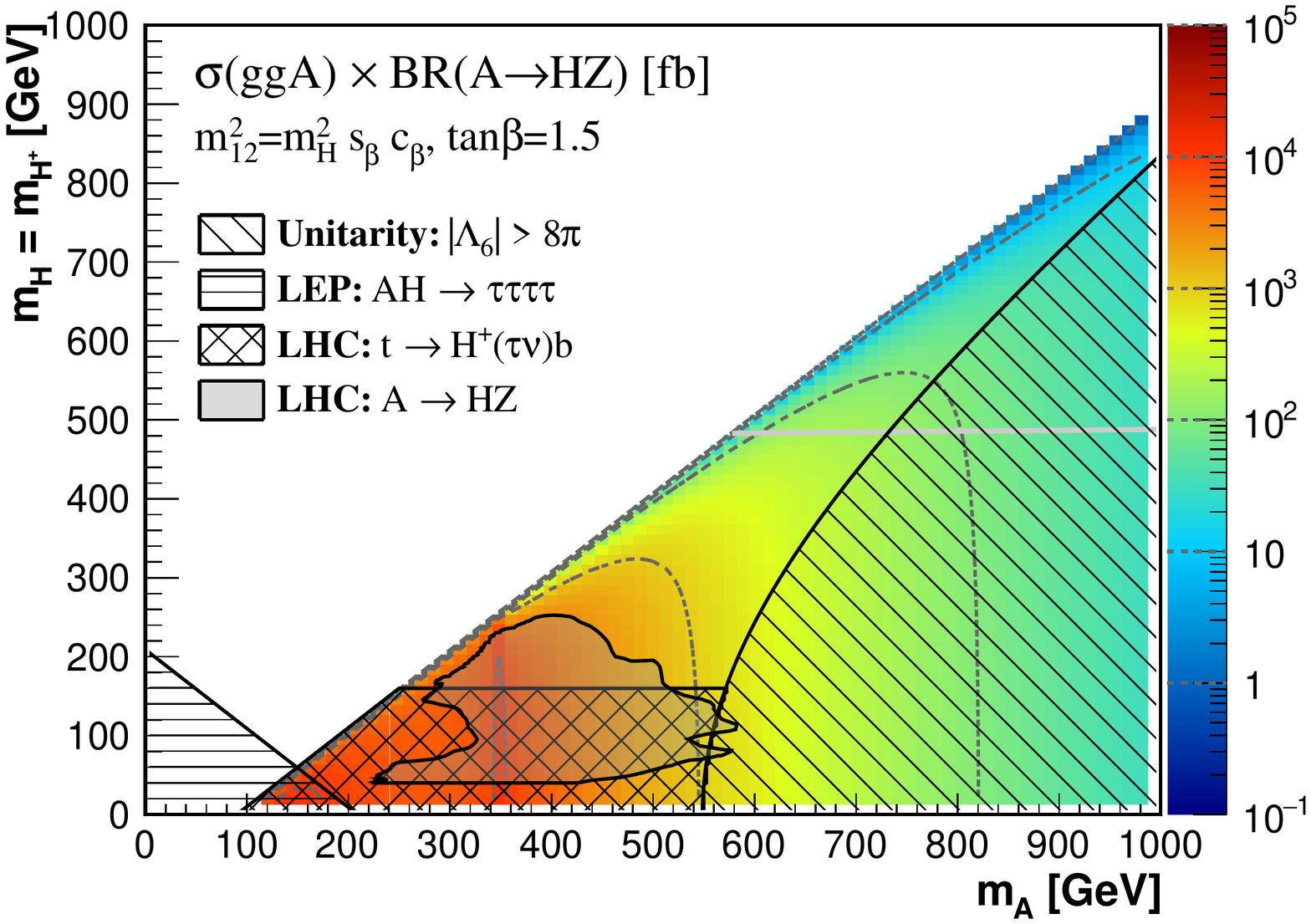}
	\includegraphics[width=0.495\textwidth]{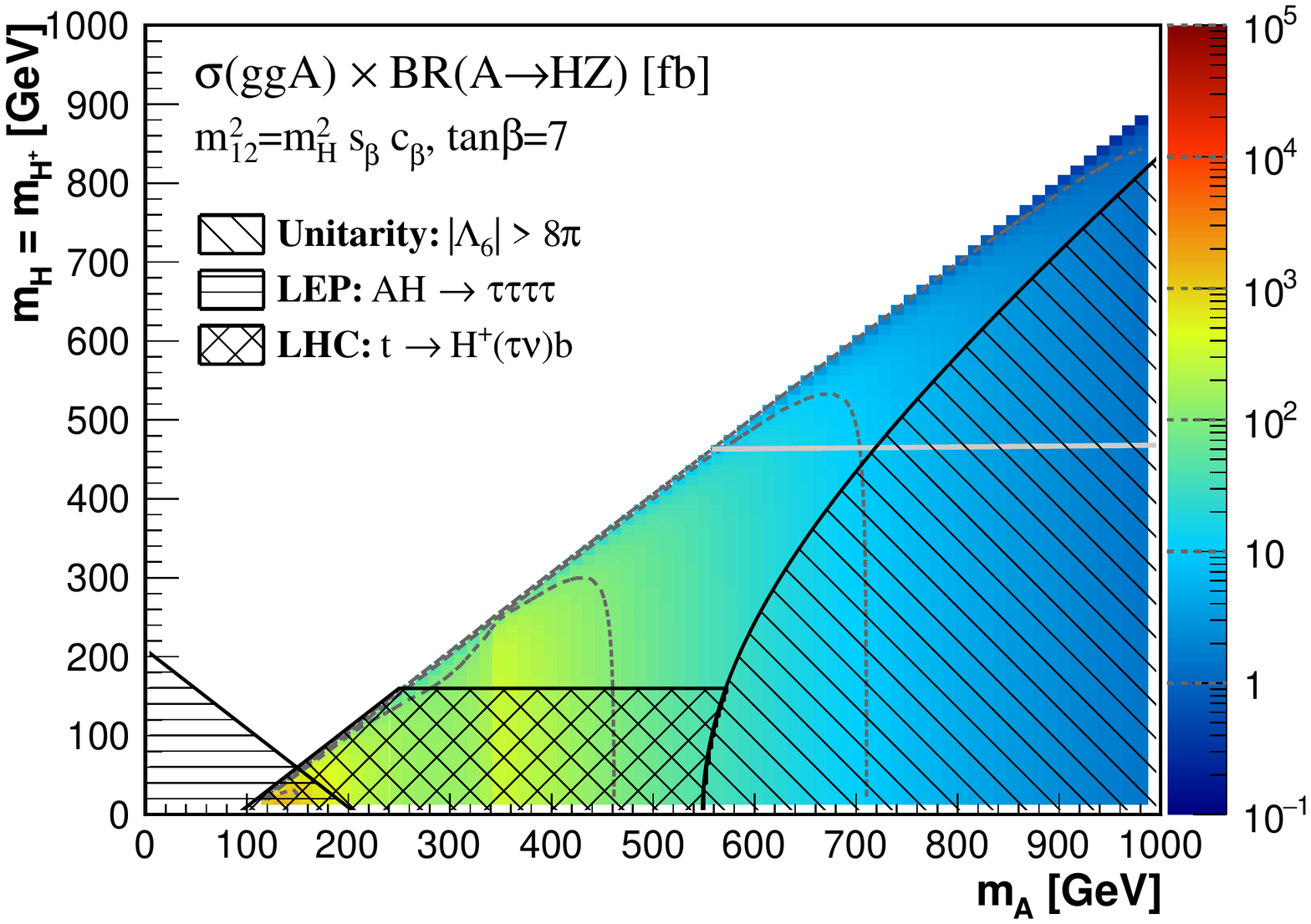}
	\includegraphics[width=0.495\textwidth]{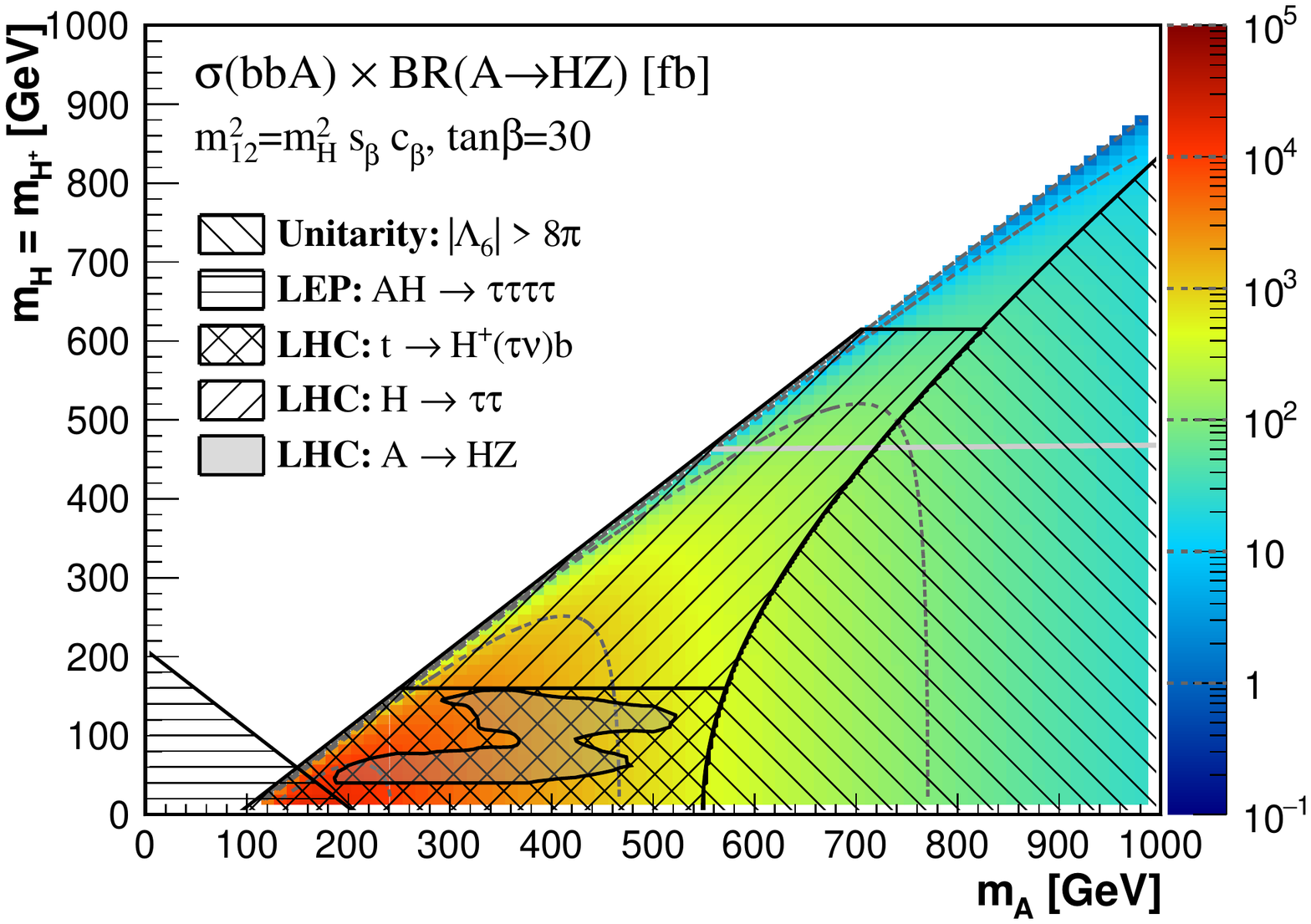}
	\includegraphics[width=0.495\textwidth]{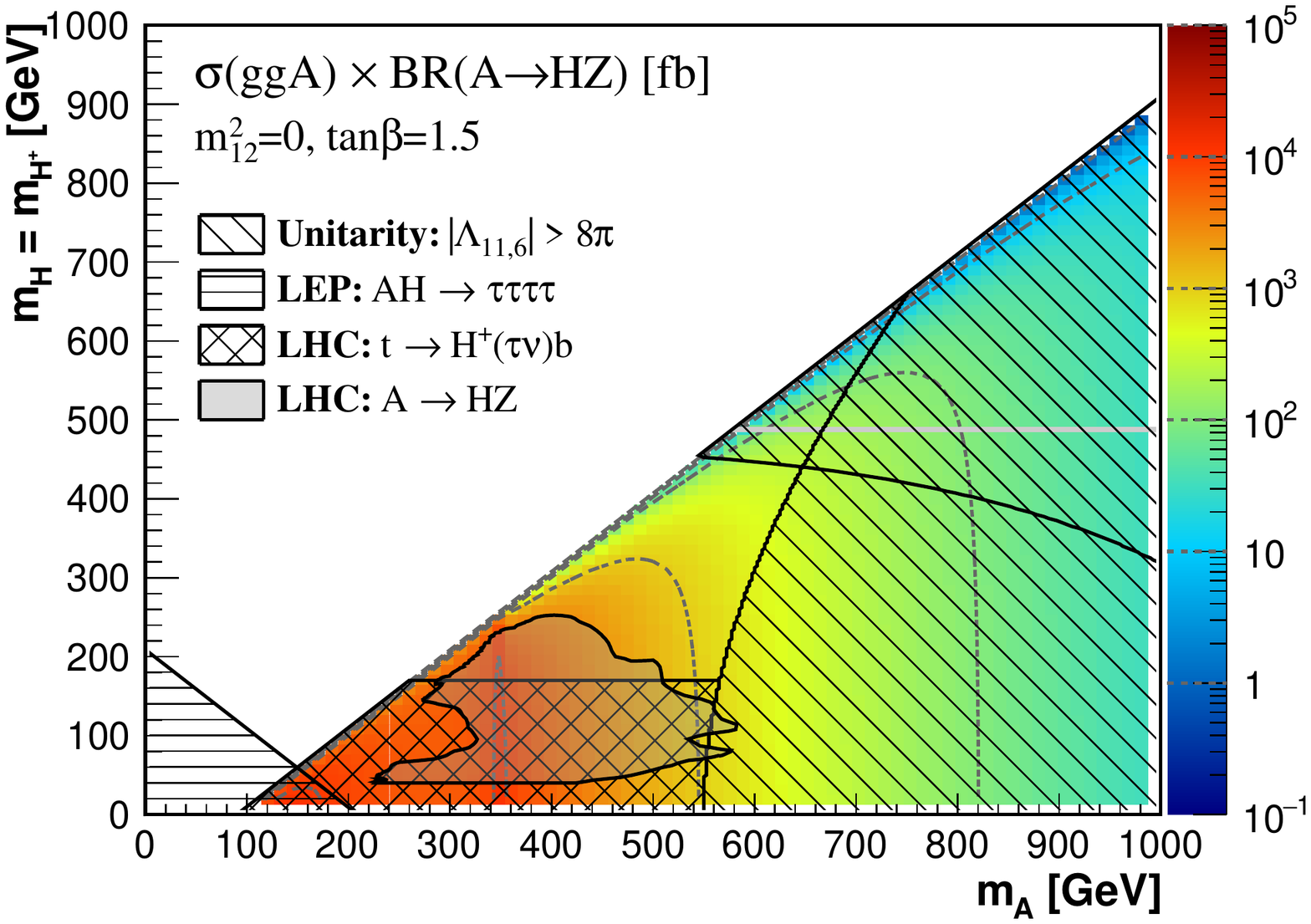}
\caption{$\sigma\times{\rm BR}$ for the exotic decay $A \rightarrow HZ$ in BP IA: $m_A$ vs $m_H=m_{H^{\pm}}$ plane 
(see caption of Figure~\ref{fig:bp-CH-vs-A} for further details). For $t_{\beta} = 1.5$, the contour line at $\sigma\times{\rm BR} =$ 10 pb around $m_A=350$ GeV is 
caused by the enhanced $gg\rightarrow A$ production cross section at the top threshold.}
\label{fig:bp-HC-vs-A}
\end{figure}

In this scenario, and for a sufficient mass splitting, 
there are two dominant exotic decay channels: $A \rightarrow H^\pm W^\mp$ and $A\rightarrow HZ$, for which we respectively show 
the $\sigma\times{\rm BR}$ in Figures~\ref{fig:bp-CH-vs-A} and~\ref{fig:bp-HC-vs-A}. In each case, we show four panels corresponding to the choices of 
$m_{12}^2$ and $t_{\beta}$ described in Table~\ref{tab:Summary}. Note that for $t_{\beta} = 1.5, \,7$ we consider the dominant 
$ggA$ production, while for $t_{\beta} = 30$ the $bbA$ production dominates and is considered instead.
For each panel, contour lines of 10, $10^2$, $10^3$ and $10^4$ fb are drawn as light grey dashed curves to guide the eye. 
Large cross sections $\sigma\times{\rm BR} \gtrsim 1$ pb are possible for $t_{\beta} \sim 1$ and $t_{\beta} \gg 1$, respectively due to the 
enhanced top and bottom Yukawa coupling contribution, even for large CP-odd scalar masses $m_A \sim 500 - 600$ GeV.
The shaded areas enclosed by an irregular curve in Figures~\ref{fig:bp-CH-vs-A} and~\ref{fig:bp-HC-vs-A} are excluded by the CMS $A\to H Z$ search~\cite{CMS:2015mba, Khachatryan:2016are}, 
which already constrains a sizable portion of parameter space and highlights the potential of such a search at LHC 13 TeV in the 
$bb\ell\ell$ and $\tau\tau\ell\ell$ final states, as a probe of both $A$ and $H$.

Hatched regions show the parameter space excluded by other experimental searches, as well as unitarity constraints.
The former exclusions are mainly due to $t \to H^{+}b$ searches, which yield a limit $m_{H^{\pm}} > m_t$, as well as  
$H\rightarrow \tau\tau$ searches for large $t_{\beta}$, which rule out $m_{H} < 600$ GeV for $t_{\beta} = 30$. We also show the 
flavour bound $m_{H^{\pm}} > 480$ GeV as a horizontal grey line for indicative purposes. 

Regarding unitarity, 
for Case 1 ($m_{12}^2=m_H^2 s_\beta c_\beta$) with  $m_A> m_H=m_{H^\pm}$ the eigenvalues 
of the scattering matrix are $|\Lambda_{i \neq 6}| v^2 = m_A^2-m_H^2 \pm \mathcal{O}(m_h^2)$ and 
$|\Lambda_{6}| v^2 = 5(m_A^2-m_H^2) +m_h^2$. The latter imposes the strongest constraint, which rules out regions 
with a very large mass splitting $m_A - m_H$ (as indicated by the hatched region in the lower-right corner of each panel in Figures~\ref{fig:bp-CH-vs-A} and~\ref{fig:bp-HC-vs-A}). 
For Case 2 ($m_{12}^2=0$) with $m_A> m_H=m_{H^+}$, the  strongest unitarity constraints come from
$|\Lambda_{6}| v^2 = 5m_A^2-3m_H^2 \pm \mathcal{O}(m_h^2)$ and 
$|\Lambda_{11}| v^2 = \frac{1}{2}m_H^2(\frac{1}{t_\beta^2}+t_\beta^2 ) + \frac{1}{2} \sqrt{ 9m_H^4 (\frac{1}{t_\beta^2} -t_\beta^2 )^2+4 m_A^4} \pm \mathcal{O}(m_h^2)$.  
In particular, $|\Lambda_{11}|$ rules out the large $m_H$ region (upper hatched region in the lower right panel 
in Figures~\ref{fig:bp-CH-vs-A} and~\ref{fig:bp-HC-vs-A}).  

Taking into account both the theoretical and experimental constraints, relatively large regions 
of $m_A$ vs. $m_H=m_{H^\pm}$ remain viable and having a sizable signal cross section for small to intermediate values of $t_{\beta}$ 
for Case I. For $t_{\beta}\gg 1$, only the region $m_A \gtrsim m_{H} = m_{H^\pm} > 600$ GeV still survives. 
For Case 2, given the unitary constraints ruling out large values of $m_H$ and $m_A$, only the region 200 GeV $< m_A < 650$ GeV
and 175 GeV $< m_H=m_{H^\pm}<450$ GeV remains viable.

\begin{figure}[h!]
 \centering
 	\includegraphics[width=0.495\textwidth]{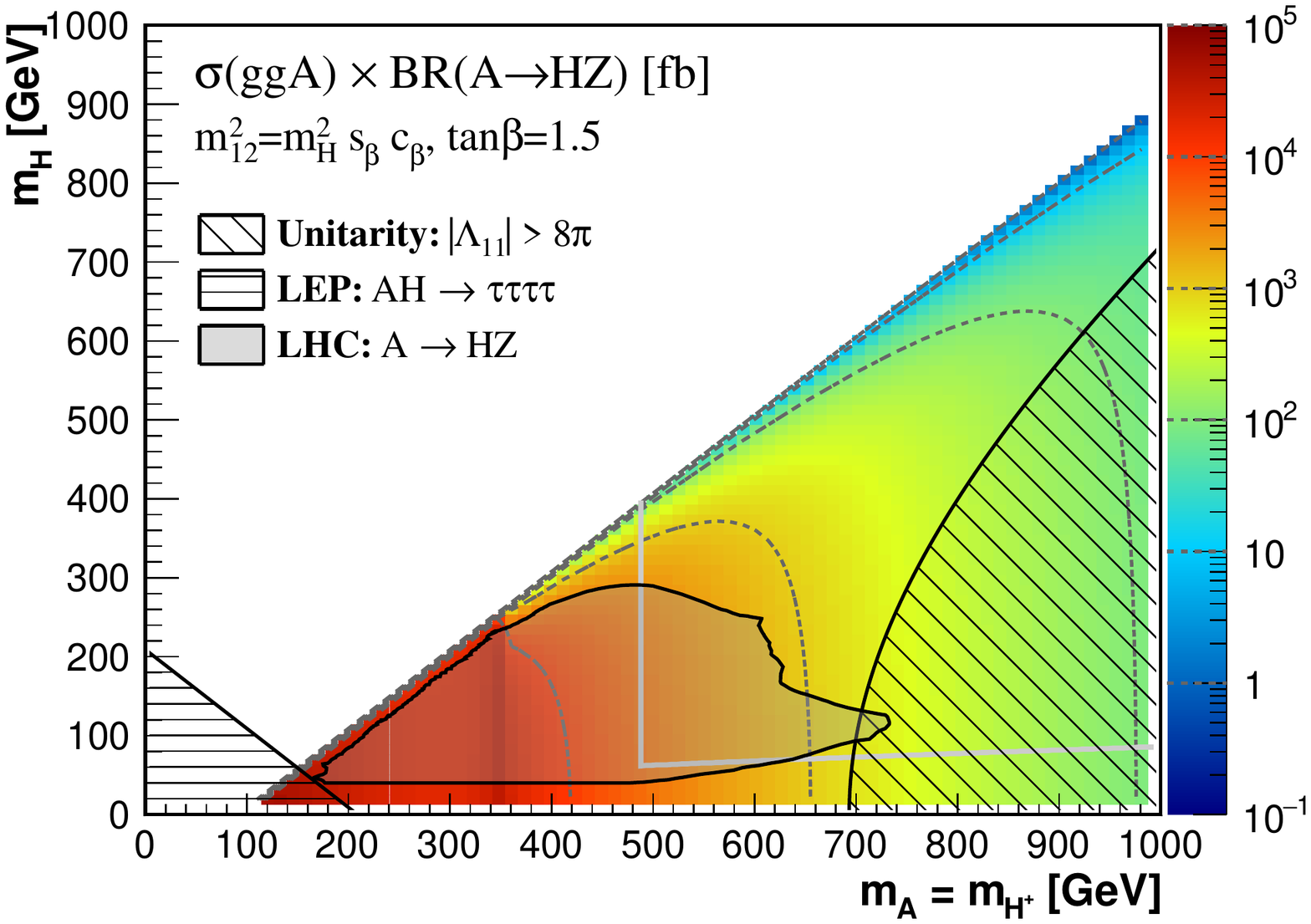}
	\includegraphics[width=0.495\textwidth]{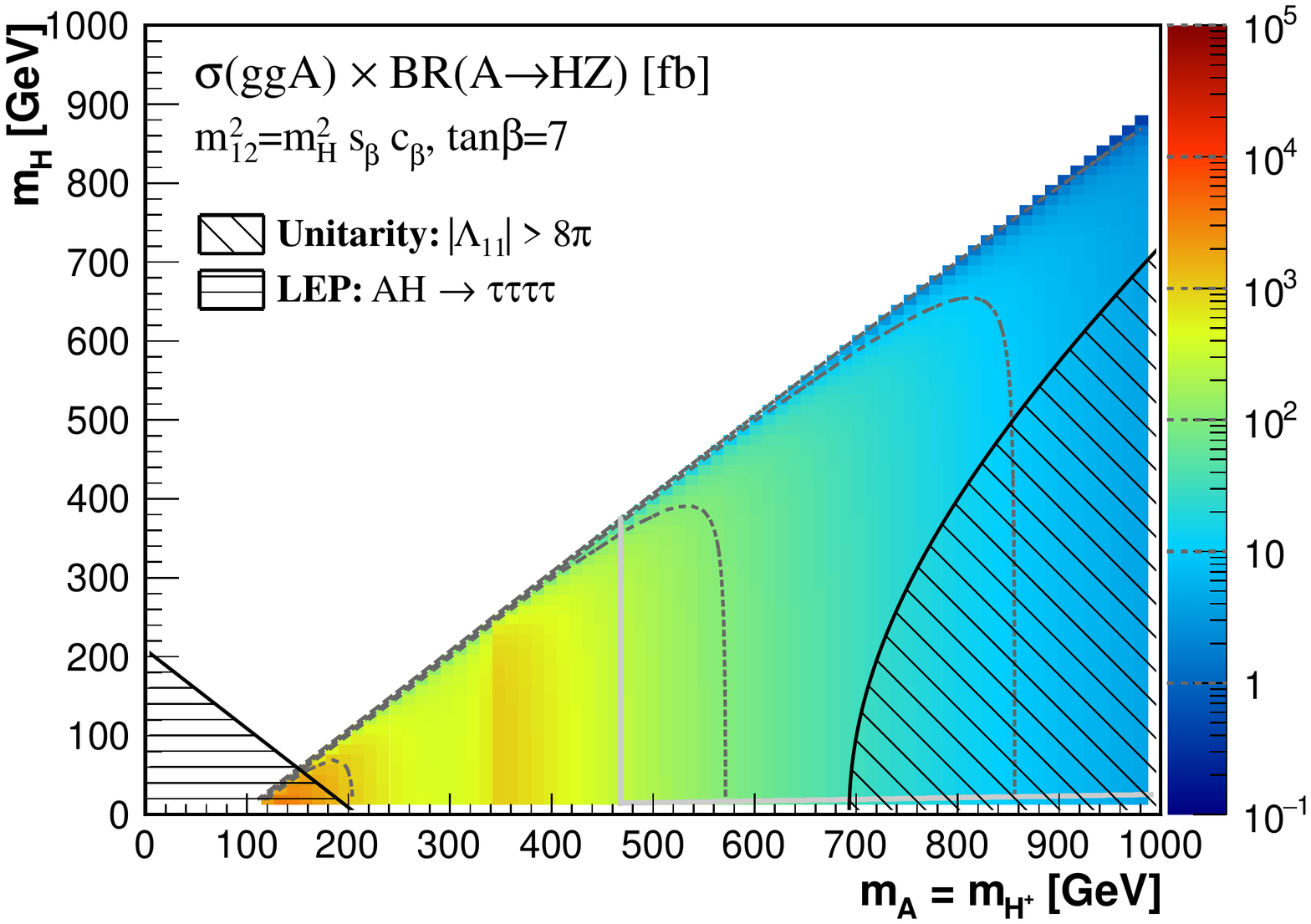}
	\includegraphics[width=0.495\textwidth]{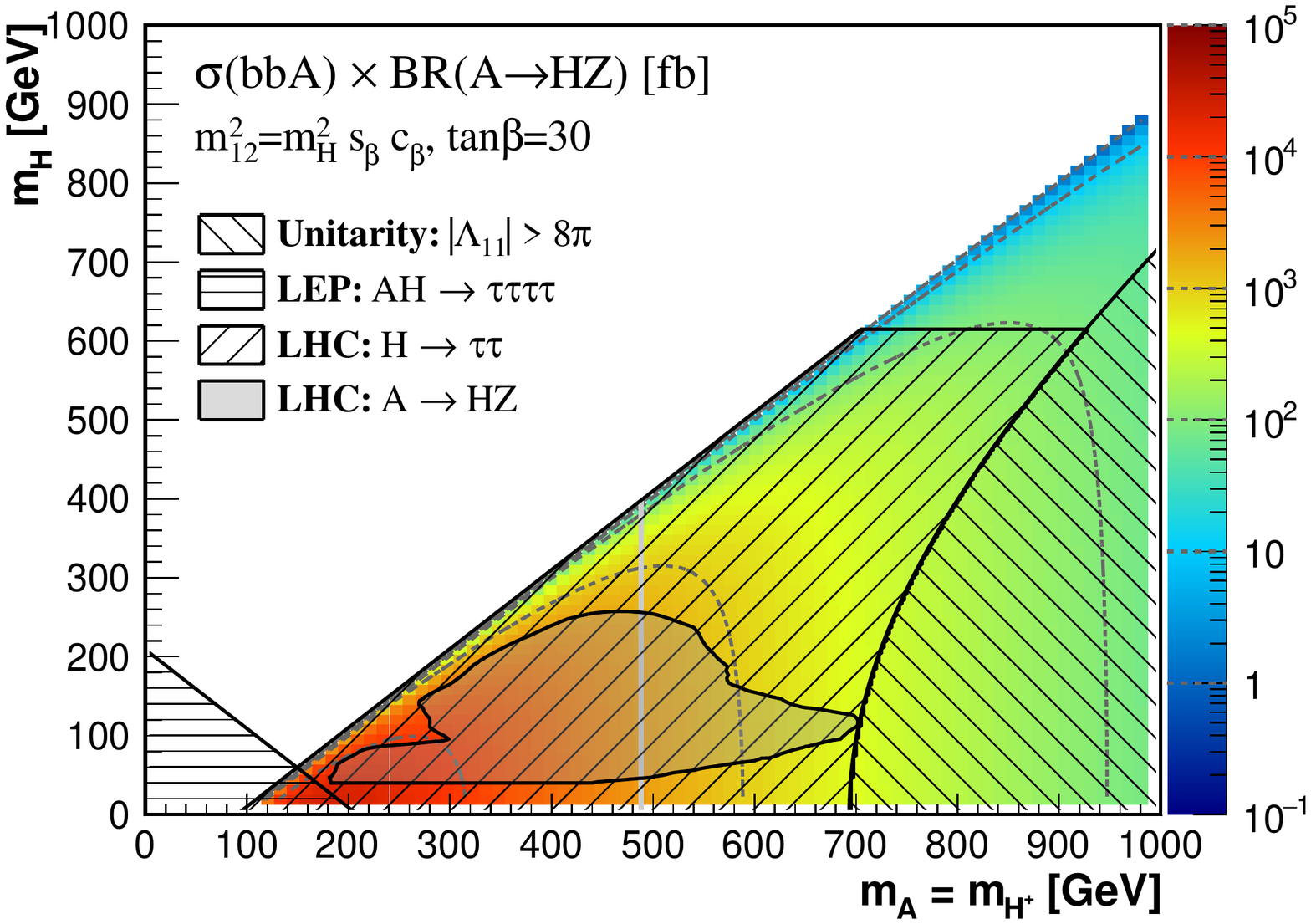}
	\includegraphics[width=0.495\textwidth]{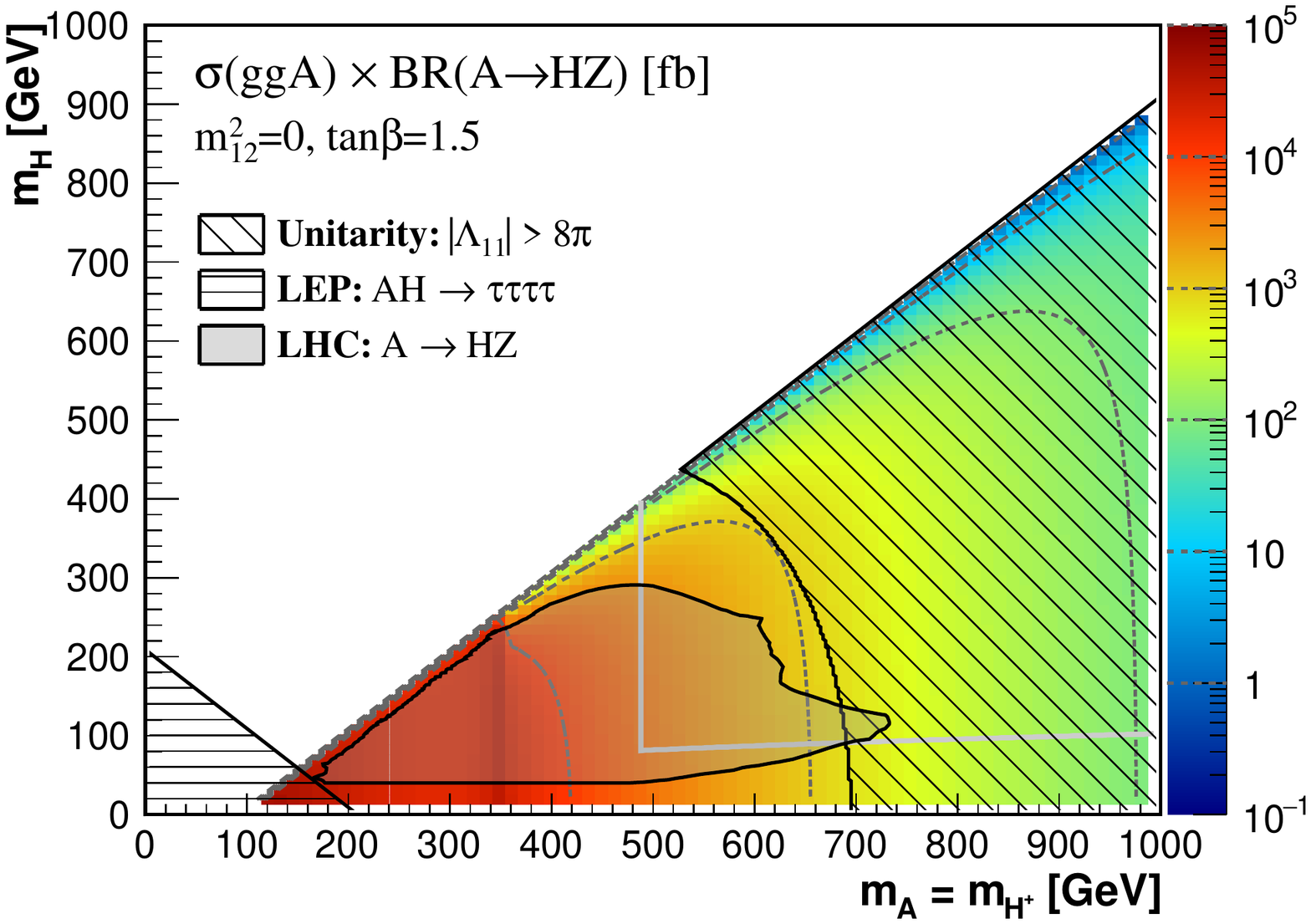}
\caption{$\sigma\times{\rm BR}$ for the exotic decay $A \rightarrow HZ$ in BP IIB: $m_A=m_{H^{\pm}}$ vs $m_H$ plane (see caption of Figure~\ref{fig:bp-CH-vs-A} 
for further details). The solid horizontal and vertical light grey lines indicate the various flavour constraints.}
\label{fig:bp-H-vs-AC}
\end{figure}
 
\subsubsection{BP IIB:  $m_H < m_A=m_{H^\pm}$}

\begin{figure}[h!]
 \centering
 	\includegraphics[width=0.495\textwidth]{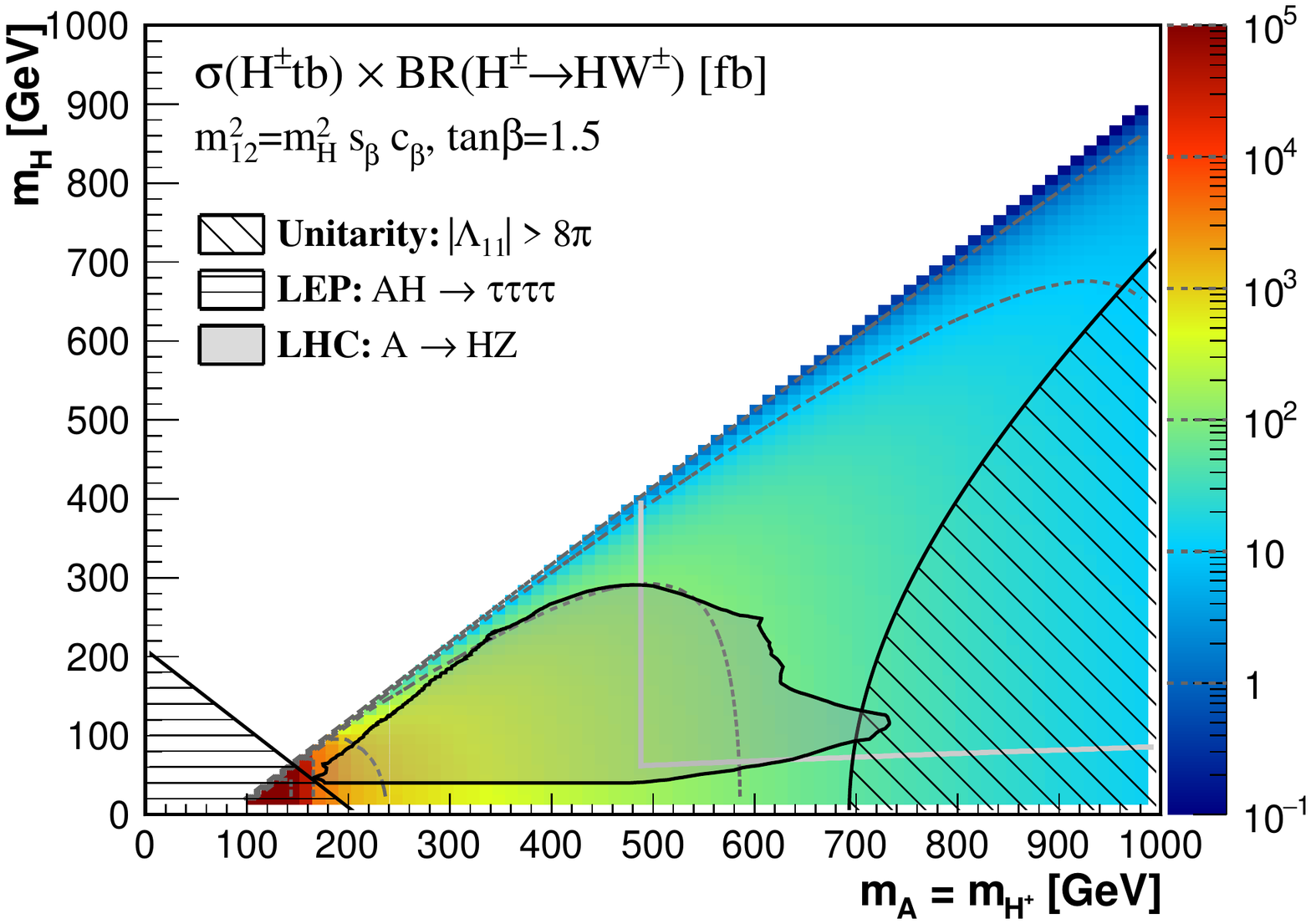}
	\includegraphics[width=0.495\textwidth]{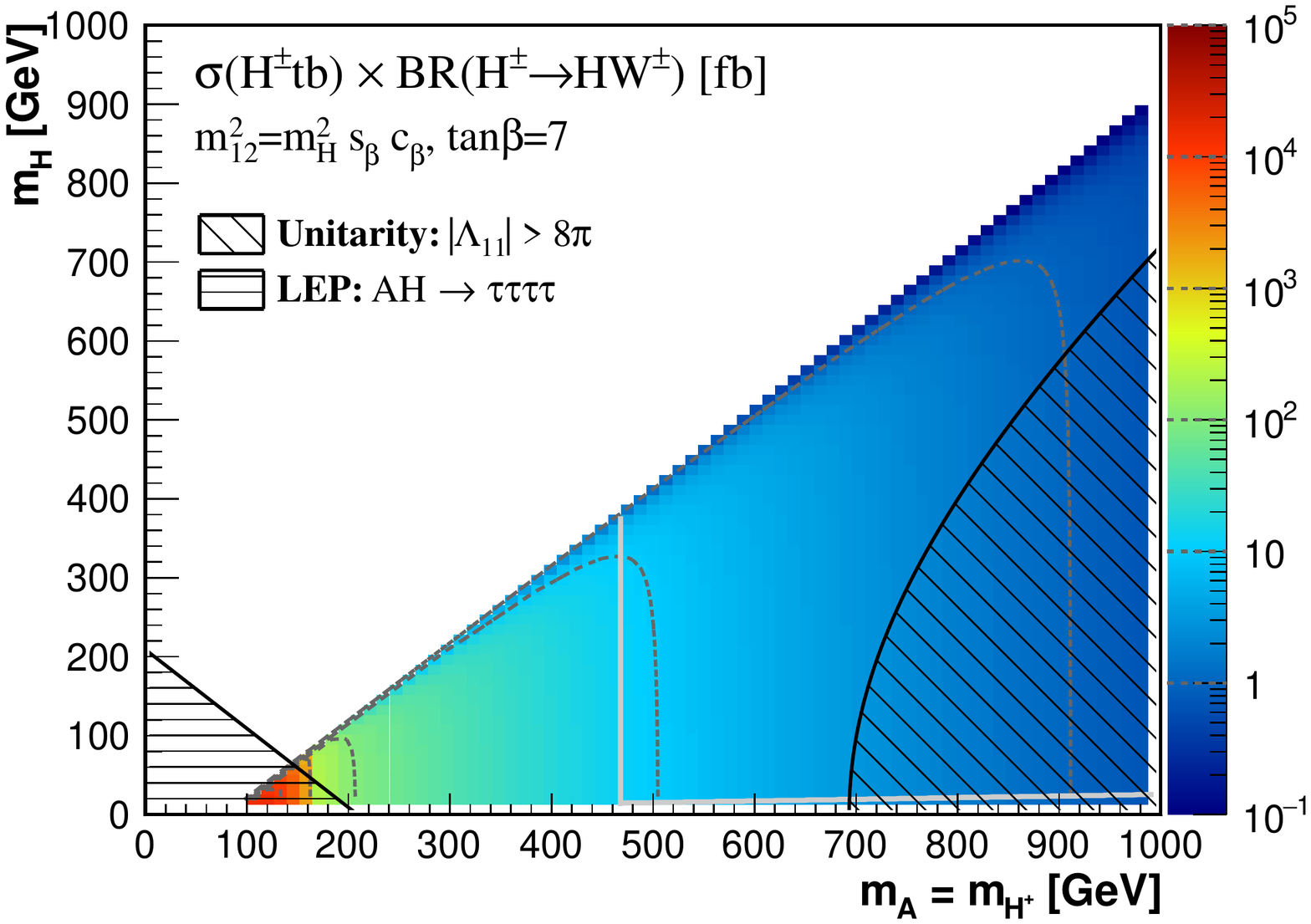}
	\includegraphics[width=0.495\textwidth]{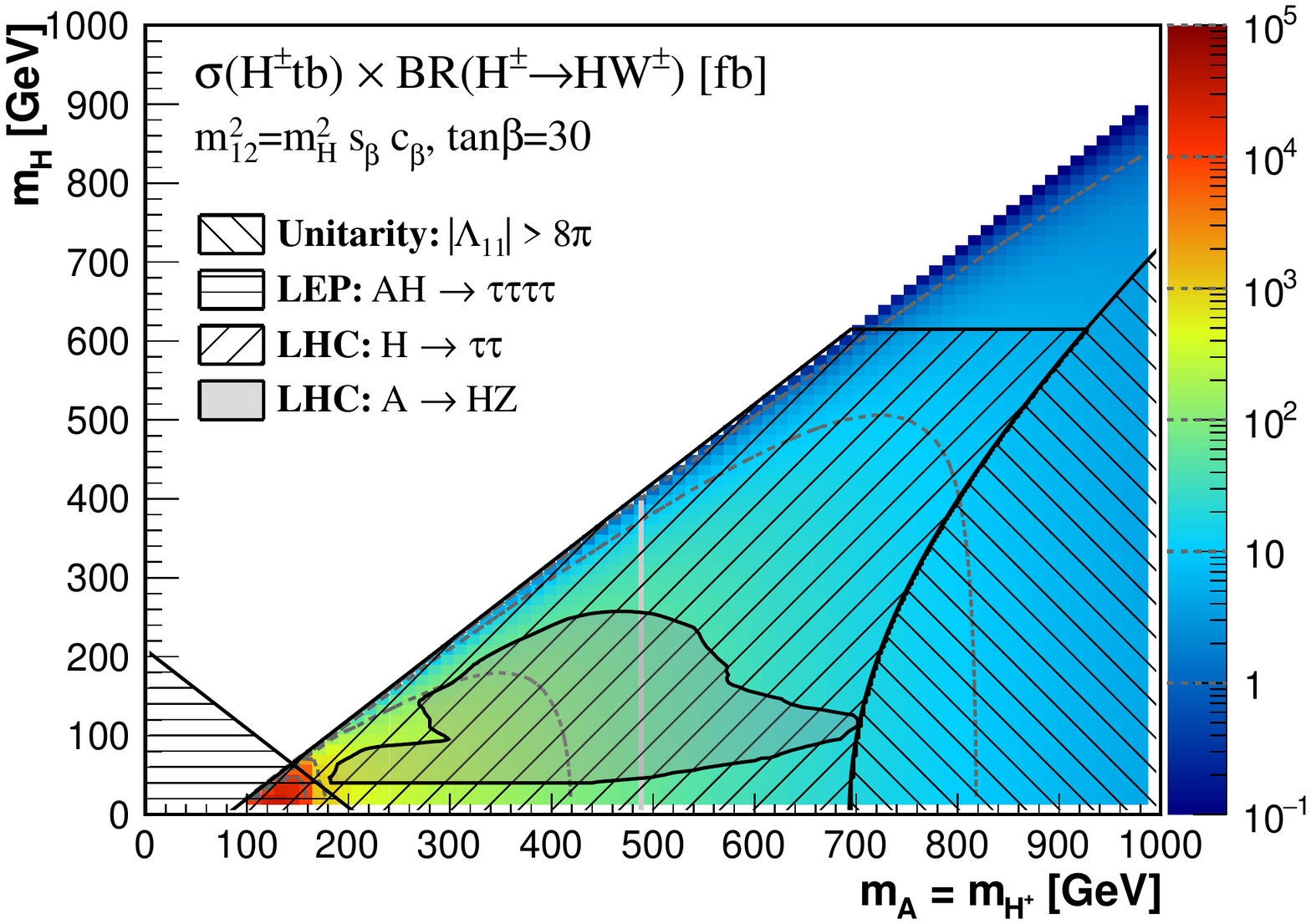}
	\includegraphics[width=0.495\textwidth]{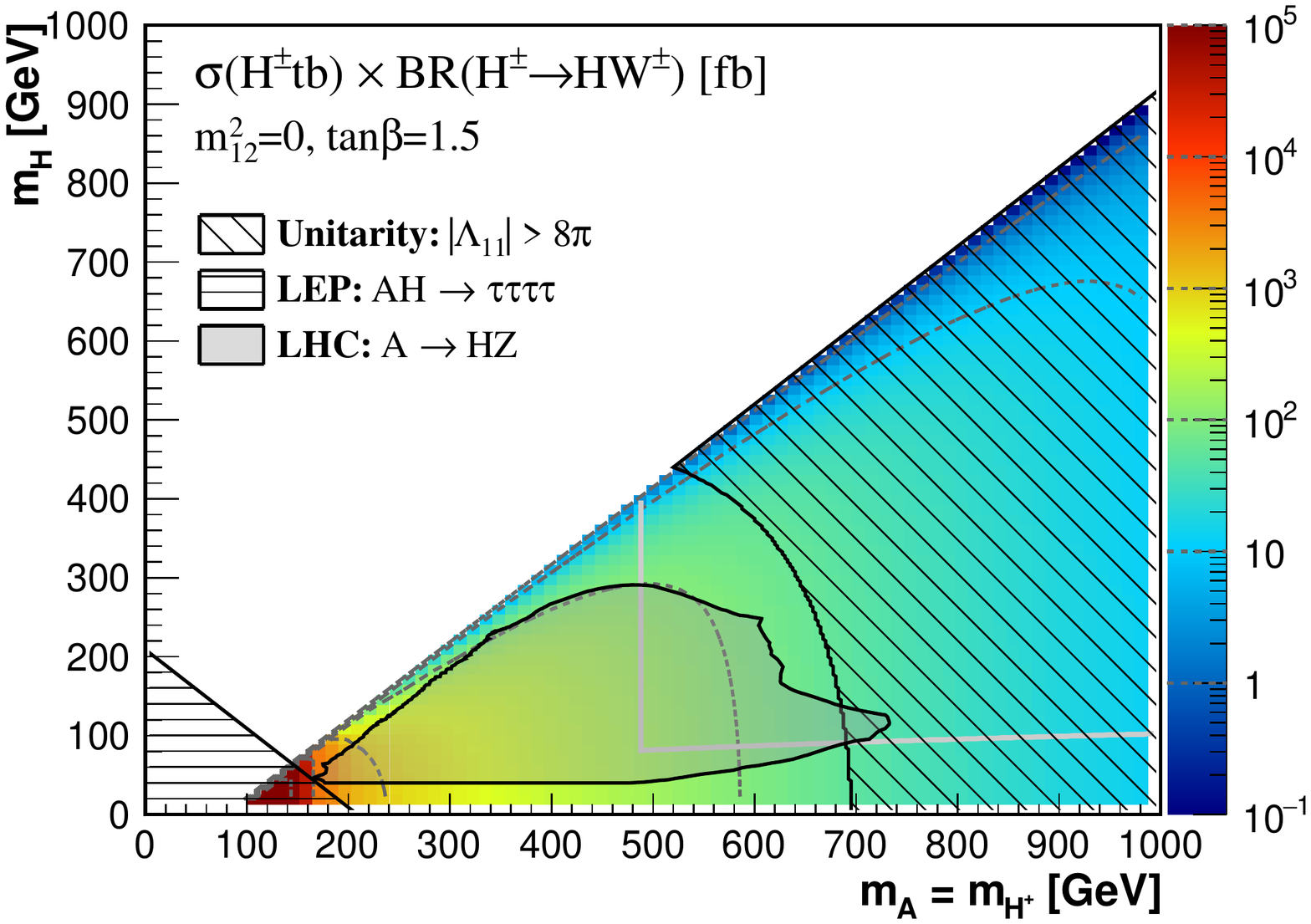}
\caption{$\sigma\times{\rm BR}$ for the exotic decay $H^\pm \rightarrow HW^\pm$ in BP IIB: $m_A=m_{H^{\pm}}$ vs. $m_H$  plane (see caption of Figure~\ref{fig:bp-CH-vs-A} 
for further details). The solid horizontal and vertical light grey lines indicate the various flavour constraints.}
\label{fig:bp-H-vs-CA}
\end{figure}

For $m_H< m_A=m_{H^\pm} $, the dominant exotic decay channels are $A\rightarrow HZ$ and $H^\pm \rightarrow HW^\pm$. 
We show the $\sigma\times{\rm BR}$ for $A \rightarrow HZ$ and $H^\pm \rightarrow HW^\pm$ respectively in Figures~\ref{fig:bp-H-vs-AC} and \ref{fig:bp-H-vs-CA}. 
The low $m_A + m_H$ region is ruled out by the LEP search $e^+e^-\rightarrow AH$ (recall the discussion in 
Section~\ref{sec:LEP_LHC}), while unitarity constraints bound large values for $m_A$, $m_H$: For Case I the strongest constraints arise from 
$|\Lambda_{2,4,5,6,11,12}| v^2 = 3(m_A^2-m_H^2) \pm \mathcal{O}(m_h^2)$, which limit $m_A^2-m_H^2$ for large $m_H$ and/or $m_A$. For Case 2, 
large values of either $m_A$ or $m_H$ are excluded, since the strongest unitarity constraint comes from
$|\Lambda_{11}| v^2 = \frac{1}{2}m_H^2(\frac{1}{t_\beta^2}+t_\beta^2 ) + \frac{1}{2}\sqrt{ 9m_H^4(\frac{1}{t_\beta^2} -t_\beta^2 )^2+4(3m_A^2-2m_H^2)^2} \pm \mathcal{O}(m_h^2)$.

For Case 1 with $t_{\beta}=1.5$ (upper left panel of Figure~\ref{fig:bp-H-vs-AC}), signal cross sections for $A \rightarrow HZ$ in excess of 10 pb are viable 
given all the constraints, while we note that the LHC Run 1 CMS $A\to H Z$ search rules out a large portion of the parameter space with $m_H < 300$ GeV and $m_A < 650$ GeV.  
Intermediate values of $t_{\beta}$ (exemplified by the $t_{\beta}=7$ case shown in the upper right panel of Figure~\ref{fig:bp-H-vs-AC}) 
only permit signal cross sections below 1 pb, due to the small gluon fusion production 
cross section (for $m_A > 600$ GeV the $\sigma \times$ BR values are in fact below 20 fb). 
For $t_{\beta}=30$ (lower left panel of Figure~\ref{fig:bp-H-vs-AC}), the current collider search of $H\to \tau\tau$ rules out $m_H<600$ GeV, 
leaving only a small corner of parameter space allowed, with signal cross sections $\sigma \times$ BR $\lesssim 100$ fb. 
For Case 2, the lower right panel of Figure~\ref{fig:bp-H-vs-AC} shows that the CMS $A \to H Z$ search constrains most of the viable parameter space, 
which may in turn be probed completely by LHC 13. 
 
While the generic features for $H^\pm \rightarrow H W^\pm$ are similar to those of $A\to HZ$, the signal cross sections are about two order 
of magnitude smaller, due to the suppressed production cross section of $pp\rightarrow H^\pm tb$. This, in addition to the  
complicated final state $H W^{+}W^{-} bb$ which results, makes this channel challenging for LHC studies at 13 TeV. 

 \subsubsection{BP IB: $m_A < m_{H^{\pm}}=m_H $}

\begin{figure}[h!]
 \centering
	\includegraphics[width=0.495\textwidth]{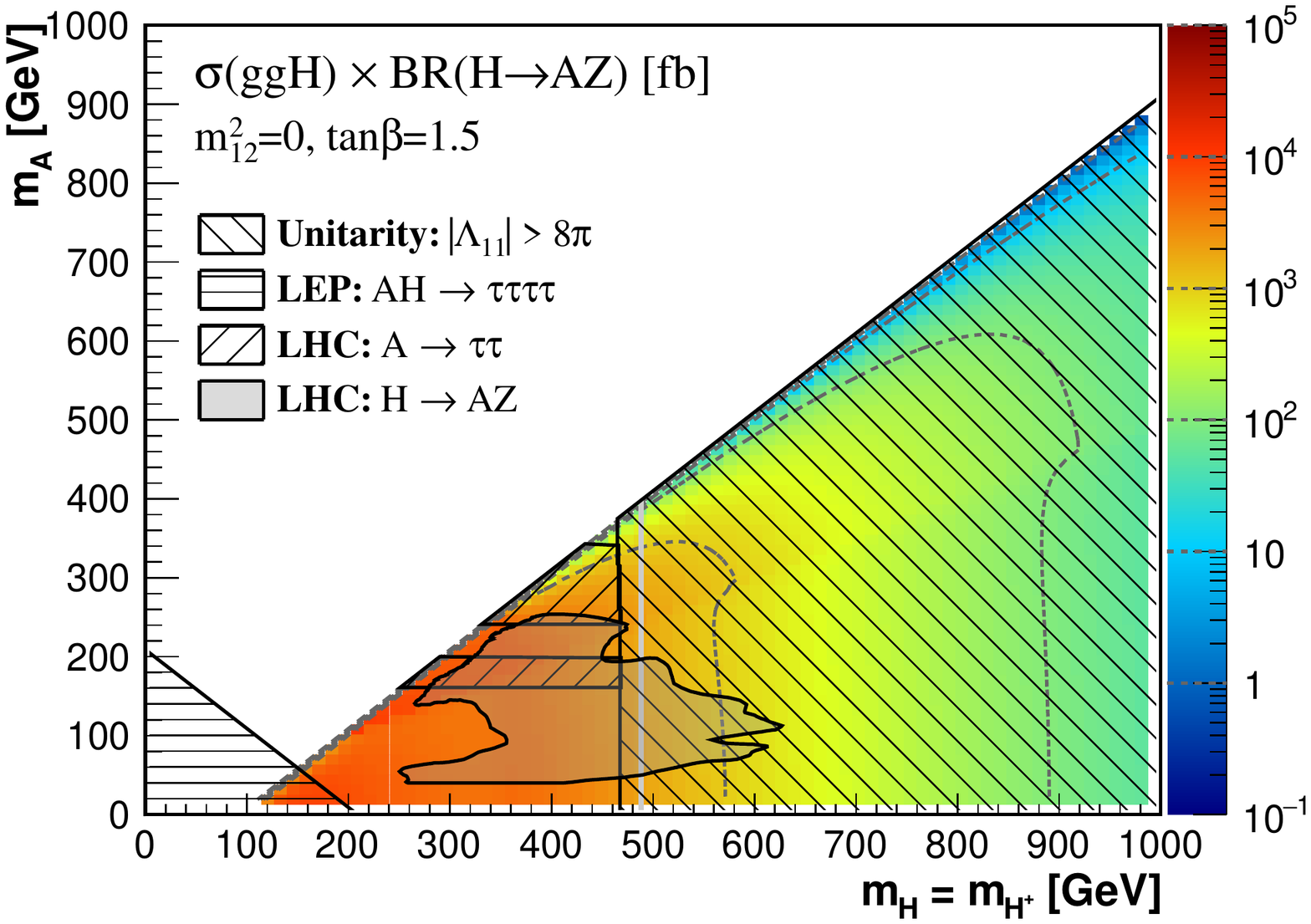}
 	\includegraphics[width=0.495\textwidth]{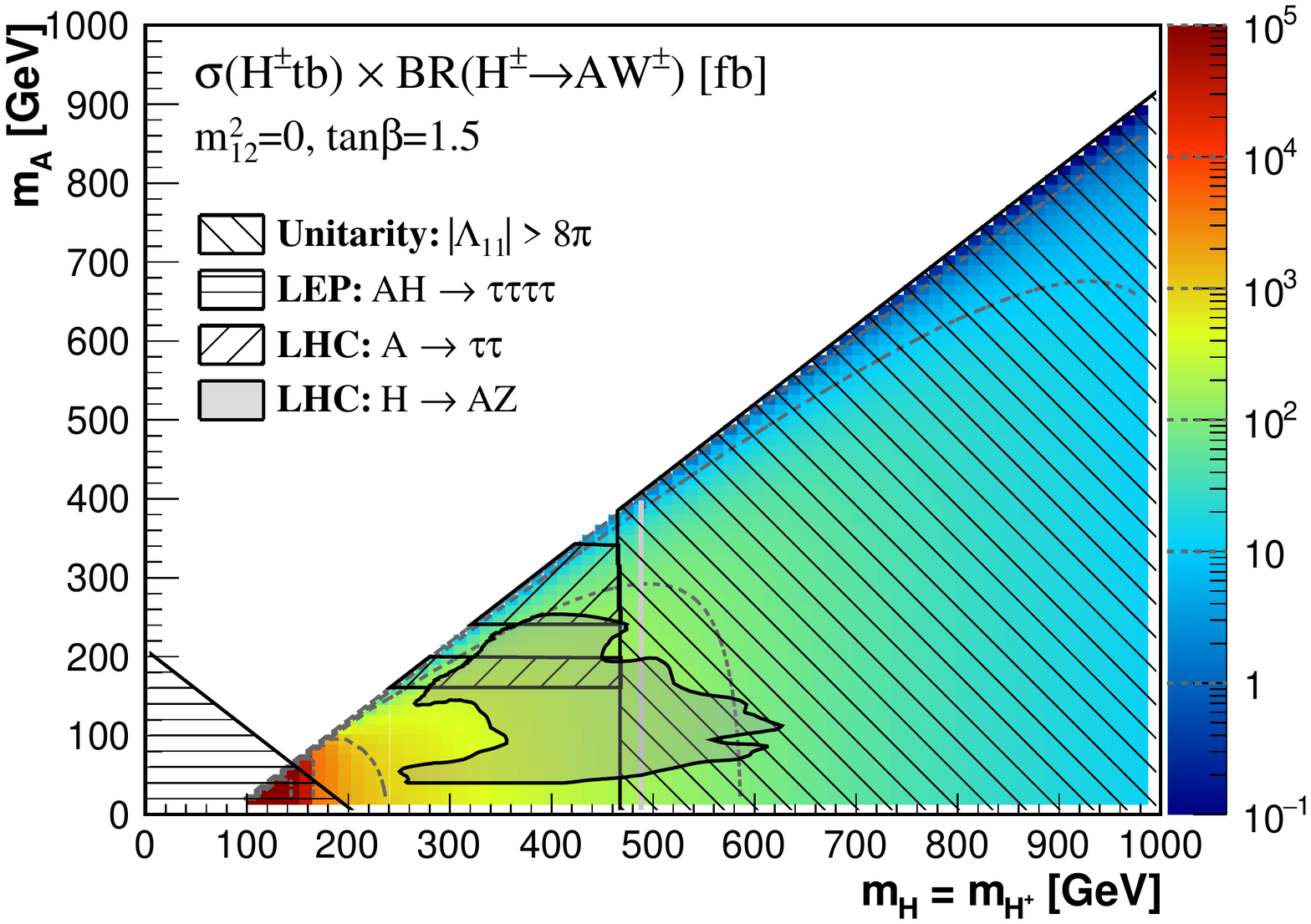}
	\includegraphics[width=0.495\textwidth]{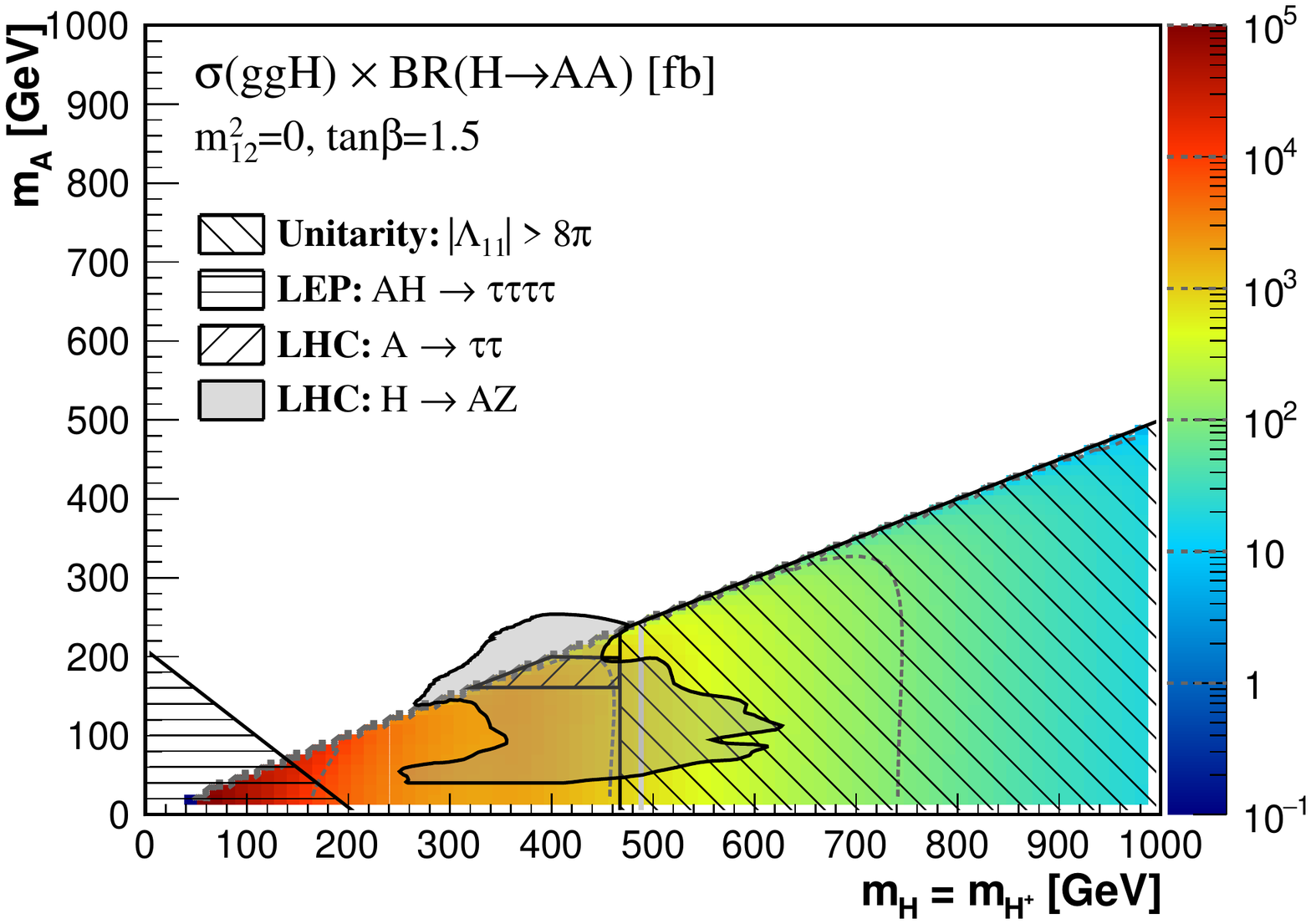}
	
	\vspace{-2mm}
\caption{$\sigma\times{\rm BR}$ for the exotic decays $H\to AZ$ (up left), $H^\pm \rightarrow AW^\pm$ (up right) and $H\to AA$ (down) in BP IB: $m_H=m_{H^{\pm}}$ vs. 
$m_A$ plane, for Case 2 ($m_{12}^2 =0$) with $t_{\beta} = 1.5$. 
Contour lines of  10, $10^2$ and $10^3$ fb are drawn as light grey dashed curves to guide the eye. 
The shaded areas enclosed by an irregular curve and hatched regions are ruled out by theoretical and experimental constraints (see text for details). 
The solid (vertical) light grey lines indicate the flavour constraint $m_{H^{\pm}} > 480$ GeV.}
\label{fig:BPIB}
\end{figure}

In this scenario, only Case 2 ($m_{12}^2=0$) is viable. The $\sigma \times$ BR for the three possible exotic decay channels   
$H \to AZ$, $H^\pm \to A W^\pm$ and $H\to AA$ is shown in Figure~\ref{fig:BPIB} for our benchmark $t_{\beta} = 1.5$. 
The $m_H>460$ GeV region is excluded by unitarity, the strongest unitarity constraint coming from 
$|\Lambda_{11}| v^2 = \frac{1}{2}m_H^2(\frac{1}{t_\beta^2}+t_\beta^2 ) +\frac{1}{2}\sqrt{ 9m_H^4(\frac{1}{t_\beta^2} -t_\beta^2 )^2+ 4 m_A^4} \pm \mathcal{O}(m_h^2)$.  

Below the unitarity limit on $m_H$, the ATLAS/CMS limits on $A\to \tau \tau$ at low $t_{\beta}$ ($ggA$ production)~\cite{Aad:2014vgg,Khachatryan:2014wca} 
combined with the bounds from the CMS $H \to A Z$ search~\cite{CMS:2015mba,Khachatryan:2016are} rule out $m_A > 40$ GeV down to $m_H \lesssim 350$ GeV.
As can be seen from Figure~\ref{fig:BPIB}, only a small region of parameter space survives the unitarity and LHC 8 TeV constraints.
We also stress that in this case including the flavour constraint $m_{H^{\pm}} > 480$ GeV would rule out this benchmark scenario completely.

\vspace{-2mm}

\subsubsection{BP IIA: $m_H > m_A=m_{H^\pm}$}

\begin{figure}[h!]
 \centering
	\includegraphics[width=0.495\textwidth]{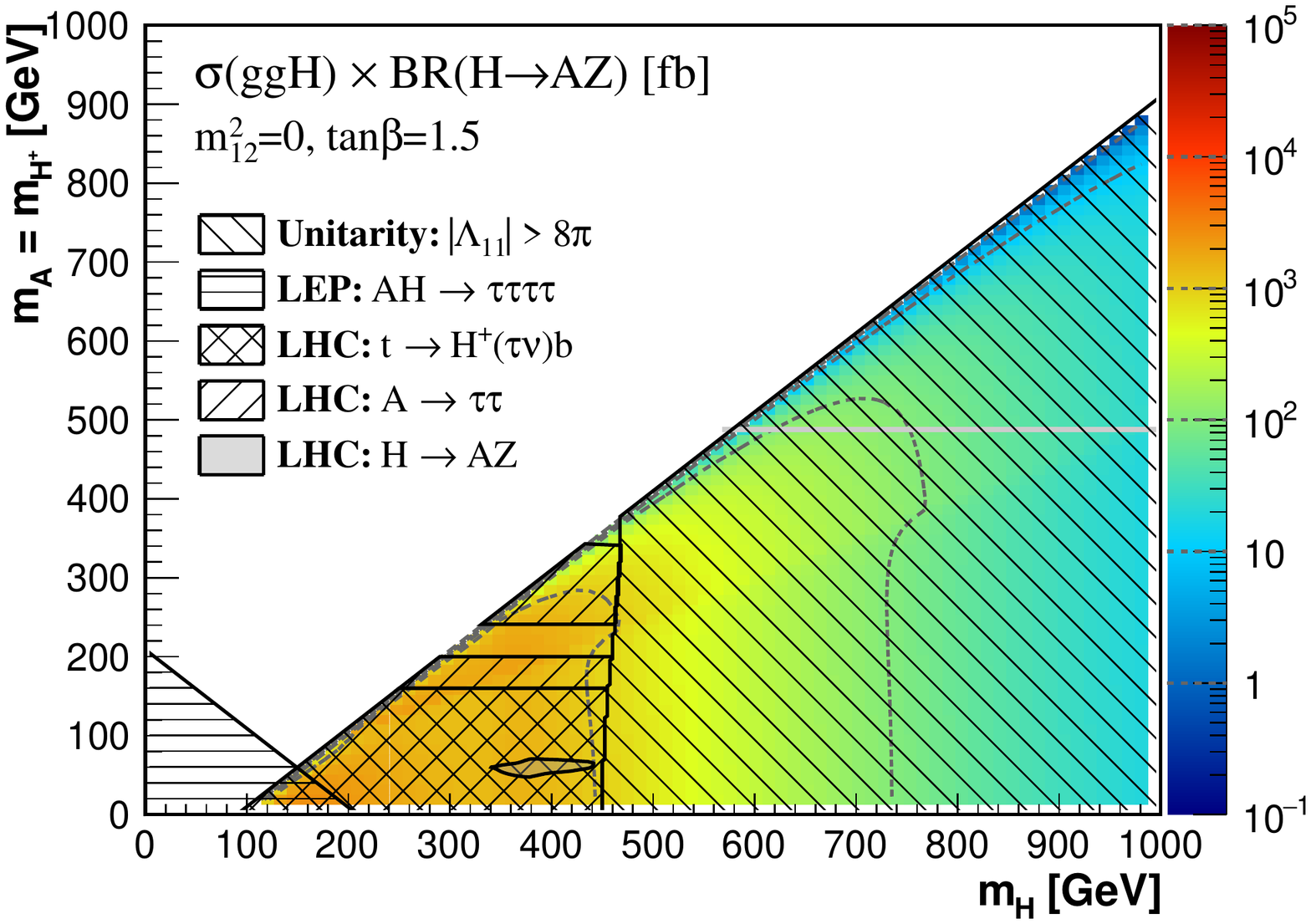}
	\includegraphics[width=0.495\textwidth]{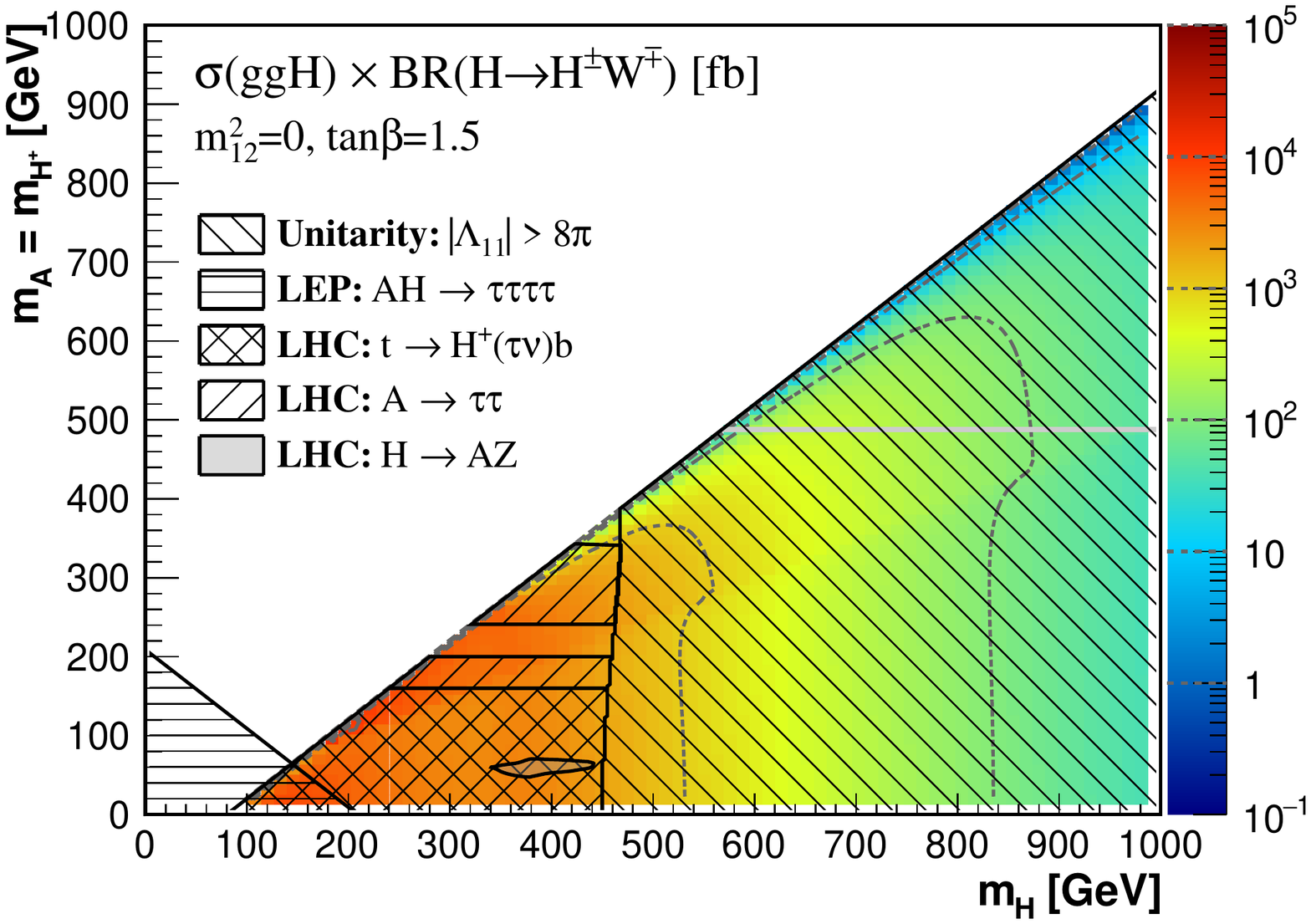}
	\includegraphics[width=0.495\textwidth]{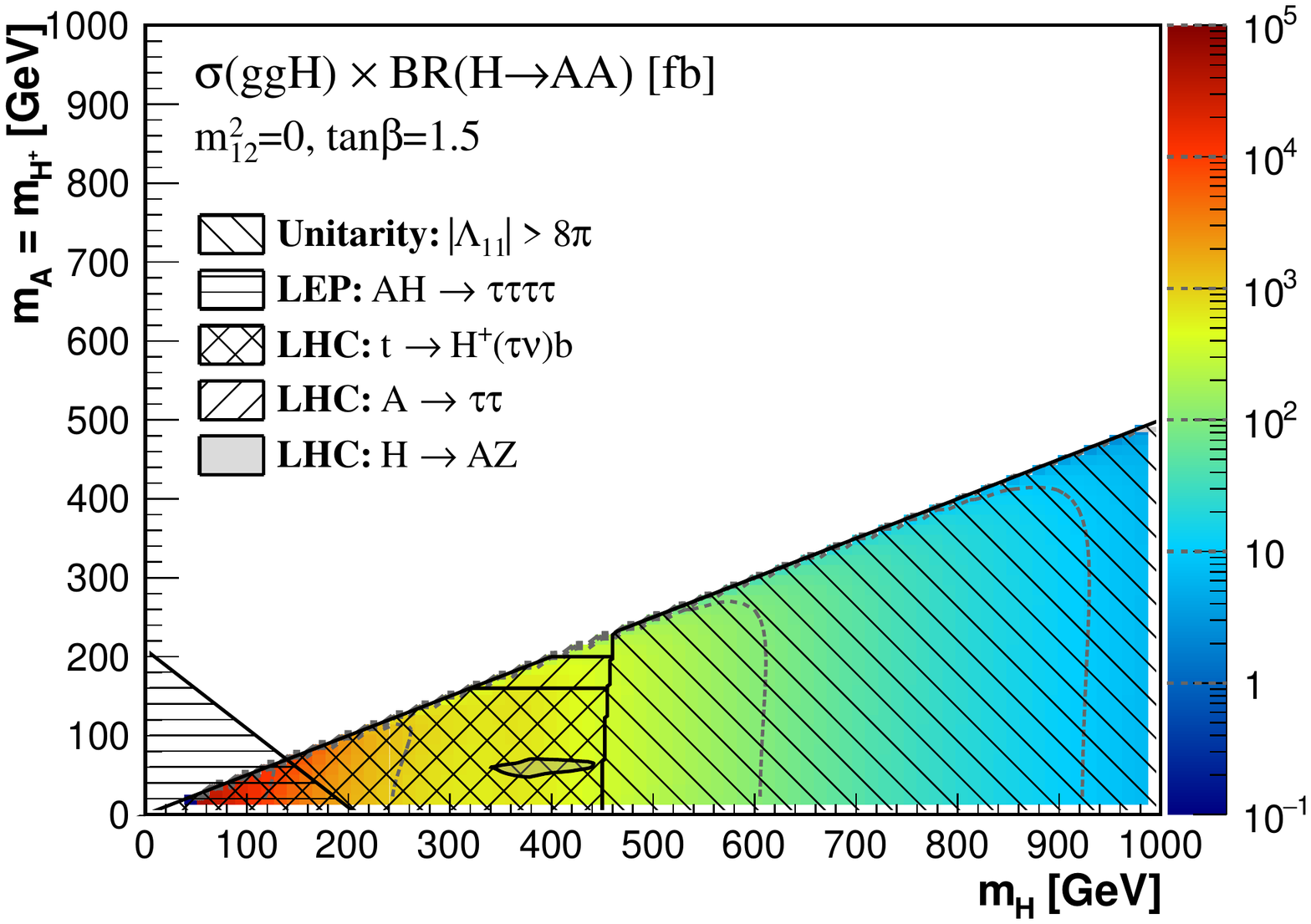}
	\includegraphics[width=0.495\textwidth]{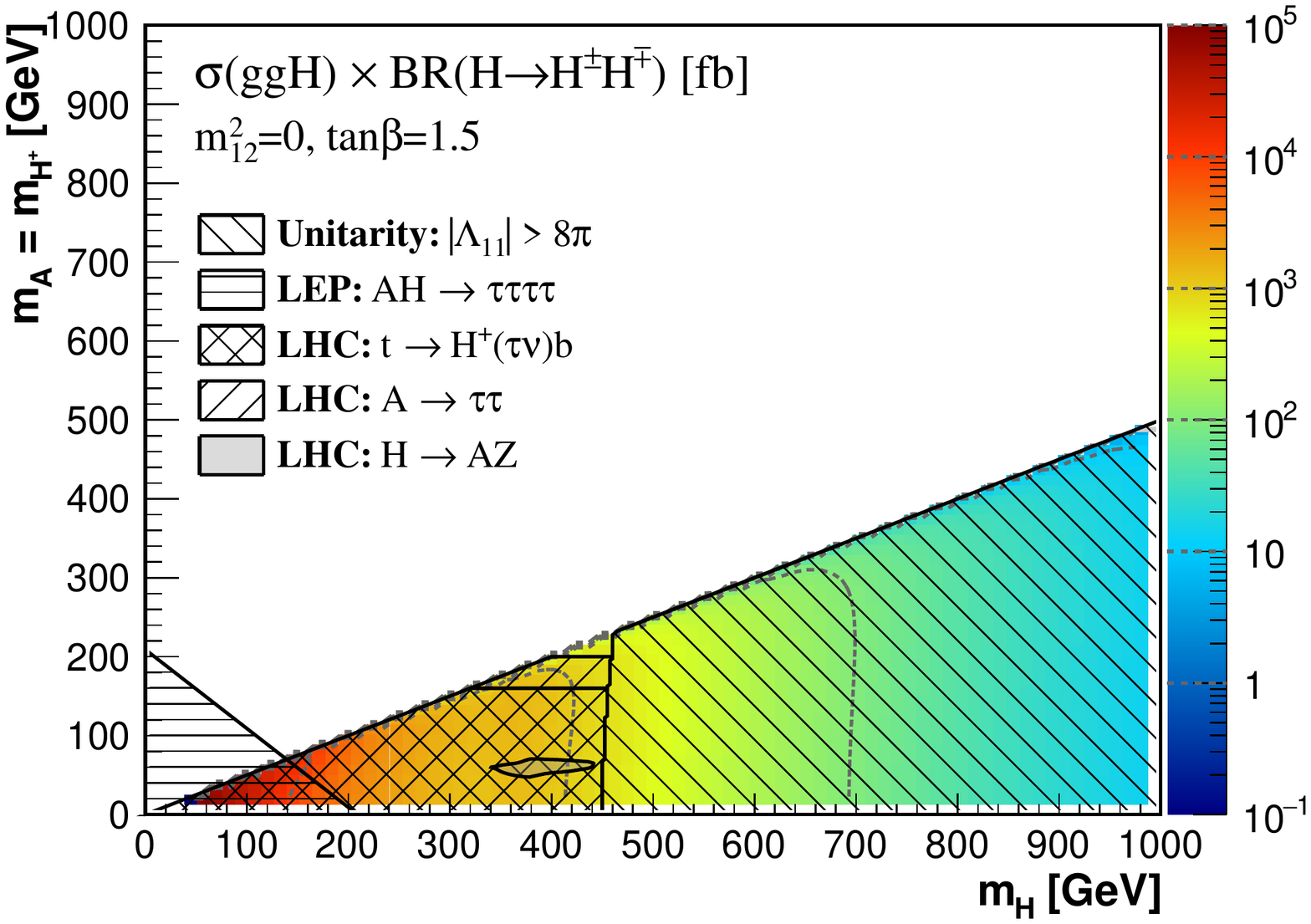}
	
	\vspace{-2mm}
	\caption{$\sigma\times{\rm BR}$ for the exotic decays $H\to AZ$ (up left), $H^\pm \rightarrow H^\pm W^\mp$ (up right), $H\to AA$ (down left) and $H\to H^+H^-$ 
(down right) in BP IIA: $m_H$ vs. $m_A=m_{H^{\pm}}$ plane (see caption of Figure~\ref{fig:BPIB} for further details).}
\label{fig:BPIIA}
\end{figure}

Four exotic Higgs decay channels, $H\to AZ$, $H \rightarrow H^\pm W^\mp$, $H\to AA$, and $H\to H^+H^-$ are possible for BP IIA
(which we recall is only allowed for Case 2), shown respectively in the four panels of Figure~\ref{fig:BPIIA}. 
Comparing to BP IIB, the additional collider search limit $m_{H^\pm} > m_t$ applies,
which overlaps with the 8 TeV LHC exclusion from $A\to \tau\tau$. This results in only a small stripe in parameter space, corresponding to 
200 GeV $< m_A=m_{H^\pm} < 240$ GeV and 300 GeV $<m_H<450$ GeV, being viable.
Moreover, we note that the decays $H\rightarrow AA$ and $H\rightarrow H^+H^-$ are essentially not kinematically allowed in the viable region, as 
shown in the lower panels of Figure~\ref{fig:BPIIA}. This benchmark scenario should indeed be possible to probe completely at LHC 13 TeV.

 \subsection{Exotic Decays into $h$ Away from Alignment}
\label{sec:nonalign}

\subsubsection{BP III: $m_A=m_H=m_{H^\pm}$ vs. $c_{\beta-\alpha}$}

\begin{figure}[h!]
 \centering
 	\includegraphics[width=0.495\textwidth]{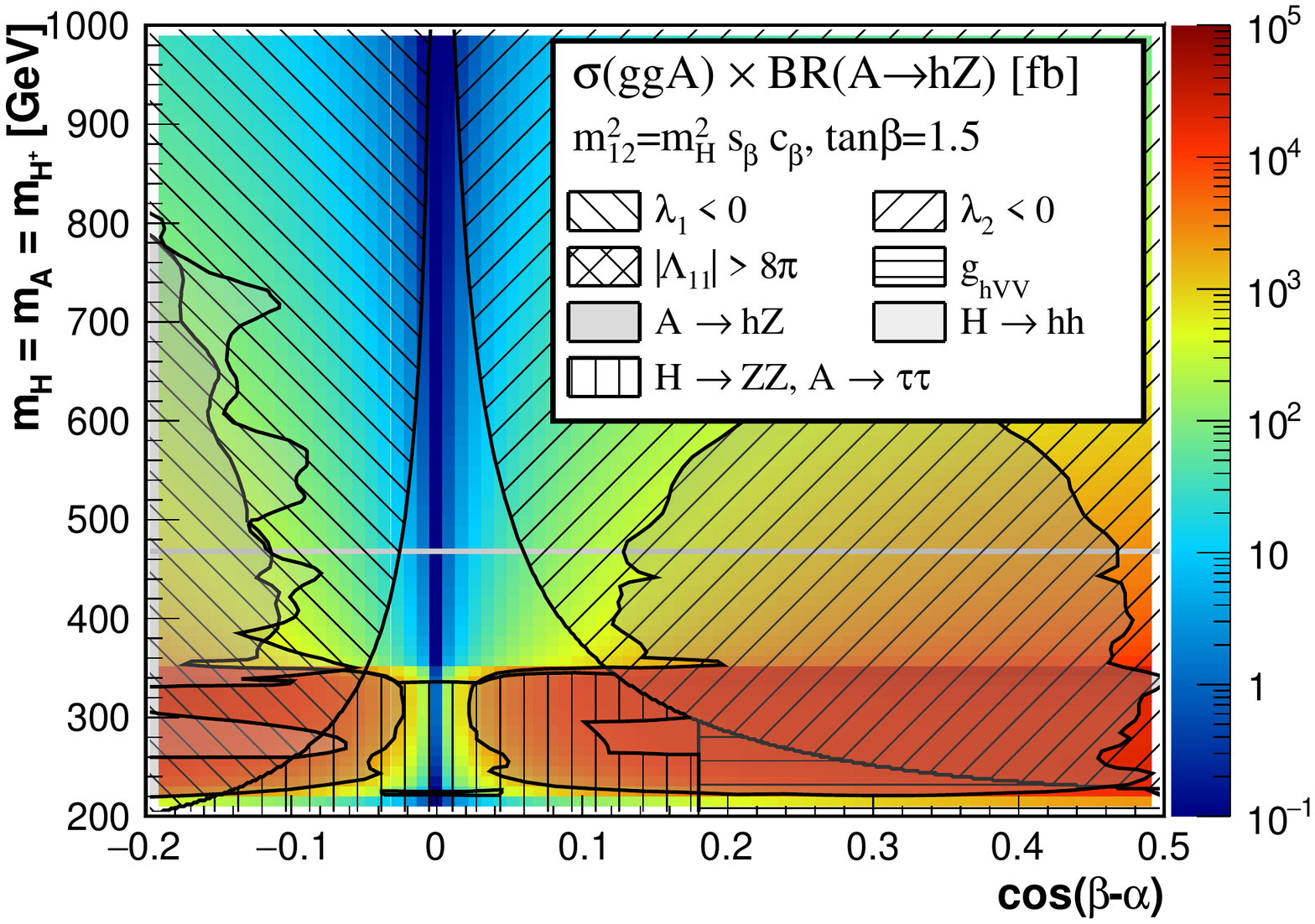}
	\includegraphics[width=0.495\textwidth]{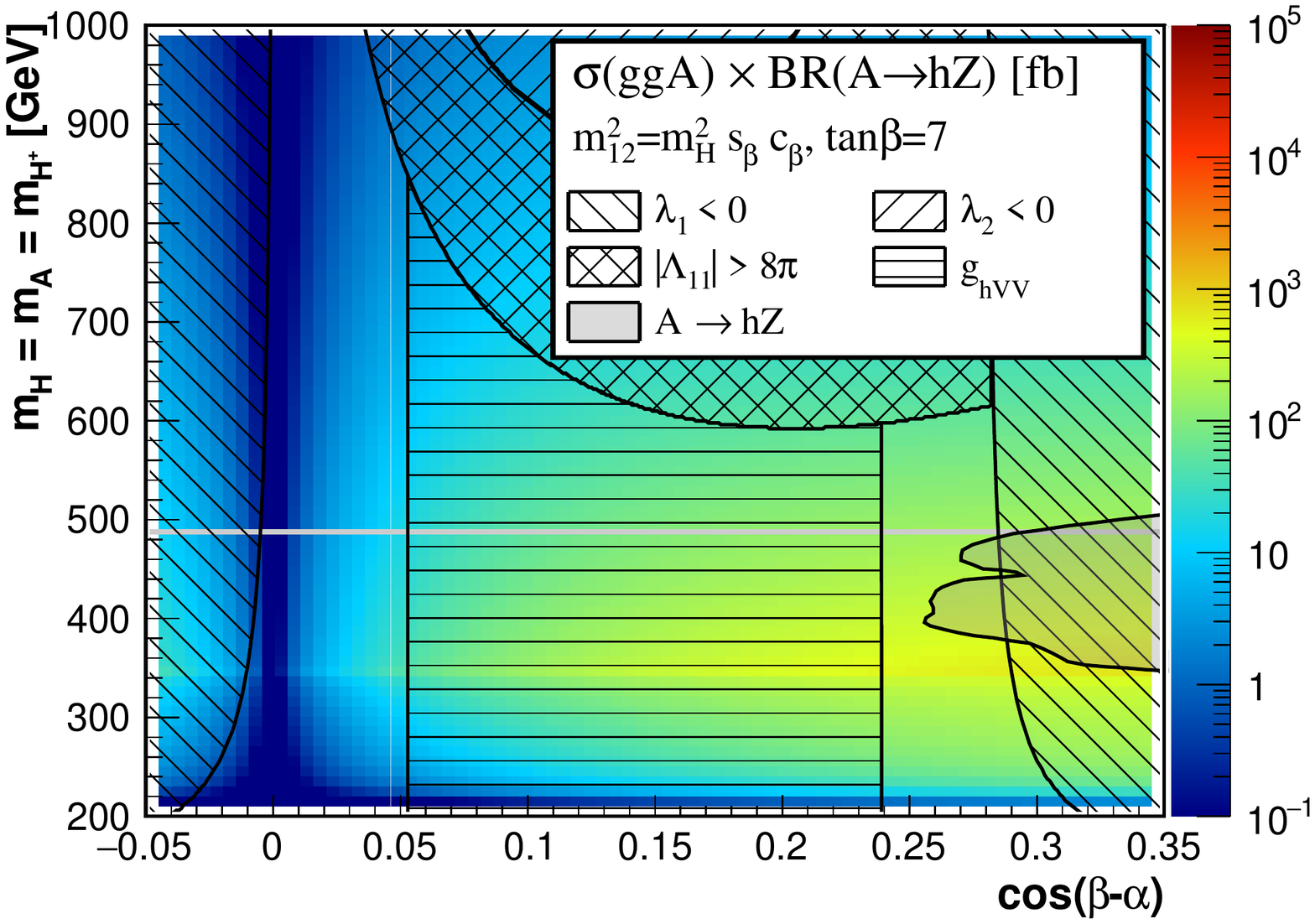}
	\includegraphics[width=0.495\textwidth]{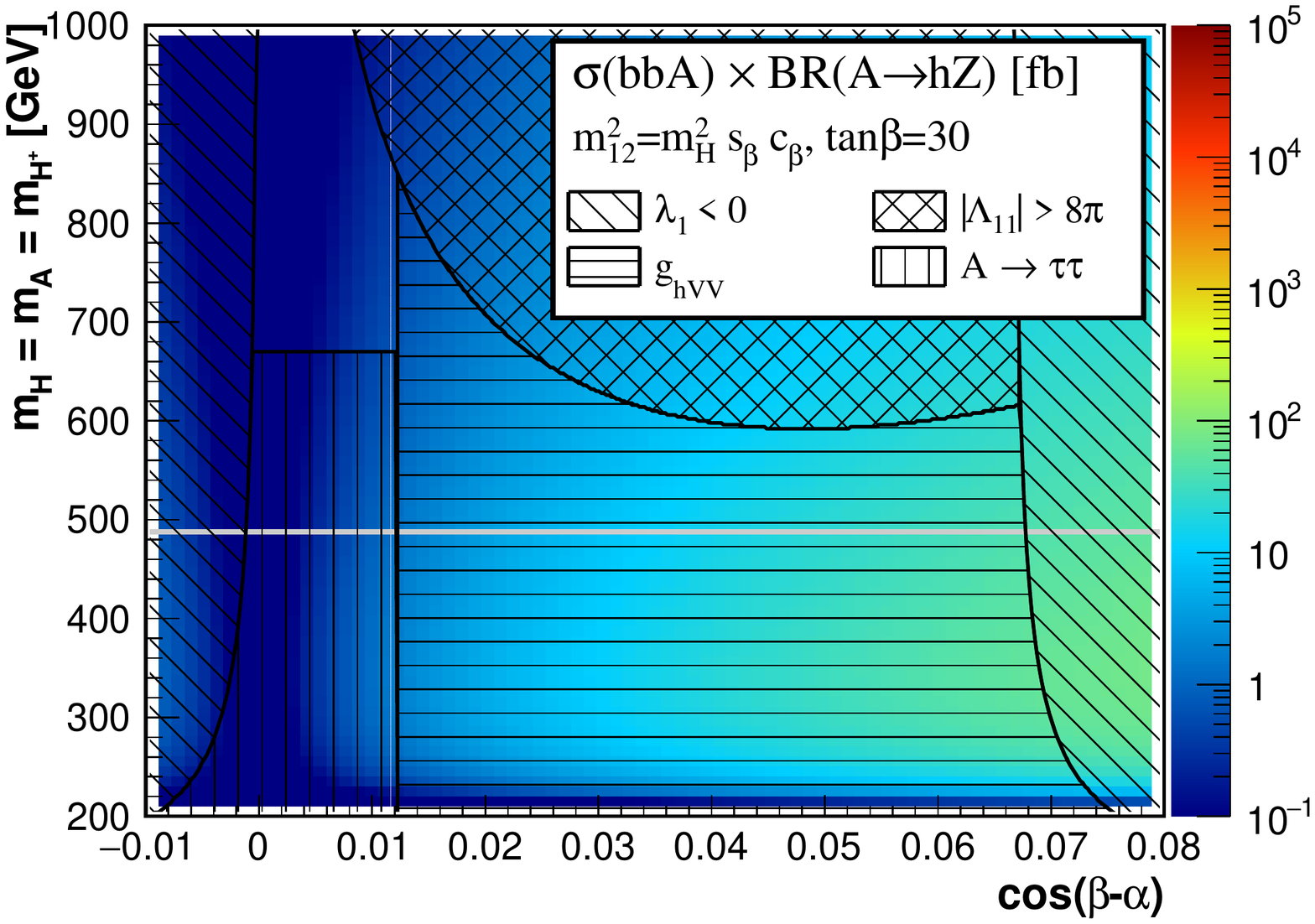}
	\includegraphics[width=0.495\textwidth]{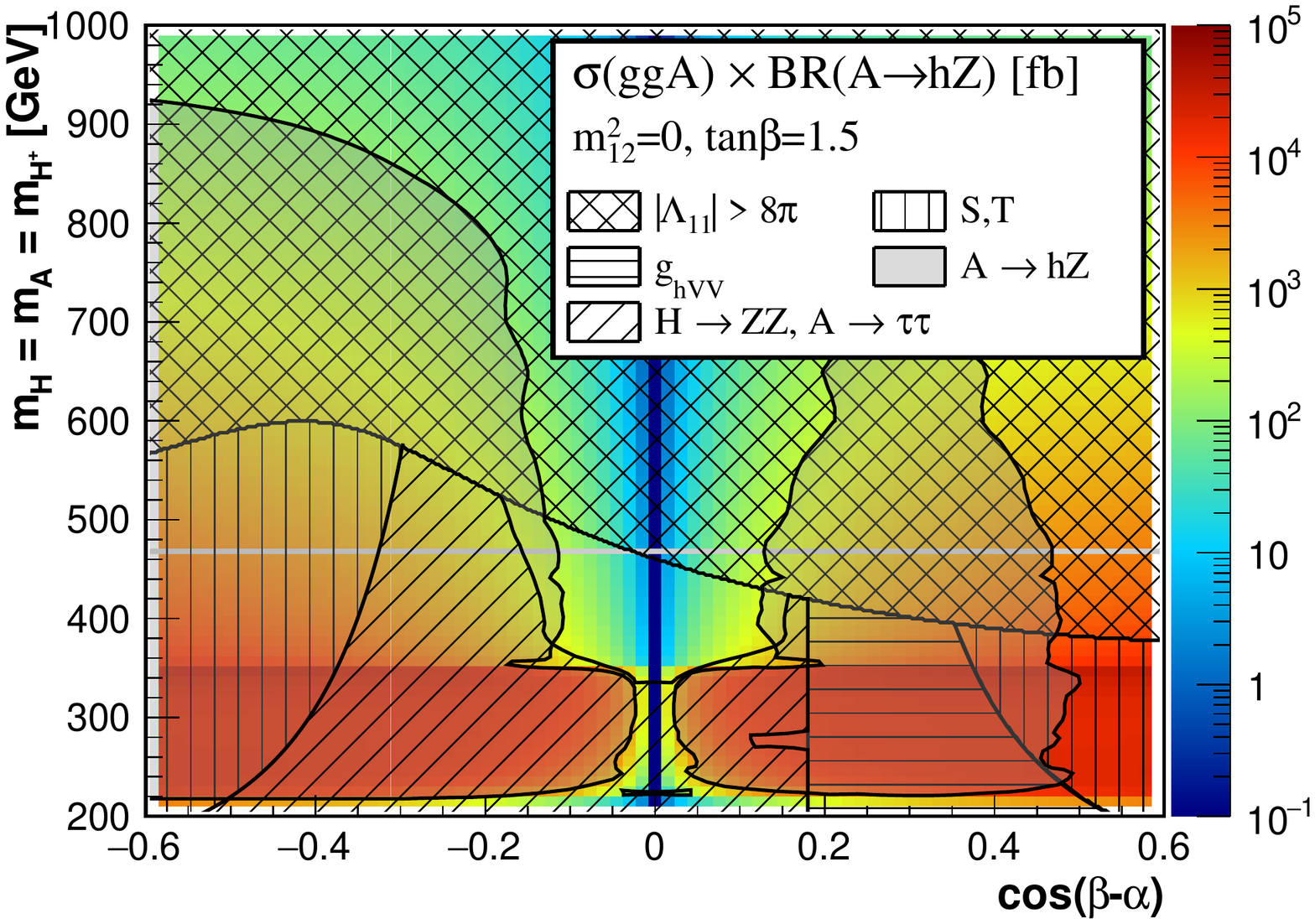}
	
\caption{$\sigma\times {\rm BR}$ for $A\to hZ$ for the gluon fusion production in Case 1, $\tan\beta=1.5$ (upper left),
7 (upper right), as well as Case 2, $\tan\beta=1.5$ (lower right), and $bbA$ associated production for Case 1, $\tan\beta=30$ (lower left) 
in BPIII: $c_{\beta-\alpha}$ vs. $m_A=m_H=m_{H^\pm}$. Hatched regions are excluded by either theoretical or experimental constraints (as indicated in the legend), 
while the shaded areas enclosed by an irregular curve indicate the parameter space constrained by LHC searches for exotic (non-SM) Higgs decays: $A\to h Z$ and $H\to h h$. 
The solid horizontal light grey lines indicate the flavour constraint $m_{H^{\pm}} > 480$ GeV.}
\label{fig:bp-ACH-vs-hsm}
\end{figure}

\begin{figure}[h!]
 \centering
 	\includegraphics[width=0.495\textwidth]{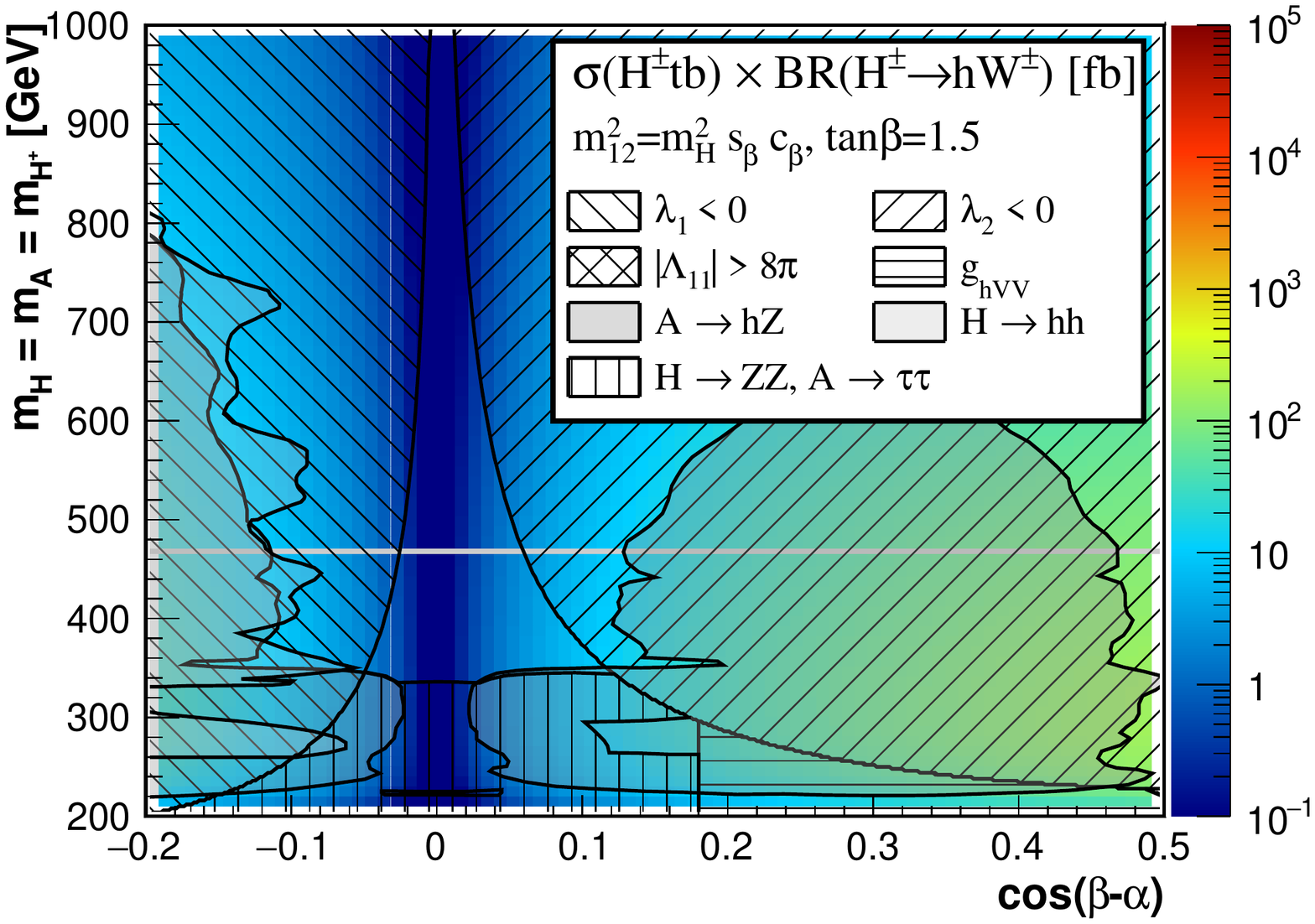}
	\includegraphics[width=0.495\textwidth]{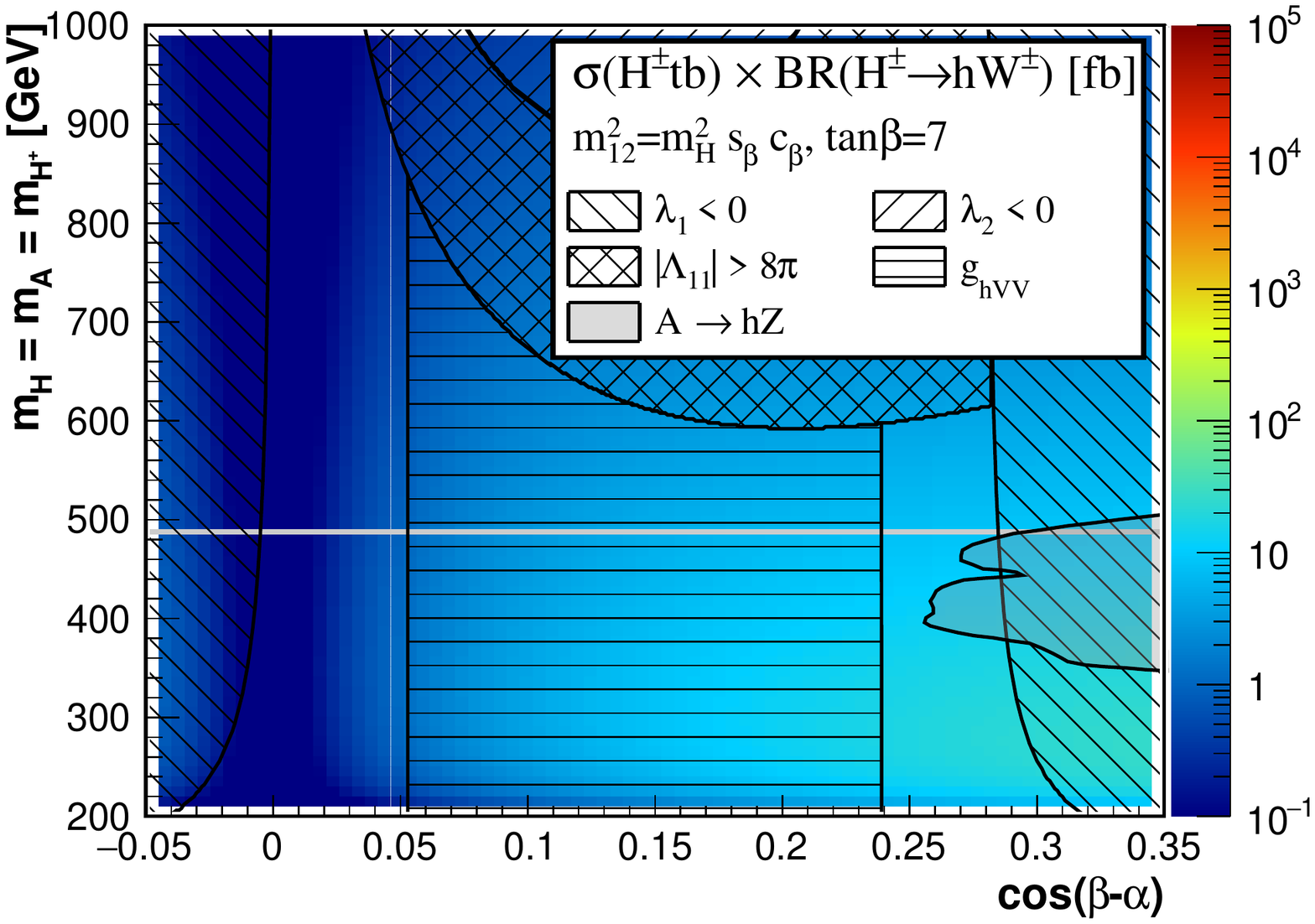}
	\includegraphics[width=0.495\textwidth]{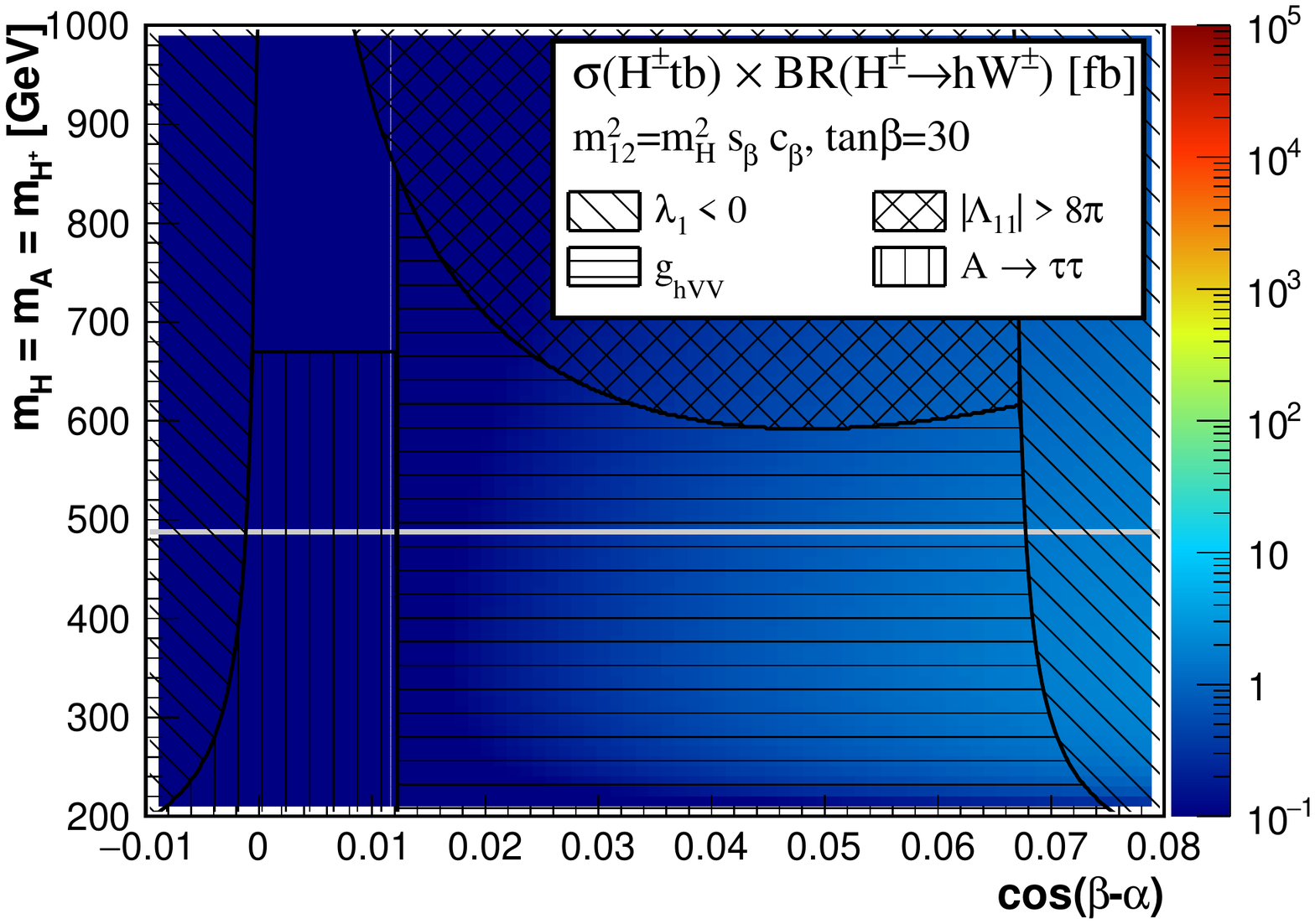}
	\includegraphics[width=0.495\textwidth]{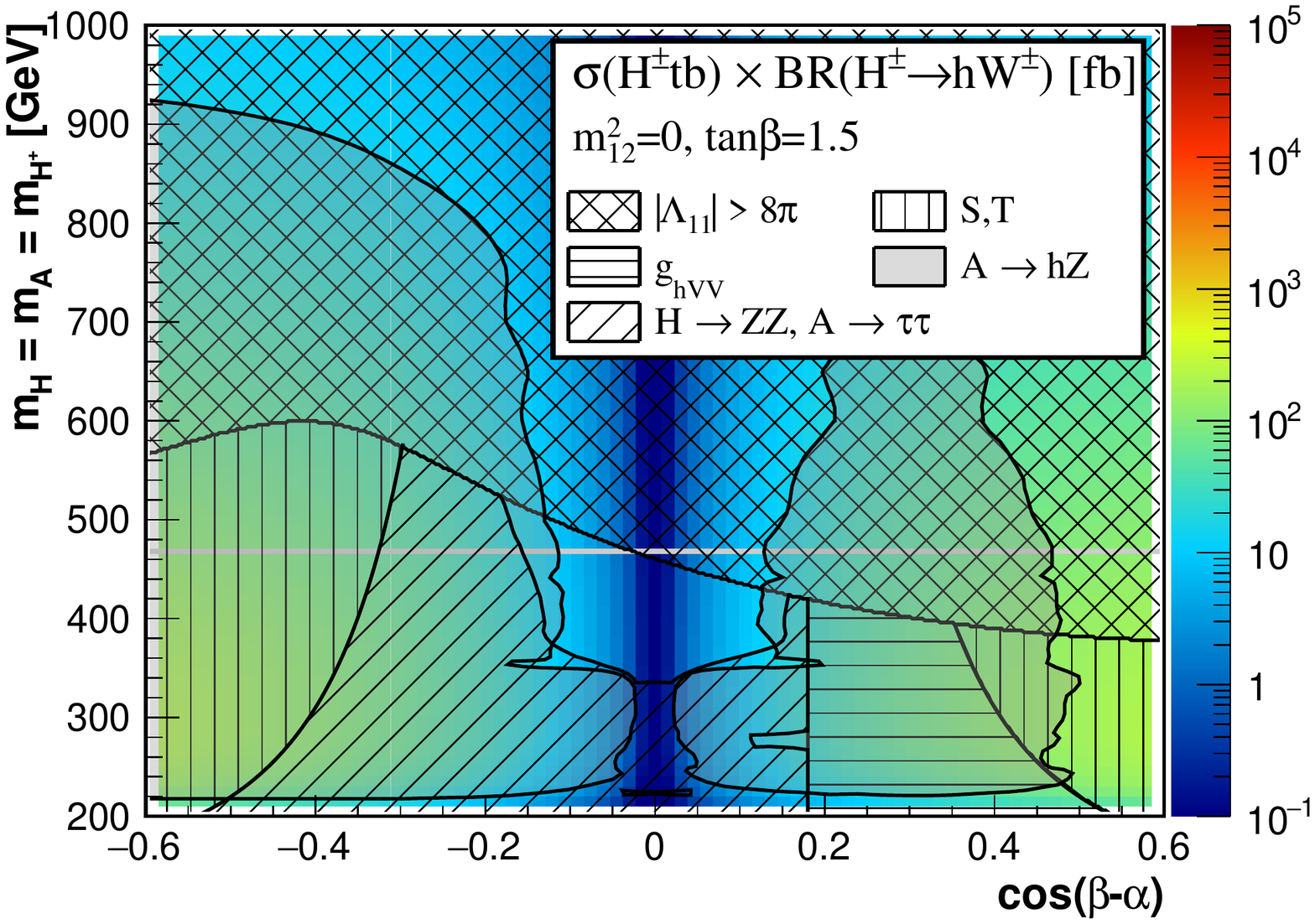}
\caption{$\sigma\times {\rm BR}$ for $H^\pm \to hW^\pm$ in BPIII: $c_{\beta-\alpha}$ vs. $m_A=m_H=m_{H^\pm}$ (see caption of Figure~\ref{fig:bp-ACH-vs-hsm} for 
further details).}
\label{fig:bp-CHA-vs-hsm}
\end{figure}

\begin{figure}[h!]
 \centering
 	\includegraphics[width=0.495\textwidth]{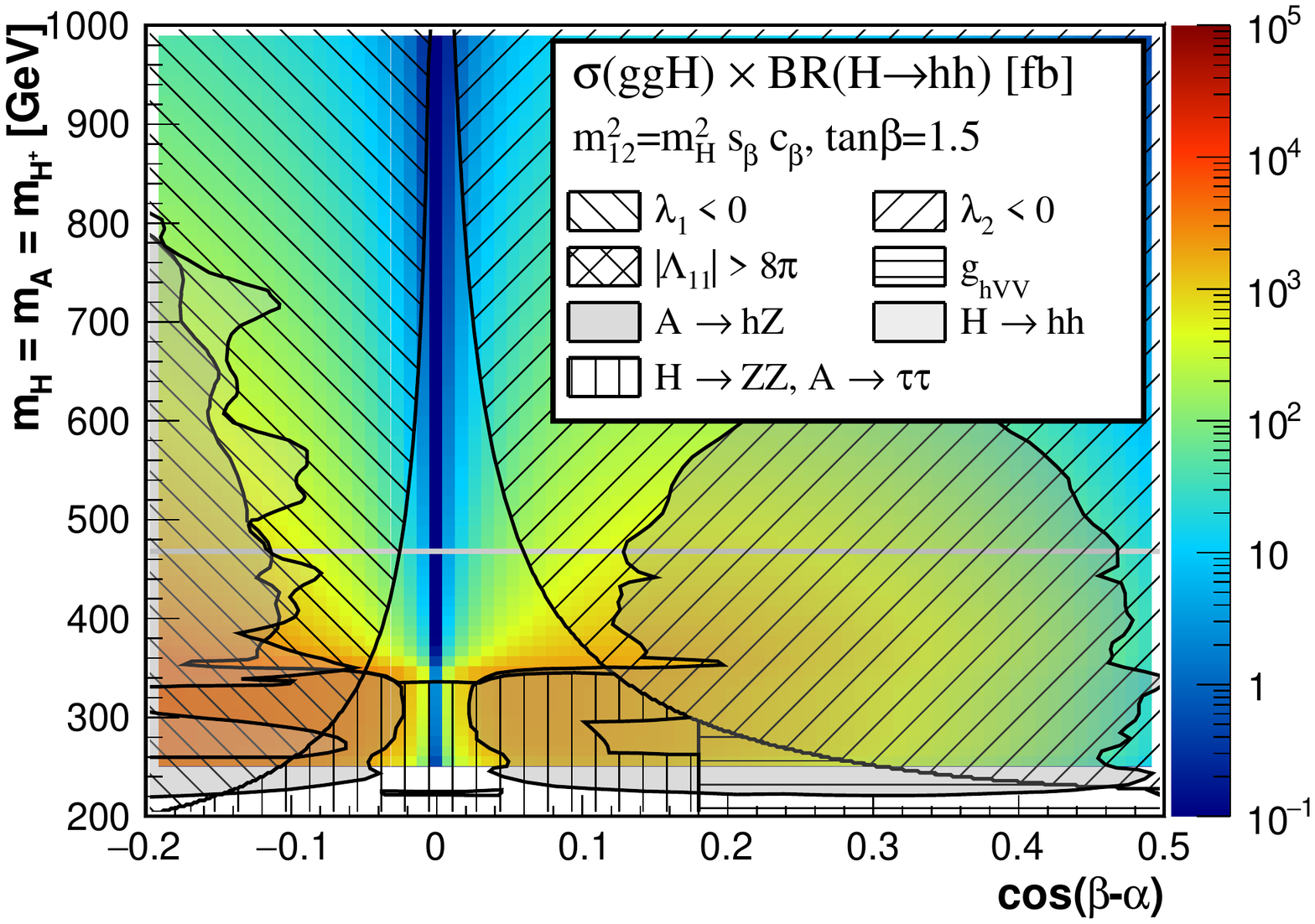}
	\includegraphics[width=0.495\textwidth]{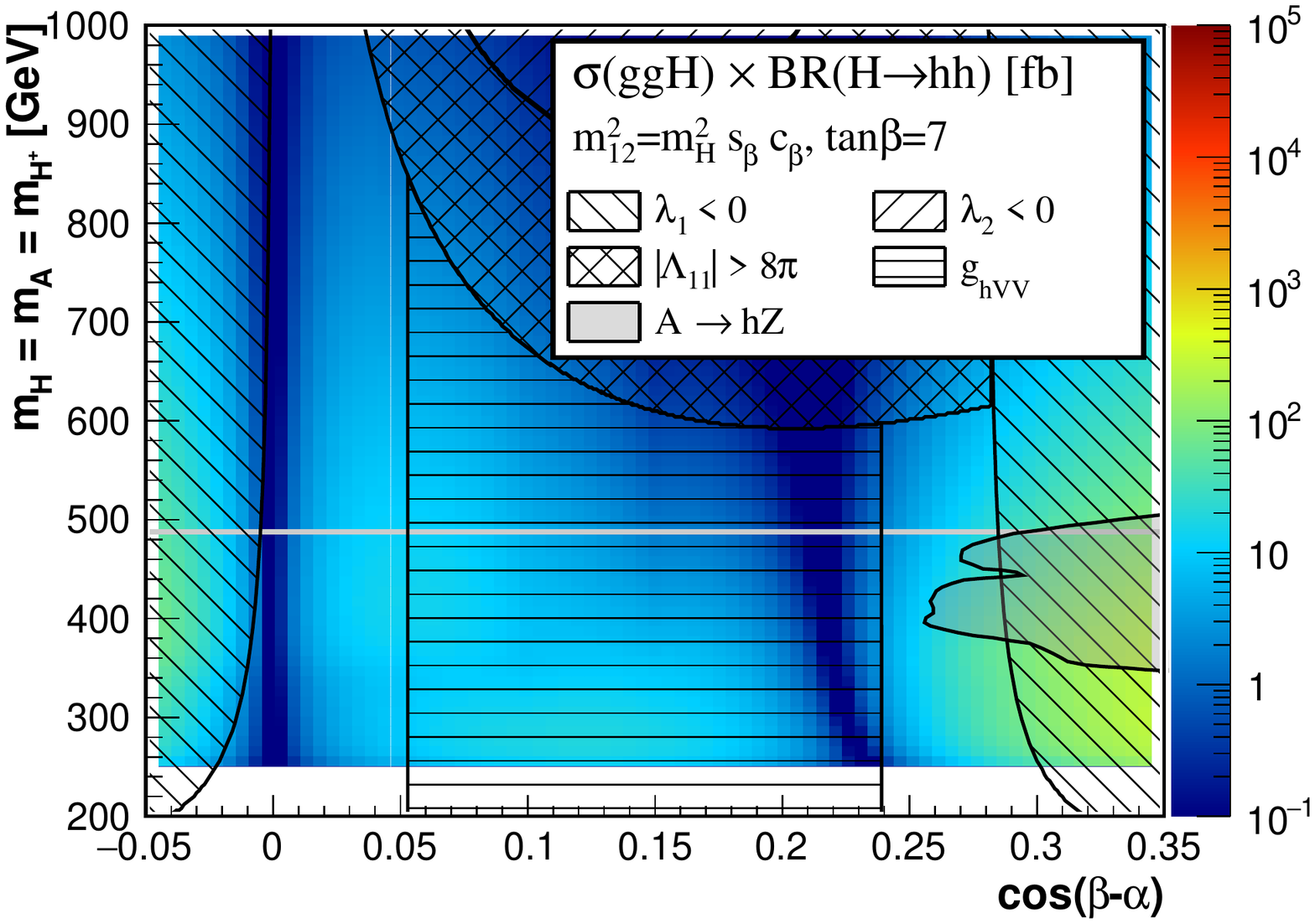}
	\includegraphics[width=0.495\textwidth]{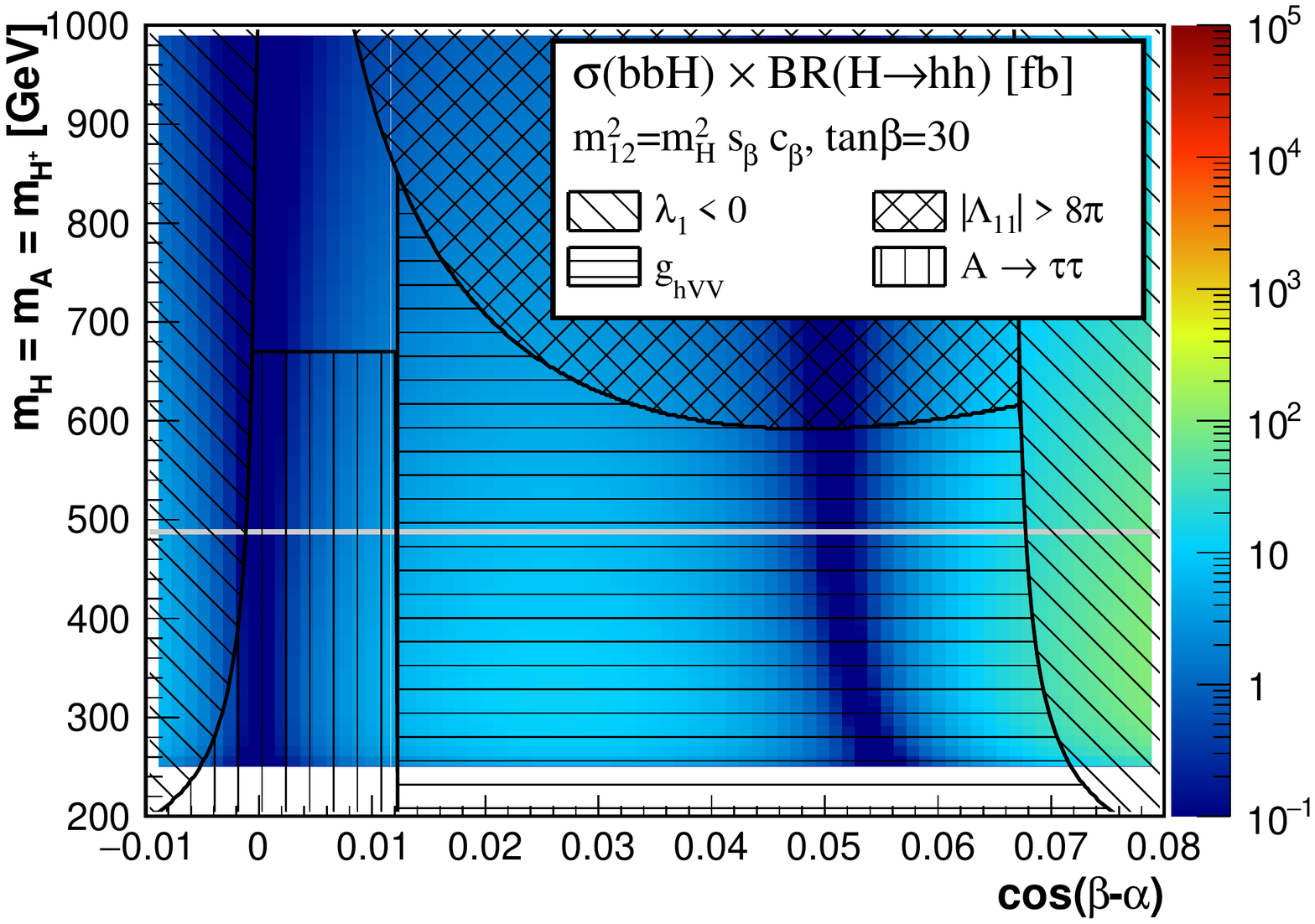}
	\includegraphics[width=0.495\textwidth]{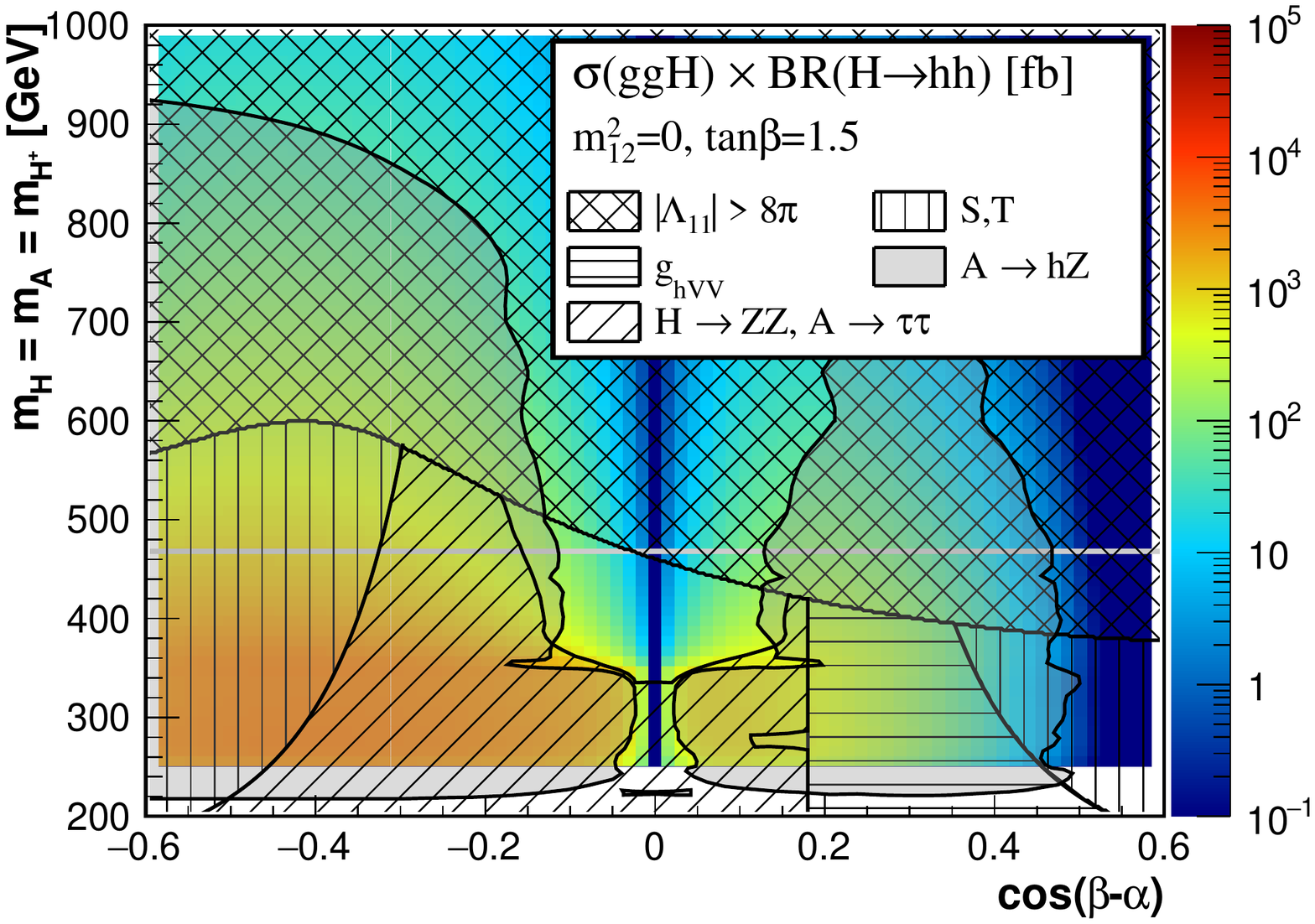}
\caption{$\sigma\times {\rm BR}$ for $H \to hh$ in  BPIII: $c_{\beta-\alpha}$ vs. $m_A=m_H=m_{H^\pm}$ (see caption of 
Figure~\ref{fig:bp-ACH-vs-hsm} for further details).}
\label{fig:bp-HAC-vs-hsm}
\end{figure}

Exotic decays with the SM-like Higgs $h$ in the final state are possible away from the alignment limit $c_{\beta-\alpha}=0$, as 
the $AhZ$, $H^\pm h W^\mp$ and $Hhh$ couplings are proportional to $c_{\beta-\alpha}$.
In Figures~\ref{fig:bp-ACH-vs-hsm},~\ref{fig:bp-CHA-vs-hsm}, and~\ref{fig:bp-HAC-vs-hsm} 
we respectively show the $\sigma\times {\rm BR}$ for $A\to hZ$, $H^\pm \rightarrow h W^\pm$ and $H\rightarrow hh$, in each case for 
Case 1 with $t_{\beta} = 1.5,\, 7,\, 30$ and Case 2 with $t_{\beta}=1.5$ in the ($c_{\beta-\alpha}$ vs $m_A=m_H=m_{H^\pm}$) plane.  

For Case 1 with $t_{\beta}=1.5$, only the region $|c_{\beta-\alpha}|\lesssim 0.2$ (close to the 
alignment limit) is viable as a result of Higgs signal strength measurements (mainly driven by the $g_{hVV}$ couplings) considering all the theoretical and experimental constraints. 
The allowed range for $c_{\beta-\alpha}$ shrinks as the masses of the heavy 2HDM scalars grow due to stability constraints, being already restricted to 
$-0.02 < c_{\beta-\alpha} < 0.06$ for $m_A=m_H=m_{H^\pm} \simeq 500$ GeV. At the same time, LHC bounds on $H\to Z Z$ and $A\to \tau\tau$ rule out
$m_A=m_H=m_{H^\pm} < 350$ GeV. For significantly higher values of $t_{\beta}$ (as our $t_{\beta} = 7,\,30$ scenarios) the stability constraints rule out 
almost completely the region $c_{\beta-\alpha}<0$, while unitarity imposes a strong constraint on $c_{\beta-\alpha}>0$ for high scalar masses 
$m_A=m_H=m_{H^\pm} > 600$ GeV. In addition, for $t_{\beta}=7$ the vacuum stability constraint rules out the region $c_{\beta-\alpha}>0.3$ while Higgs signal strengths rule 
out the region $ 0.05 < c_{\beta-\alpha} < 0.24$. For $t_{\beta} = 30$, Higgs signal strengths rule out $c_{\beta-\alpha} \gtrsim 0.01$, while 
$A \to \tau \tau$ searches restrict the allowed parameter space to $m_A=m_H=m_{H^\pm} > 650$ GeV, leaving only a very narrow stripe as viable parameter space.
For Case 2 with $t_{\beta}=1.5$, satisfying the constraints from $ H\to Z Z$ and $A\to \tau\tau$ requires $m_A=m_H=m_{H^\pm} > 350$ GeV
and $|c_{\beta-\alpha}|\lesssim 0.2$, while unitarity imposes an upper bound on the scalar masses in the range $450$ GeV -- $550$ GeV depending on $c_{\beta-\alpha}$.
As shown in Figure~\ref{fig:bp-ACH-vs-hsm}, the cross sections for $A\to hZ$ in the allowed region of parameter space could reach 1 pb or 
higher for $t_{\beta} = 1.5$ both in Case 1 and 2. For $t_{\beta} = 7$ (Case 1) the cross section for $A\to hZ$ is still sizable in the allowed region 
$0.24 < c_{\beta-\alpha} < 0.3$, reaching values $\sim 100$ fb. For $t_{\beta} = 30$ the signal cross section is however very small due to the suppressed
branching ratio BR($A\to h Z$) close to the alignment limit.
The signal cross sections for $H^\pm \rightarrow h W^\pm$ shown in Figure~\ref{fig:bp-CHA-vs-hsm} follow a trend similar to those for 
$A\to h Z$, but being typically a factor 10 -- 100 smaller due to the suppressed production cross section for $H^{\pm}$ above $m_t$ (see Appendix~\ref{sec:xs}).
Finally for $H\to hh$ the signal cross sections, shown in Figure~\ref{fig:bp-HAC-vs-hsm}, are about factor of 10 smaller than those of $A \rightarrow hZ$, 
and an additional suppression of the branching ratio BR($H\rightarrow hh$) occurs for certain values of $c_{\beta-\alpha}$ 
(e.g.~$c_{\beta-\alpha} \sim 0.22$ for $t_{\beta} = 7$ and $c_{\beta-\alpha} \sim 0.052$ for $t_{\beta} = 30$, as seen from Figure~\ref{fig:bp-HAC-vs-hsm}).

\section{Conclusions}
\label{sec:conclusion}

In the 2HDM, other than decaying to pairs of SM quarks, leptons, and gauge bosons, the exotic decays of heavy Higgses into two 
lighter Higgses or one light Higgs and a SM gauge boson are likely to dominate once they are kinematically open. 
While the collider search bounds for heavy Higgses based on conventional search modes $WW$, $ZZ$, $\gamma\gamma$, $bb$ and $\tau\tau$ for 
neutral Higgses, and $\tau\nu$ and $cs$ modes for charged Higgses would be 
relaxed once those exotic modes are open, the exotic decay modes offer new discovery channels in large regions of the 2HDM parameter space.  
 
Away from the 2HDM alignment limit, exotic decays into the SM-like 125 GeV Higgs boson $h$, namely 
$H \to hh$, $A \to h Z$ and $H^{\pm} \to h W^{\pm}$, are potentially important, and there is already an ongoing ATLAS and CMS 
search programme for $A\to h Z$~\cite{Aad:2015wra, Khachatryan:2015lba} and $H\to h h$~\cite{Aad:2014yja,CMS:2014ipa,Khachatryan:2015yea}.  
In contrast, close to the alignment limit, as favoured by measurements of Higgs signal strengths, exotic decays among the new 2HDM scalars 
become particularly relevant. The experimental searches based on those channels, however, have just started 
with $H/A \rightarrow AZ/HZ$~\cite{CMS:2015mba,Khachatryan:2016are}.  
In this work, we carefully examine the exotic Higgs decay channels in the 2HDM, both in the presence of a  hierarchy between Higgses and 
away from alignment when this hierarchy is not present. By taking into account the various theoretical and experimental constraints, 
we propose 2HDM benchmark plane scenarios for LHC searches at 13 TeV: 
 \begin{itemize}
 \item{BP IA}: $m_A > m_H = m_{H^\pm}$, with $A \rightarrow HZ, \ H^\pm W^\mp$. 
\item{BP IB}: $m_A< m_H=m_{H^\pm} $, with $H\rightarrow AZ,\ AA$ and $H^\pm \rightarrow A W^\pm$.
\item{BP IIA}: $m_H> m_A=m_{H^\pm}$, with $H \rightarrow AZ,\ H^\pm W^\mp,\ AA,\ H^+H^-$.
\item{BP IIB}: $m_H< m_A=m_{H^\pm}  $, with $A \rightarrow HZ$ and $H^\pm \rightarrow HW^\pm$.
\item{BP III}: $m_A=m_H=m_{H^\pm}$ vs. $c_{\beta-\alpha}$, with $A \rightarrow hZ$, $H^\pm \rightarrow hW^\pm$, and $H \rightarrow hh$.
 \end{itemize}
 
\noindent In each case, we analyze the allowed regions of parameter space and the LHC 13 TeV $\sigma\times{\rm BR}$ 
for the relevant exotic Higgs decay modes in those regions. 

\vspace{2mm}

To summarize, exotic Higgs decays provide new discovery avenues for heavy Higgses. In turn, the exploration of the proposed benchmarks 
via these decays could help to understand the structure of the electroweak symmetry breaking sector beyond the SM.


\begin{figure}[h!]
 \centering
 	 \includegraphics[width=0.495\textwidth]{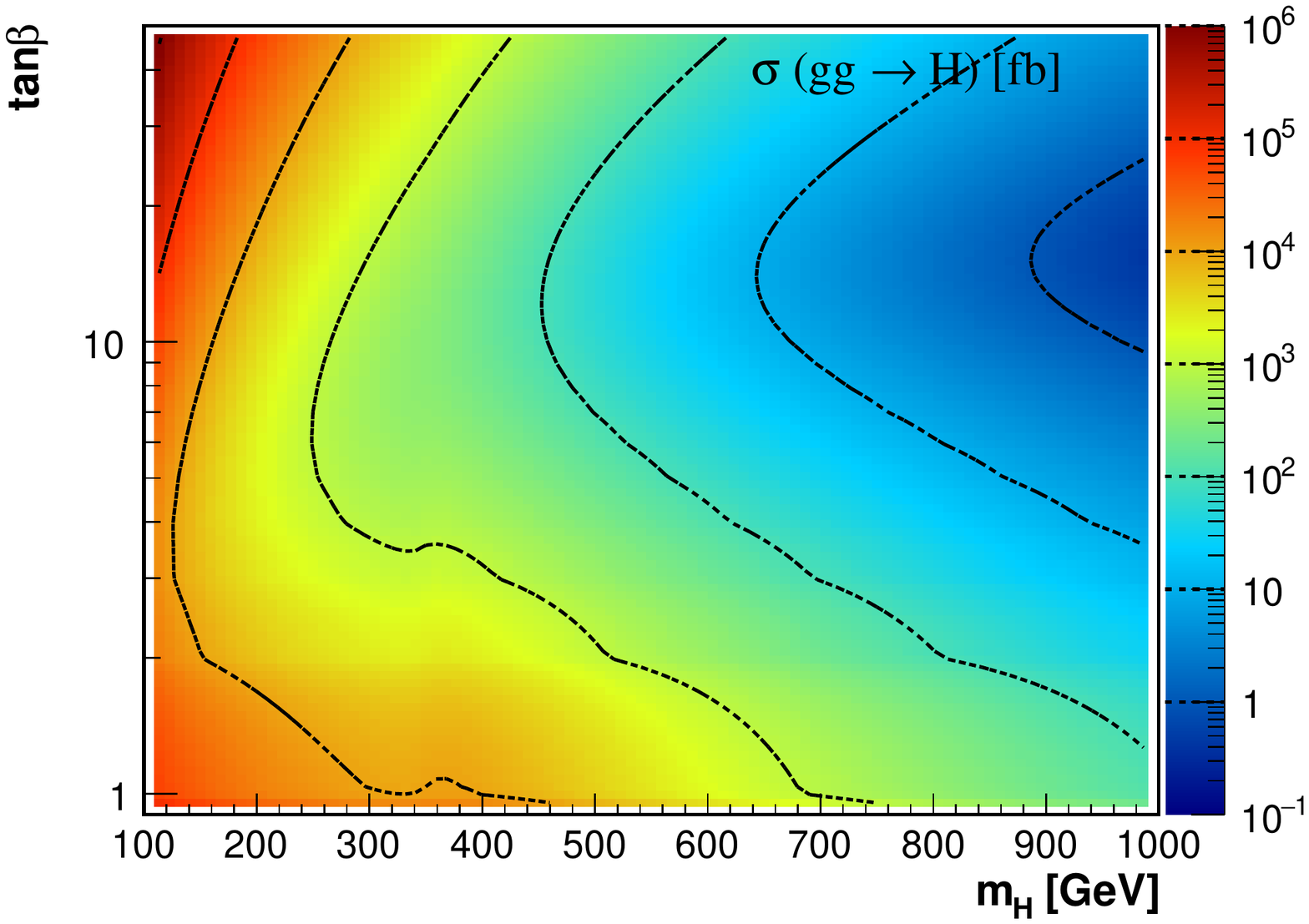}
 	 \includegraphics[width=0.495\textwidth]{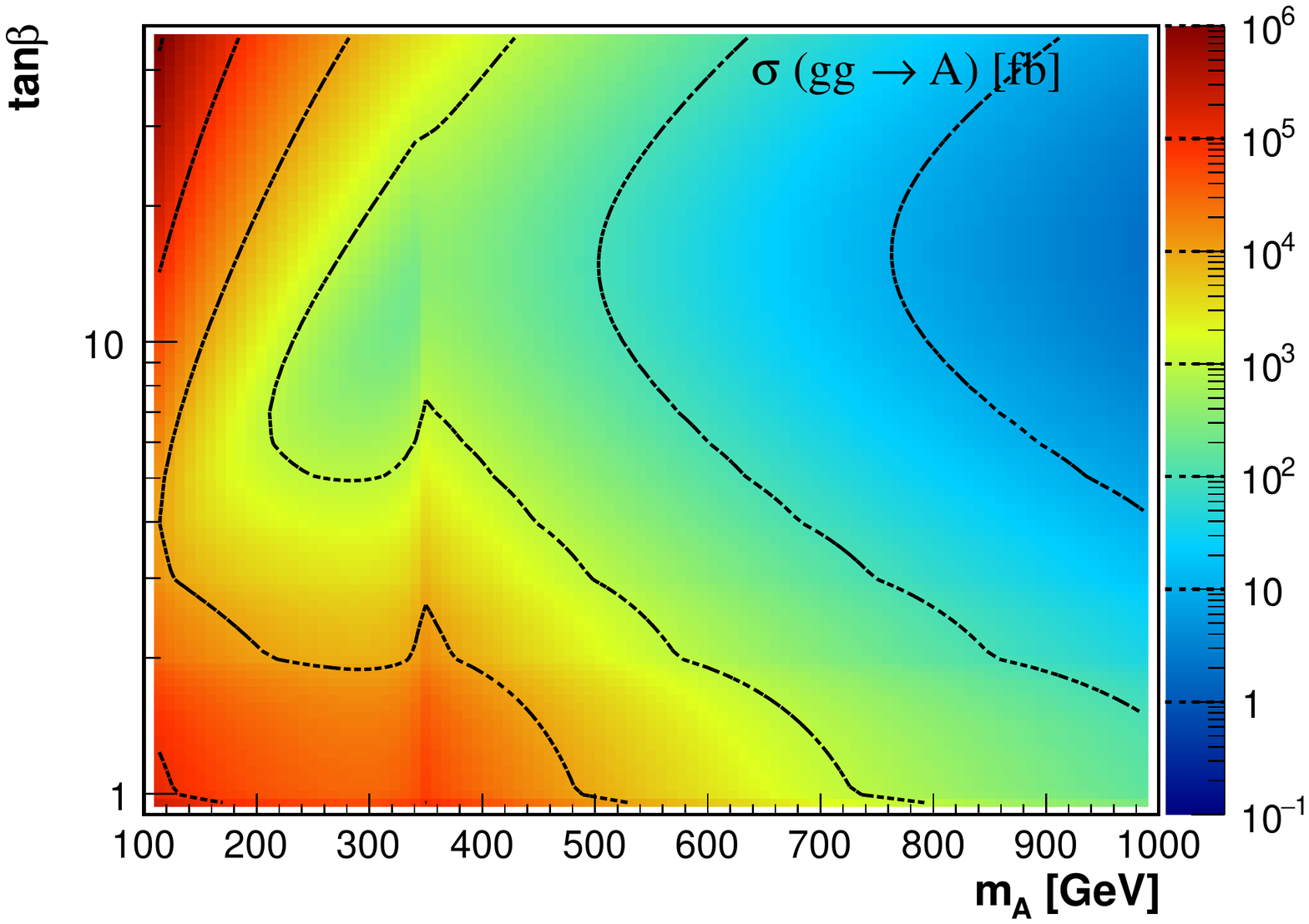}
 	 \includegraphics[width=0.495\textwidth]{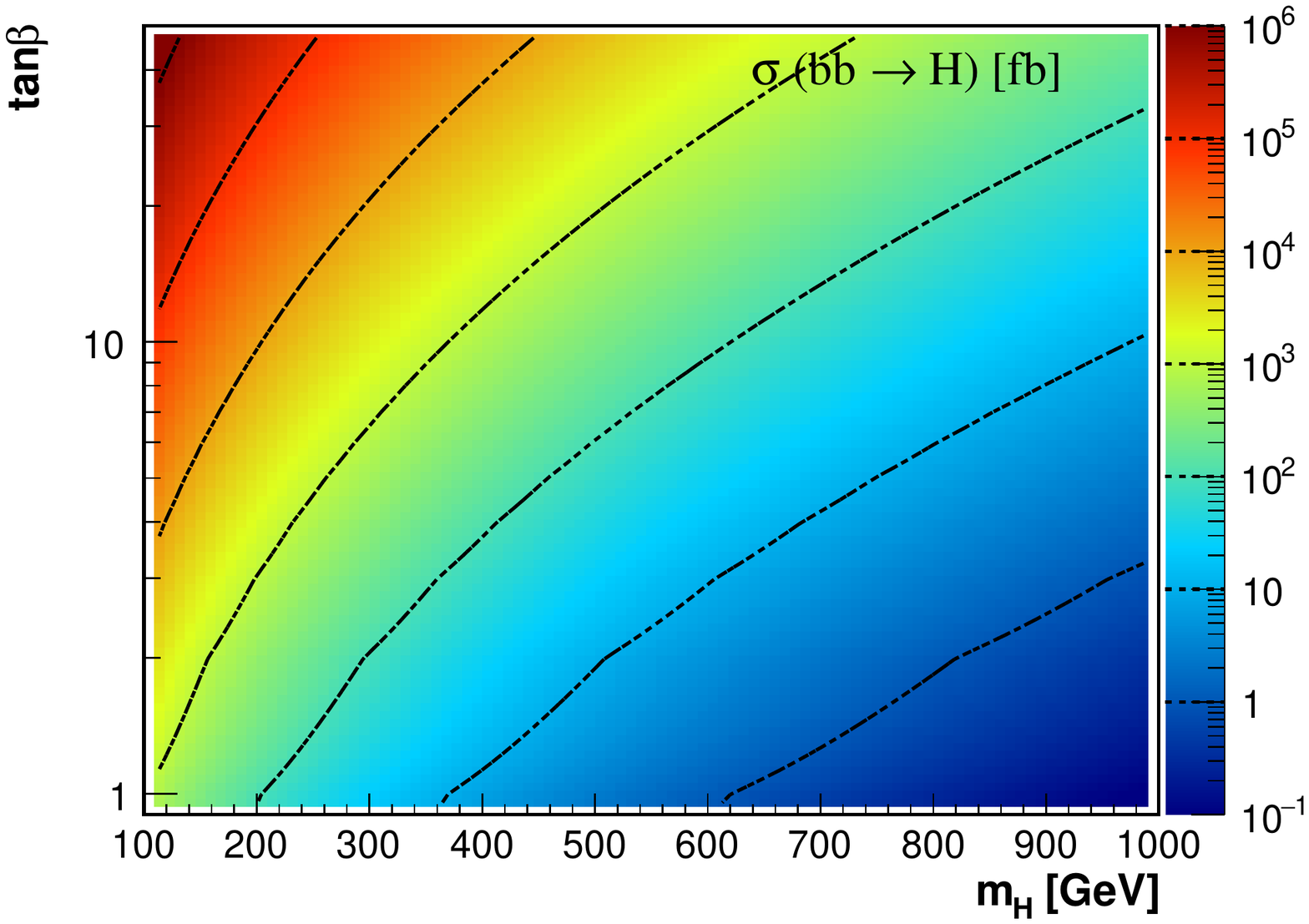}
 	 \includegraphics[width=0.495\textwidth]{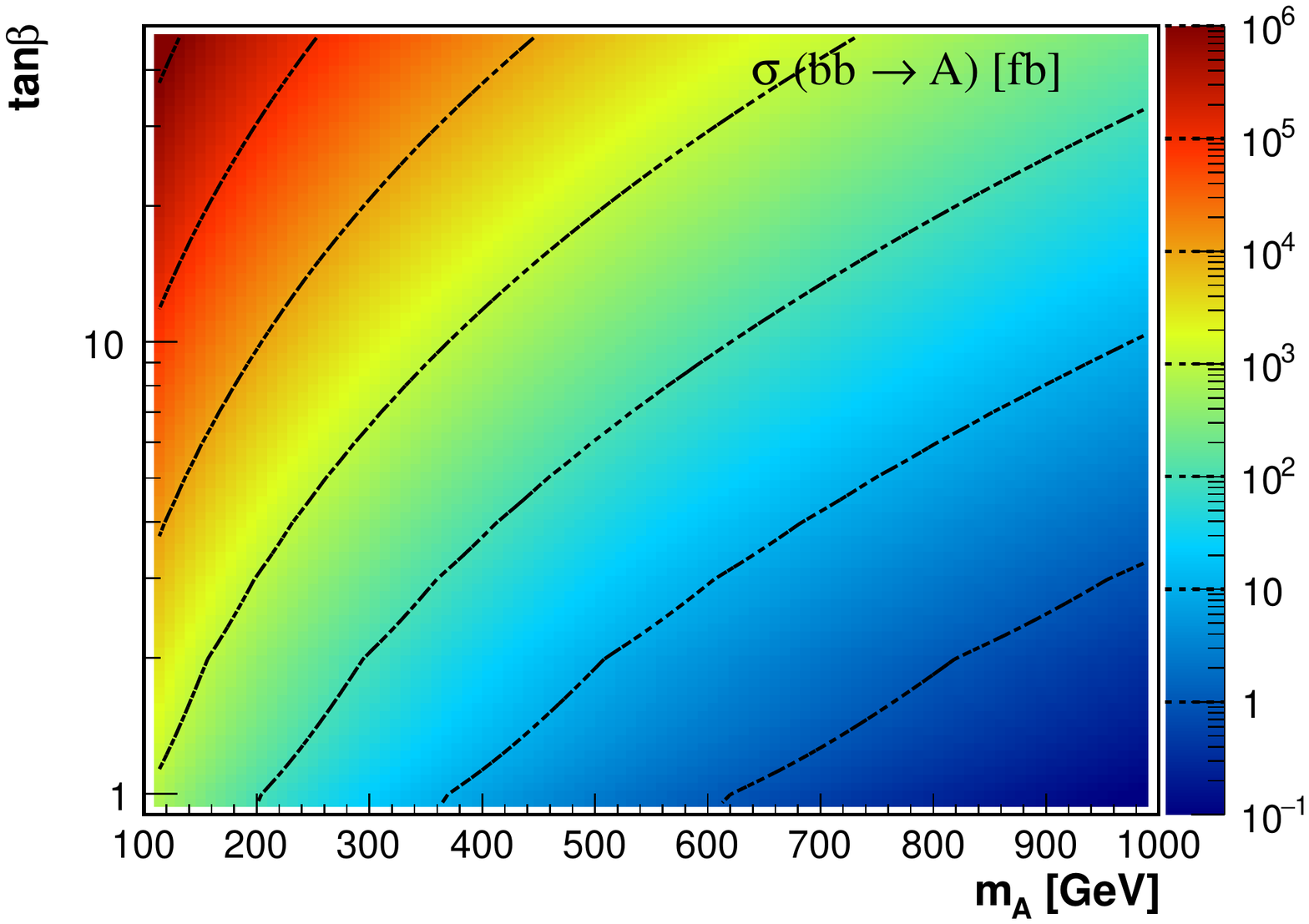}
 	 \includegraphics[width=0.495\textwidth]{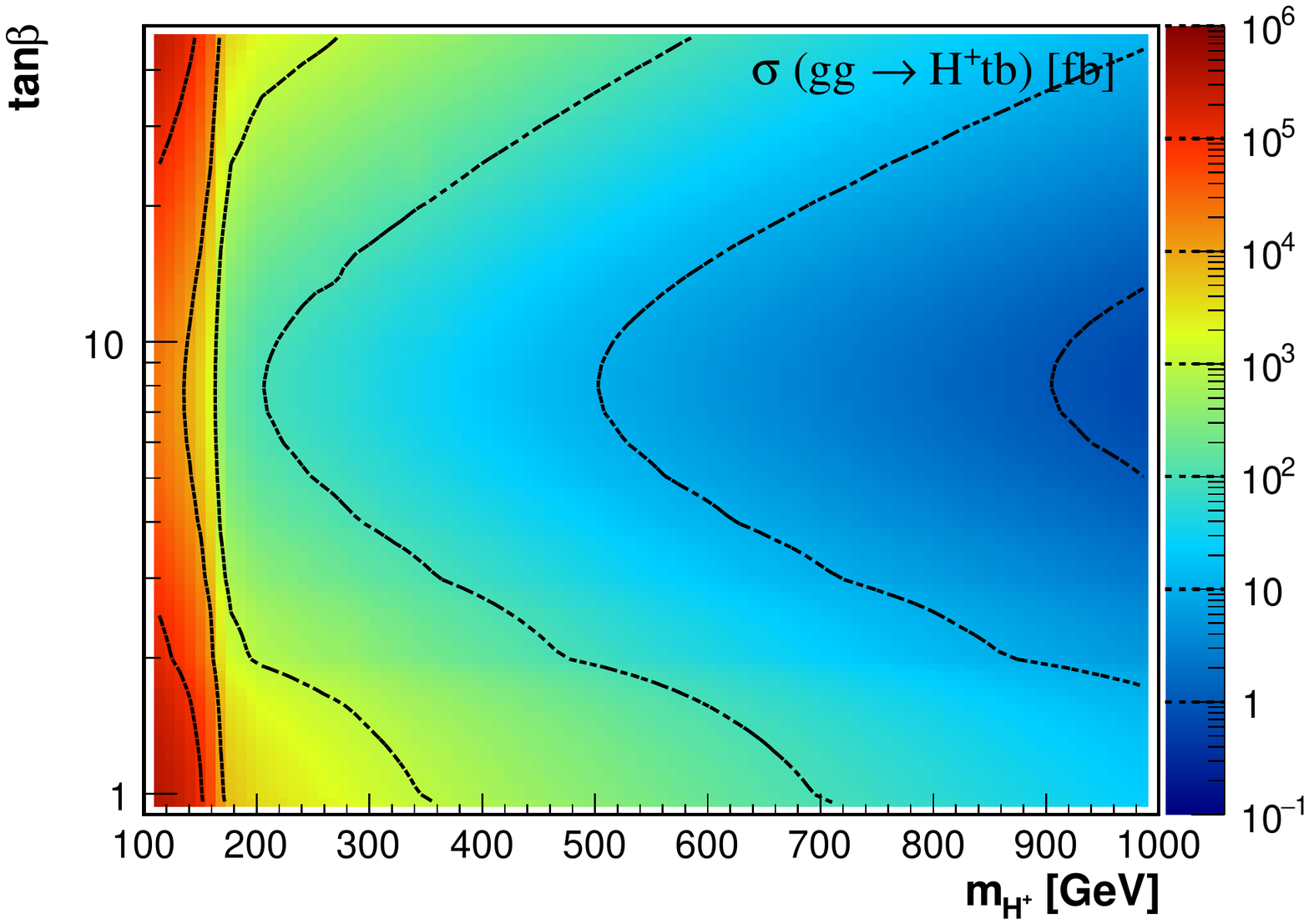}
\caption{Production cross section for $H$, $A$ and $H^+$ at LHC 13 TeV.
The contour lines indicate the cross section of 1, 10, $10^2$, $10^3$, $10^4$, $10^5$ and $10^6$ fb.}
\label{fig:xs-1}
\end{figure}

\acknowledgments
We would like to thank Baradhwaj Coleppa, Tao Han, Tao Liu, Ken Mimasu and Adarsh Pyarelal for helpful discussions.   
We also thank the Munich Institute for Astro- and Particle Physics (MIAPP) of the DFG cluster of excellence 
``Origin and Structure of the Universe" for support and hospitality during the initial stages of this work.
The work of S.S.~and F.K.~was supported by US Department of Energy under Grant~DE-FG02-04ER-41298. 
F.K.~ also acknowledges support from the Fermilab Graduate Student Research Program in Theoretical Physics operated by Fermi Research Alliance, 
LLC under Contract No. DE-AC02-07CH11359 with the United States Department of Energy.
J.M.N. is supported by the People Programme (Marie Curie Actions) of the European Union Seventh 
Framework Programme (FP7/2007-2013) under REA grant agreement PIEF-GA-2013-625809.

\appendix

\section{Production Cross Sections and Branching Ratios of 2HDM Higgses}
\label{sec:pro_decay2}

\subsection{2HDM Production Cross Sections}
\label{sec:xs}

In Figure~\ref{fig:xs-1}, we show the gluon fusion production cross section for $H$ (upper left panel) and $A$ (upper right panel), 
$bb$-associated production cross section for $H$ (middle left panel) and $A$ (middle right panel), 
and $tbH^\pm$ production cross section (bottom) for the charged scalar (details are given in Section~\ref{sec:pro_decay}).

\begin{figure}[h!]
 \centering
 	\includegraphics[width=0.495\textwidth]{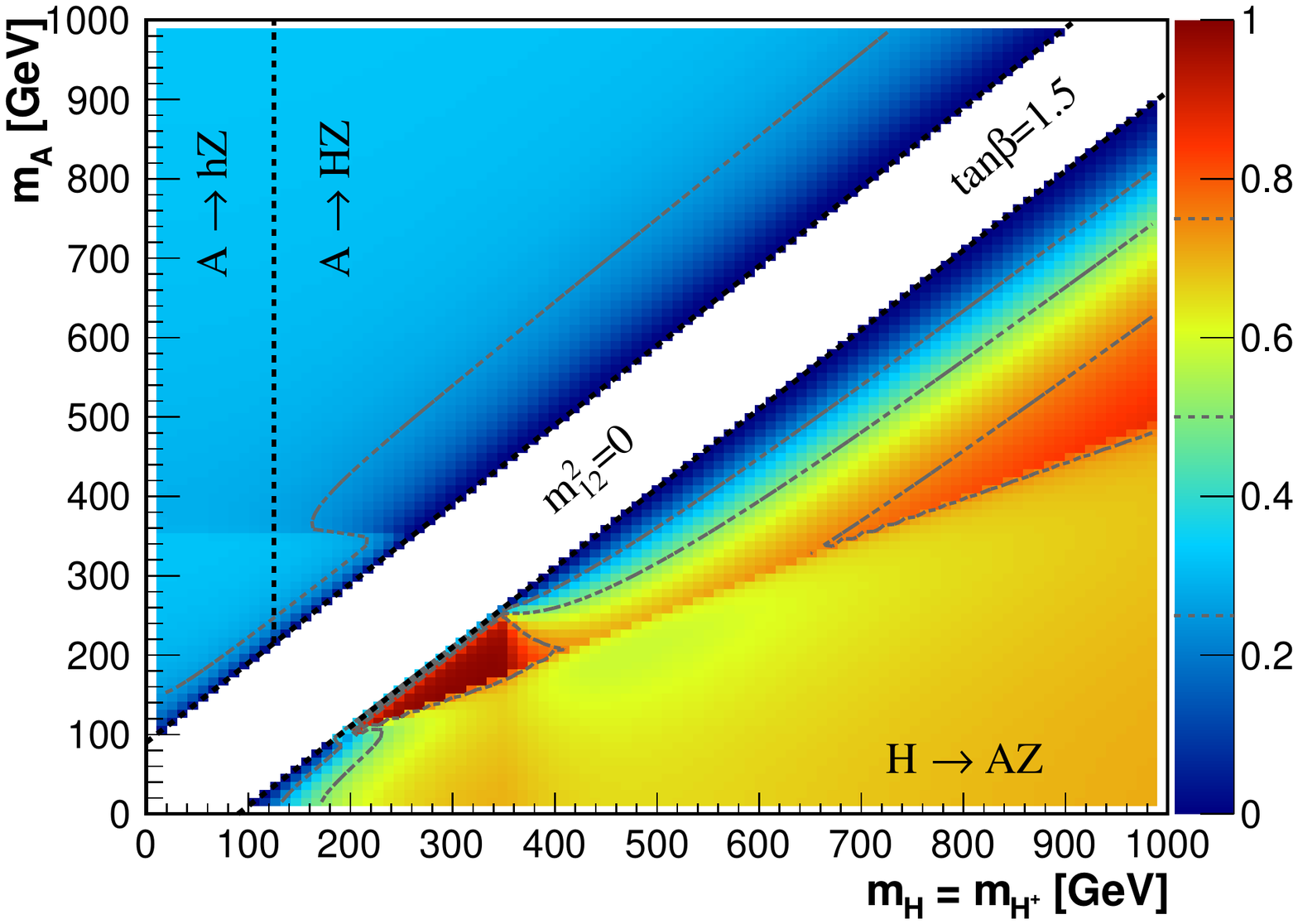}
	\includegraphics[width=0.495\textwidth]{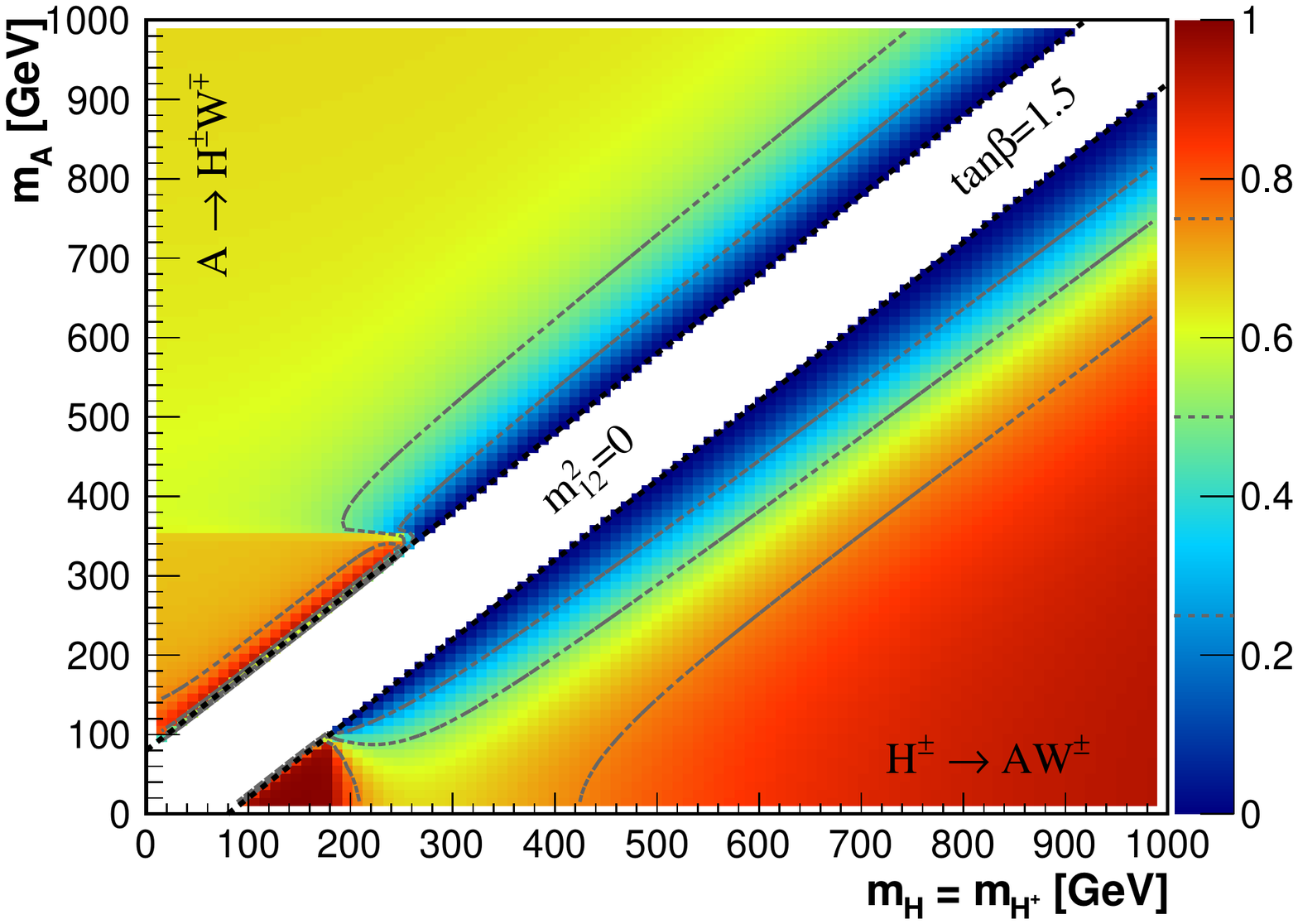}
	\includegraphics[width=0.495\textwidth]{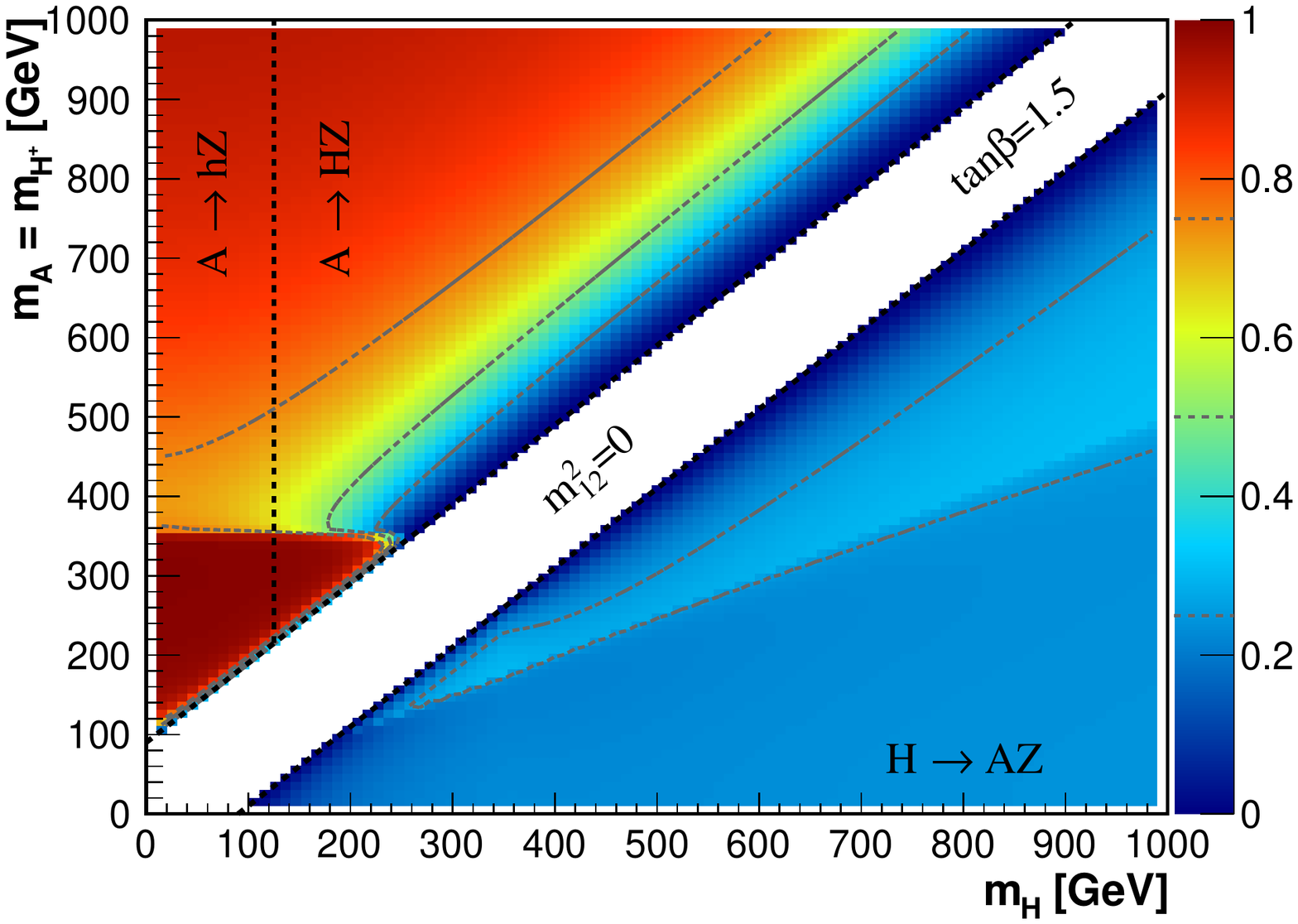}
	\includegraphics[width=0.495\textwidth]{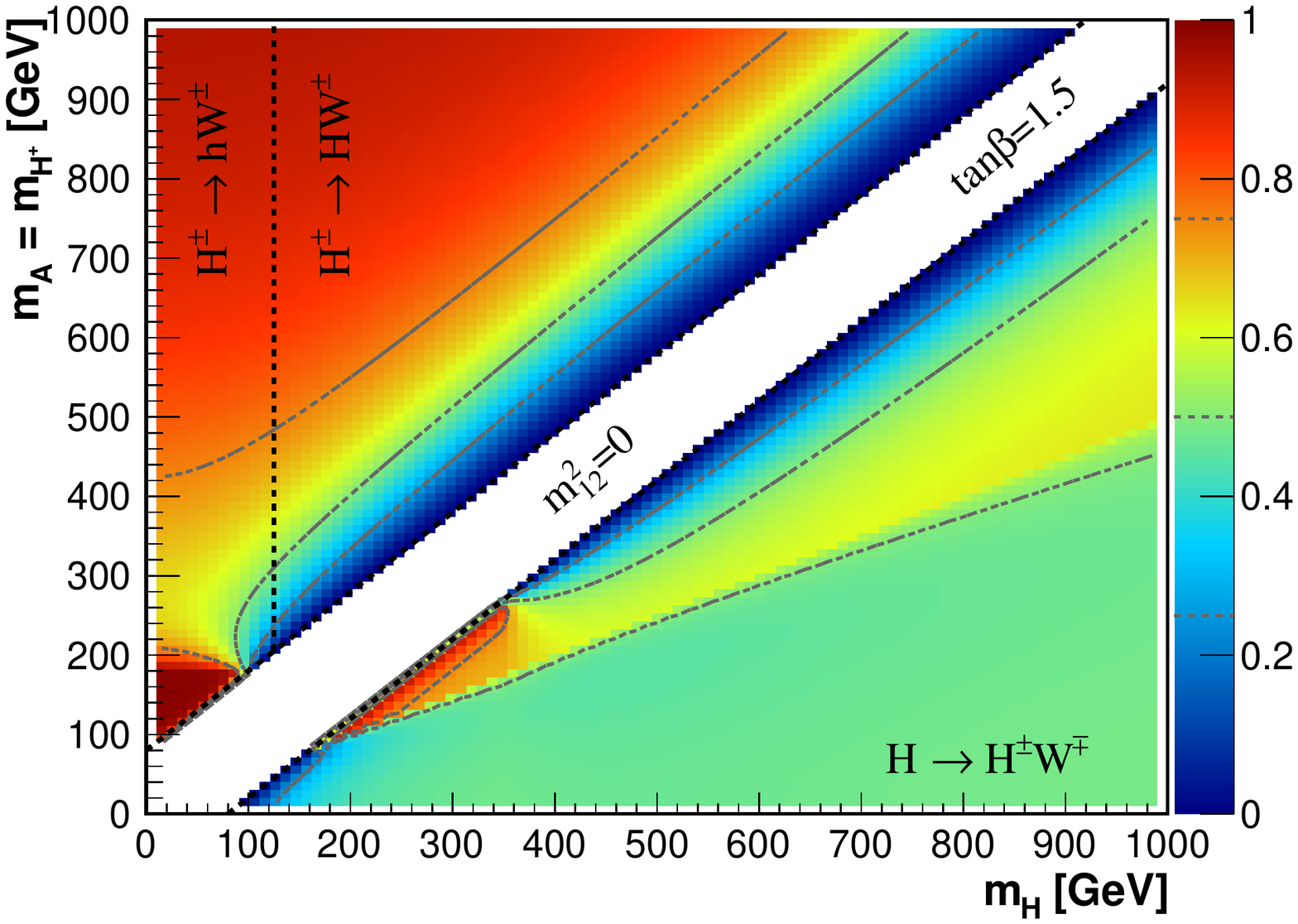}
\caption{Left: Exotic decay BR $H/A \rightarrow AZ/H(h)Z$ for Plane I (top) and Plane II (bottom) for Case 2 ($m_{12}^2 = 0$) with $t_{\beta} = 1.5$. 
For $A\to H(h)Z$, we consider $A\rightarrow HZ$ for $m_H> m_h = 125$ GeV and $A \to hZ$ for $m_h < m_H = 125$ GeV (so that the BR into the non-SM like Higgs boson is 
shown in each case).
Right: Exotic decay BR $A/H^\pm \rightarrow H^\pm W^\mp/A W^\pm$ for Plane I (top) and $H/H^\pm \rightarrow H^\pm W^\mp/H W^\pm$ for Plane II (bottom), 
for Case 2 ($m_{12}^2 = 0$) with $t_{\beta} = 1.5$.}
\label{fig:exotic-A-vs-HC}
\end{figure}

\subsection{2HDM Branching Ratios for Exotic Higgs Decays}
\label{sec:br}

For illustration, we show in Figure~\ref{fig:exotic-A-vs-HC} the branching ratios of $H_{a} \to H_{b} V$ (with $H_{a,b} = H,\,A,\,H^{\pm}$ and $V = W^{\pm},\,Z$)
for Plane I (top) and Plane II (bottom) for Case 2 ($m_{12}^2 = 0$) with $t_{\beta} = 1.5$ (being the scenario allowed for the four benchmarks BP IA, BP IB, BP IIA and BP IIB).
The decay branching ratios for $H/A \rightarrow AZ/H(h)Z$ are shown on the left panels of Figure~\ref{fig:exotic-A-vs-HC}, while 
those for $A/H^\pm \rightarrow H^\pm W^\mp/A W^\pm$ (Plane I) and $H/H^\pm \rightarrow H^\pm W^\mp/H W^\pm$ (Plane II) 
are shown on the right panels. 

 \bibliographystyle{JHEP}

\begin{thebibliography}{100}
 
 
 
  \bibitem{Aad:2015gba}
  G.~Aad {\it et al.} [ATLAS Collaboration],
  Eur.\ Phys.\ J.\ C {\bf 76} (2016) 1,  6
  [arXiv:1507.04548 [hep-ex]].
 
 
 \bibitem{Khachatryan:2014jba}
  V.~Khachatryan {\it et al.} [CMS Collaboration],
  Eur.\ Phys.\ J.\ C {\bf 75} (2015) 5,  212
  [arXiv:1412.8662 [hep-ex]].
 
 \bibitem{Djouadi:2005gj}
  A.~Djouadi,
  Phys.\ Rept.\  {\bf 459} (2008) 1
  [hep-ph/0503173].
 
 \bibitem{Composite}
  J.~Mrazek, A.~Pomarol, R.~Rattazzi, M.~Redi, J.~Serra and A.~Wulzer,
  Nucl.\ Phys.\ B {\bf 853} (2011) 1
  [arXiv:1105.5403 [hep-ph]];
  E.~Bertuzzo, T.~S.~Ray, H.~de Sandes and C.~A.~Savoy,
  JHEP {\bf 1305} (2013) 153
  [arXiv:1206.2623 [hep-ph]].
 
 \bibitem{EWBG}
  J.~M.~Cline, K.~Kainulainen and A.~P.~Vischer,
  Phys.\ Rev.\ D {\bf 54} (1996) 2451
  [hep-ph/9506284];
  J.~M.~Cline and P.~A.~Lemieux,
  Phys.\ Rev.\ D {\bf 55} (1997) 3873
  [hep-ph/9609240];
  L.~Fromme, S.~J.~Huber and M.~Seniuch,
  JHEP {\bf 0611} (2006) 038
  [hep-ph/0605242];
  G.~C.~Dorsch, S.~J.~Huber and J.~M.~No,
  JHEP {\bf 1310} (2013) 029
  [arXiv:1305.6610 [hep-ph]]. 
 
 \bibitem{Celis:2013rcs}
  A.~Celis, V.~Ilisie and A.~Pich,
  JHEP {\bf 1307} (2013) 053
  [arXiv:1302.4022 [hep-ph]].
 
 \bibitem{Grinstein:2013npa}
  B.~Grinstein and P.~Uttayarat,
  JHEP {\bf 1306} (2013) 094
   [JHEP {\bf 1309} (2013) 110]
  [arXiv:1304.0028 [hep-ph]].
 
 
 \bibitem{Coleppa:2013dya} 
  B.~Coleppa, F.~Kling and S.~Su,
  JHEP {\bf 1401}, 161 (2014)
  [arXiv:1305.0002 [hep-ph]].
 
  \bibitem{Chen:2013rba}
  C.~Y.~Chen, S.~Dawson and M.~Sher,
  Phys.\ Rev.\ D {\bf 88} (2013) 015018
   [Phys.\ Rev.\ D {\bf 88} (2013) 039901]
  [arXiv:1305.1624 [hep-ph]].
 
 \bibitem{Eberhardt:2013uba}
  O.~Eberhardt, U.~Nierste and M.~Wiebusch,
  JHEP {\bf 1307} (2013) 118
  [arXiv:1305.1649 [hep-ph]].
 
 \bibitem{Dumont:2014wha}
  B.~Dumont, J.~F.~Gunion, Y.~Jiang and S.~Kraml,
  Phys.\ Rev.\ D {\bf 90} (2014) 035021
  [arXiv:1405.3584 [hep-ph]].
 
 \bibitem{Bernon:2014vta}
  J.~Bernon, B.~Dumont and S.~Kraml,
  Phys.\ Rev.\ D {\bf 90} (2014) 071301
  [arXiv:1409.1588 [hep-ph]].
 
 \bibitem{Craig:2015jba}
  N.~Craig, F.~D'Eramo, P.~Draper, S.~Thomas and H.~Zhang,
  JHEP {\bf 1506} (2015) 137
  [arXiv:1504.04630 [hep-ph]].
  
 
 \bibitem{Bernon:2015qea}
  J.~Bernon, J.~F.~Gunion, H.~E.~Haber, Y.~Jiang and S.~Kraml,
  Phys.\ Rev.\ D {\bf 92} (2015) 7, 075004
  [arXiv:1507.00933 [hep-ph]]. 
  
  \bibitem{Dorsch:2016tab}
  G.~C.~Dorsch, S.~J.~Huber, K.~Mimasu and J.~M.~No,
  arXiv:1601.04545 [hep-ph].
  
 
 
\bibitem{Coleppa:2013xfa} 
  B.~Coleppa, F.~Kling and S.~Su,
  arXiv:1308.6201 [hep-ph].

\bibitem{Coleppa:2014hxa} 
  B.~Coleppa, F.~Kling and S.~Su,
  JHEP {\bf 1409}, 161 (2014)
  [arXiv:1404.1922 [hep-ph]].
  
  \bibitem{Dorsch:2014qja}
  G.~C.~Dorsch, S.~J.~Huber, K.~Mimasu and J.~M.~No,
  Phys.\ Rev.\ Lett.\  {\bf 113} (2014) 21,  211802
  [arXiv:1405.5537 [hep-ph]].
  
  
\bibitem{Coleppa:2014cca} 
  B.~Coleppa, F.~Kling and S.~Su,
  JHEP {\bf 1412}, 148 (2014)
  [arXiv:1408.4119 [hep-ph]].
  
  
 \bibitem{Li:2015lra}
  T.~Li and S.~Su,
  JHEP {\bf 1511} (2015) 068
  [arXiv:1504.04381 [hep-ph]]. 
  
  
\bibitem{Kling:2015uba} 
  F.~Kling, A.~Pyarelal and S.~Su,
  JHEP {\bf 1511}, 051 (2015)
  [arXiv:1504.06624 [hep-ph]].
  
  \bibitem{CMS:2015mba}
  CMS Collaboration [CMS Collaboration],
  CMS-PAS-HIG-15-001.
 
 \bibitem{Khachatryan:2016are}
  V.~Khachatryan {\it et al.} [CMS Collaboration],
  arXiv:1603.02991 [hep-ex].
 
\bibitem{Gunion:1989we} 
  J.~F.~Gunion, H.~E.~Haber, G.~L.~Kane and S.~Dawson,
  Front.\ Phys.\  {\bf 80}, 1 (2000).
 
 
 \bibitem{Bernon:2015wef}
  J.~Bernon, J.~F.~Gunion, H.~E.~Haber, Y.~Jiang and S.~Kraml,
  arXiv:1511.03682 [hep-ph].
 
 
  \bibitem{Gunion:2002zf} 
  J.~F.~Gunion and H.~E.~Haber,
  Phys.\ Rev.\ D {\bf 67}, 075019 (2003)
  [hep-ph/0207010].
 
 
 \bibitem{Abbiendi:2013hk}
  G.~Abbiendi {\it et al.} [ALEPH and DELPHI and L3 and OPAL and LEP Collaborations],
  Eur.\ Phys.\ J.\ C {\bf 73} (2013) 2463
  [arXiv:1301.6065 [hep-ex]].
 
 
 \bibitem{Bernon:2014nxa}
  J.~Bernon, J.~F.~Gunion, Y.~Jiang and S.~Kraml,
  Phys.\ Rev.\ D {\bf 91} (2015) 7,  075019
  [arXiv:1412.3385 [hep-ph]].
 
 
 \bibitem{Glashow:1976nt} 
  S.~L.~Glashow and S.~Weinberg,
  Phys.\ Rev.\ D {\bf 15}, 1958 (1977). 
 
  \bibitem{Branco:2011iw} 
  G.~C.~Branco, P.~M.~Ferreira, L.~Lavoura, M.~N.~Rebelo, M.~Sher and J.~P.~Silva,
  Phys.\ Rept.\  {\bf 516}, 1 (2012)
  [arXiv:1106.0034 [hep-ph]].  
 
 
 \bibitem{Aad:2015pla}
  G.~Aad {\it et al.} [ATLAS Collaboration],
  JHEP {\bf 1511} (2015) 206
  [arXiv:1509.00672 [hep-ex]].
  
  \bibitem{Ferreira:2014naa} 
  P.~M.~Ferreira, J.~F.~Gunion, H.~E.~Haber and R.~Santos,
  Phys.\ Rev.\ D {\bf 89}, no. 11, 115003 (2014)
  [arXiv:1403.4736 [hep-ph]].  
 
 
 \bibitem{Ginzburg:2005dt}
  I.~F.~Ginzburg and I.~P.~Ivanov,
  Phys.\ Rev.\ D {\bf 72} (2005) 115010
  [hep-ph/0508020].
 
 

\bibitem{Grinstein:2015rtl} 
  B.~Grinstein, C.~W.~Murphy and P.~Uttayarat,
  arXiv:1512.04567 [hep-ph].
 
 
 \bibitem{Baak:2014ora}
  M.~Baak {\it et al.} [Gfitter Group Collaboration],
  Eur.\ Phys.\ J.\ C {\bf 74} (2014) 3046
  [arXiv:1407.3792 [hep-ph]].
 
 \bibitem{Gorbahn:2015gxa}
  M.~Gorbahn, J.~M.~No and V.~Sanz,
  JHEP {\bf 1510} (2015) 036
  [arXiv:1502.07352 [hep-ph]].
 
 \bibitem{Amhis:2014hma}
  Y.~Amhis {\it et al.} [Heavy Flavor Averaging Group (HFAG) Collaboration],
  arXiv:1412.7515 [hep-ex].
 
\bibitem{Mahmoudi:2008tp}
  F.~Mahmoudi,
  Comput.\ Phys.\ Commun.\  {\bf 180} (2009) 1579
  [arXiv:0808.3144 [hep-ph]]. 
 
\bibitem{Mahmoudi:2009zz}
  F.~Mahmoudi,
  Comput.\ Phys.\ Commun.\  {\bf 180} (2009) 1718.
 
 
 \bibitem{Lees:2012xj}
  J.~P.~Lees {\it et al.} [BaBar Collaboration],
  Phys.\ Rev.\ Lett.\  {\bf 109} (2012) 101802
  [arXiv:1205.5442 [hep-ex]].
 
 \bibitem{Huschle:2015rga}
  M.~Huschle {\it et al.} [Belle Collaboration],
  Phys.\ Rev.\ D {\bf 92} (2015) no.7,  072014
  [arXiv:1507.03233 [hep-ex]].
  
  \bibitem{Abdesselam:2016cgx}
  A.~Abdesselam {\it et al.} [Belle Collaboration],
  arXiv:1603.06711 [hep-ex].
 
\bibitem{Misiak:2015xwa}
  M.~Misiak {\it et al.},
  Phys.\ Rev.\ Lett.\  {\bf 114} (2015) 22,  221801
  [arXiv:1503.01789 [hep-ph]]. 
  
  
\bibitem{Han:2013mga} 
  T.~Han, T.~Li, S.~Su and L.~T.~Wang,
  JHEP {\bf 1311}, 053 (2013)
  [arXiv:1306.3229 [hep-ph]].
 
 
 \bibitem{Schael:2006cr}
  S.~Schael {\it et al.} [ALEPH and DELPHI and L3 and OPAL and LEP Working Group for Higgs Boson Searches Collaborations],
  Eur.\ Phys.\ J.\ C {\bf 47} (2006) 547
  [hep-ex/0602042].
 
 \bibitem{Aad:2014vgg}
  G.~Aad {\it et al.} [ATLAS Collaboration],
  JHEP {\bf 1411} (2014) 056
  [arXiv:1409.6064 [hep-ex]].
  
  \bibitem{Khachatryan:2014wca}
  V.~Khachatryan {\it et al.} [CMS Collaboration],
  JHEP {\bf 1410} (2014) 160
  [arXiv:1408.3316 [hep-ex]].
 
 
 \bibitem{ATLAS:2013nma}
  [ATLAS Collaboration],
  ATLAS-CONF-2013-013.
 
 
 \bibitem{ATLAS:2014aga}
  G.~Aad {\it et al.} [ATLAS Collaboration],
  Phys.\ Rev.\ D {\bf 92} (2015) 1,  012006
  [arXiv:1412.2641 [hep-ex]].
 
 
 \bibitem{Khachatryan:2015cwa}
  V.~Khachatryan {\it et al.} [CMS Collaboration],
  JHEP {\bf 1510} (2015) 144
  [arXiv:1504.00936 [hep-ex]]. 
 
 
 \bibitem{Aad:2015wra}
  G.~Aad {\it et al.} [ATLAS Collaboration],
  Phys.\ Lett.\ B {\bf 744} (2015) 163
  [arXiv:1502.04478 [hep-ex]].
 
 \bibitem{Khachatryan:2015lba}
  V.~Khachatryan {\it et al.} [CMS Collaboration],
  Phys.\ Lett.\ B {\bf 748} (2015) 221
  [arXiv:1504.04710 [hep-ex]].
 
 \bibitem{Aad:2014yja}
  G.~Aad {\it et al.} [ATLAS Collaboration],
  Phys.\ Rev.\ Lett.\  {\bf 114} (2015) 8,  081802
  [arXiv:1406.5053 [hep-ex]].
 
 \bibitem{CMS:2014ipa}
  CMS Collaboration [CMS Collaboration],
  CMS-PAS-HIG-13-032.
 
 
 \bibitem{Khachatryan:2015yea}
  V.~Khachatryan {\it et al.} [CMS Collaboration],
  Phys.\ Lett.\ B {\bf 749} (2015) 560
  [arXiv:1503.04114 [hep-ex]].
 
 
 \bibitem{Aad:2014kga}
  G.~Aad {\it et al.} [ATLAS Collaboration],
  JHEP {\bf 1503} (2015) 088
  [arXiv:1412.6663 [hep-ex]].
 
 \bibitem{Khachatryan:2015qxa}
  V.~Khachatryan {\it et al.} [CMS Collaboration],
  JHEP {\bf 1511} (2015) 018
  [arXiv:1508.07774 [hep-ex]].

\bibitem{Das:2015mwa} 
  D.~Das and I.~Saha,
  Phys.\ Rev.\ D {\bf 91}, no. 9, 095024 (2015)
  [arXiv:1503.02135 [hep-ph]].
 
\bibitem{Das:2015qva} 
  D.~Das,
  Int.\ J.\ Mod.\ Phys.\ A {\bf 30}, no. 26, 1550158 (2015)
  [arXiv:1501.02610 [hep-ph]].
 
 
 \bibitem{Harlander:2012pb}
  R.~V.~Harlander, S.~Liebler and H.~Mantler,
  Comput.\ Phys.\ Commun.\  {\bf 184} (2013) 1605
  [arXiv:1212.3249 [hep-ph]].
 
 \bibitem{ChargedHiggsXS}
   M.~Flechl, R.~Klees, M.~Kramer, M.~Spira and M.~Ubiali,
   Phys.\ Rev.\ D {\bf 91}, no. 7, 075015 (2015)
   [arXiv:1409.5615 [hep-ph]];
   S.~Heinemeyer {\it et al.} [LHC Higgs Cross Section Working Group Collaboration],
   arXiv:1307.1347 [hep-ph];
   S.~Dittmaier, M.~Kramer, M.~Spira and M.~Walser,
   Phys.\ Rev.\ D {\bf 83}, 055005 (2011)
   [arXiv:0906.2648 [hep-ph]];
   E.~L.~Berger, T.~Han, J.~Jiang and T.~Plehn,
   Phys.\ Rev.\ D {\bf 71}, 115012 (2005)
   [hep-ph/0312286].
 

 \bibitem{Czakon:2011xx} 
   M.~Czakon and A.~Mitov,
   Comput.\ Phys.\ Commun.\  {\bf 185}, 2930 (2014)
   [arXiv:1112.5675 [hep-ph]].
   
   
 
 \bibitem{Eriksson:2009ws}
  D.~Eriksson, J.~Rathsman and O.~St\aa l,
  Comput.\ Phys.\ Commun.\  {\bf 181} (2010) 189
  [arXiv:0902.0851 [hep-ph]].  
   
   
 \bibitem{Haber:2015pua}
  H.~E.~Haber and O.~St\aa l,
  Eur.\ Phys.\ J.\ C {\bf 75} (2015) no.10,  491
  [arXiv:1507.04281 [hep-ph]].
 
\bibitem{Gao:2016ued} 
  C.~Gao, M.~A.~Luty, M.~Mulhearn, N.~A.~Neill and Z.~Wang,
  arXiv:1604.03108 [hep-ph].
  
\bibitem{Bauer:2015fxa} 
  M.~Bauer, M.~Carena and K.~Gemmler,
  JHEP {\bf 1511}, 016 (2015)
  [arXiv:1506.01719 [hep-ph]].
  
 
 
%
%
%
%
%
%
%
%
%
%
%
%
%
%
%
%
%
%
%
%
%
%
%
%
%
%
%
%
%
%
%
%
%
%
%


  \end{thebibliography}

\end{document}